\documentclass[prd,preprint,nofootinbib]{revtex4-1} 

\usepackage{latexsym,verbatim}
\usepackage{amssymb}
\usepackage{amsfonts}
\usepackage{amsmath,color}
\usepackage[draft]{graphicx}
\usepackage{bm}
\usepackage[utf8]{inputenc}
\usepackage[T1]{fontenc}

\usepackage{hyperref}

\hypersetup{
    colorlinks=true,
    linkcolor=red,
    citecolor=red,
    filecolor=magenta,      
    urlcolor=cyan,
    pdftitle={A tale  of  analogies...},
    pdfpagemode=FullScreen,}

\def\mb#1{\mathbf{#1}}

\def\ber{\begin{eqnarray}}
\def\eer{\end{eqnarray}}
\def\beq{\begin{equation}}
\def\eeq{\end{equation}}

\def\rmd{{\rm d}}

\def\ed{\end{document}}

\def\dtau#1{\frac{\mathrm{d} #1}{\mathrm{d}\tau}}
\def\dttau#1{\frac{\mathrm{d} ^{2}#1}{\mathrm{d}\tau^{2}}}

\def\bfbeta{\pmb{\beta}}
   \let\square=\dal
\newcommand{\ppar}[2]{\frac{\partial #1}{\partial #2}}

\makeatletter
\let\jnl@style=\rm
\def\ref@jnl#1{{\jnl@style#1}}

\def\aj{\ref@jnl{AJ}}                   
\def\actaa{\ref@jnl{Acta Astron.}}      
\def\araa{\ref@jnl{ARA\&A}}             
\def\apj{\ref@jnl{ApJ}}                 
\def\apjl{\ref@jnl{ApJ}}                
\def\apjs{\ref@jnl{ApJS}}               
\def\ao{\ref@jnl{Appl.~Opt.}}           
\def\apss{\ref@jnl{Ap\&SS}}             
\def\aap{\ref@jnl{A\&A}}                
\def\aapr{\ref@jnl{A\&A~Rev.}}          
\def\aaps{\ref@jnl{A\&AS}}              
\def\azh{\ref@jnl{AZh}}                 
\def\baas{\ref@jnl{BAAS}}               
\def\bac{\ref@jnl{Bull. astr. Inst. Czechosl.}}
\def\caa{\ref@jnl{Chinese Astron. Astrophys.}}
\def\cjaa{\ref@jnl{Chinese J. Astron. Astrophys.}}
\def\icarus{\ref@jnl{Icarus}}           
\def\jcap{\ref@jnl{J. Cosmology Astropart. Phys.}}
\def\jrasc{\ref@jnl{JRASC}}             
\def\memras{\ref@jnl{MmRAS}}            
\def\mnras{\ref@jnl{MNRAS}}             
\def\na{\ref@jnl{New A}}                
\def\nar{\ref@jnl{New A Rev.}}          
\def\pra{\ref@jnl{Phys.~Rev.~A}}        
\def\prb{\ref@jnl{Phys.~Rev.~B}}        
\def\prc{\ref@jnl{Phys.~Rev.~C}}        
\def\prd{\ref@jnl{Phys.~Rev.~D}}        
\def\pre{\ref@jnl{Phys.~Rev.~E}}        
\def\prl{\ref@jnl{Phys.~Rev.~Lett.}}    
\def\pasa{\ref@jnl{PASA}}               
\def\pasp{\ref@jnl{PASP}}               
\def\pasj{\ref@jnl{PASJ}}               
\def\rmxaa{\ref@jnl{Rev. Mexicana Astron. Astrofis.}}%
\def\qjras{\ref@jnl{QJRAS}}             
\def\skytel{\ref@jnl{S\&T}}             
\def\solphys{\ref@jnl{Sol.~Phys.}}      
\def\sovast{\ref@jnl{Soviet~Ast.}}      
\def\ssr{\ref@jnl{Space~Sci.~Rev.}}     
\def\zap{\ref@jnl{ZAp}}                 
\def\nat{\ref@jnl{Nature}}              
\def\iaucirc{\ref@jnl{IAU~Circ.}}       
\def\aplett{\ref@jnl{Astrophys.~Lett.}} 
\def\apspr{\ref@jnl{Astrophys.~Space~Phys.~Res.}}
\def\bain{\ref@jnl{Bull.~Astron.~Inst.~Netherlands}}
\def\fcp{\ref@jnl{Fund.~Cosmic~Phys.}}  
\def\gca{\ref@jnl{Geochim.~Cosmochim.~Acta}}   
\def\grl{\ref@jnl{Geophys.~Res.~Lett.}} 
\def\jcp{\ref@jnl{J.~Chem.~Phys.}}      
\def\jgr{\ref@jnl{J.~Geophys.~Res.}}    
\def\jqsrt{\ref@jnl{J.~Quant.~Spec.~Radiat.~Transf.}}
\def\memsai{\ref@jnl{Mem.~Soc.~Astron.~Italiana}}
\def\nphysa{\ref@jnl{Nucl.~Phys.~A}}   
\def\physrep{\ref@jnl{Phys.~Rep.}}   
\def\physscr{\ref@jnl{Phys.~Scr}}   
\def\planss{\ref@jnl{Planet.~Space~Sci.}}   
\def\procspie{\ref@jnl{Proc.~SPIE}}   

\makeatother

\begin{document}

\author{Matteo Luca Ruggiero}
\email{matteoluca.ruggiero@unito.it}
\affiliation{Dipartimento di Matematica ``G.Peano'', Universit\`a degli studi di Torino, Via Carlo Alberto 10, 10123 Torino, Italy}
\affiliation{INFN - LNL , Viale dell'Universit\`a 2, 35020 Legnaro (PD), Italy}

\author{Davide Astesiano}
\email{dastesiano@hi.is}
\affiliation{Science Institute, University of Iceland,
Dunhaga 3, 107 , Reykjav\'{\i}k, Iceland }

\date{\today}

\title{A tale  of  analogies: a review on gravitomagnetic effects, rotating sources, observers and all that}
\begin{abstract}
Gravitoelectromagnetic analogies are somewhat ubiquitous in General Relativity, and they are often used to explain peculiar effects of Einstein's theory of gravity in terms of familiar results from classical electromagnetism. Perhaps, the best known of these analogy pertains to the similarity between the equations of electromagnetism and those of the linearized theory of General Relativity. But the analogy is somewhat deeper and ultimately rooted in the splitting of spacetime, which is preliminary to the definition of the measurement process in General Relativity.
In this paper we review the various approaches that lead to the introduction of a magnetic-like part of the gravitational interaction, briefly called \textit{gravitomagnetic} and, then, we provide a survey of the recent  developments both from the theoretical and experimental viewpoints.
\end{abstract}

\maketitle

\section{Introduction}\label{sec:intro}

The close similarity between Newton's and Coulomb's laws prompted to investigate further analogies between electromagnetism and gravitation:  as reported by \citet{McDonald:1997fd}, J.C. Maxwell himself considered the possibility that gravity could be described by a vector field, but he was puzzled by the (negative) sign of  the energy of static gravitational configurations. A further step was the Maxwellian-like theory of gravity formulated by O.  Heaviside \cite{heaviside1894electromagnetic}, which has recently received interest for pedagogical purposes \cite{hermann}: in  this theory, the magnetic part of the gravitational field originates from mass currents. This analogy obtained a natural formulation in the framework of Relativity: in fact, since we know that magnetic interactions can be explained using electrostatics and Special Relativity (SR), it is expected that if we want to put toghether Newtonian gravity and Lorentz invariance, the presence of a magnetic-like component of the gravitational field or, for short, a \textit{gravitomagnetic field}, is mandatory. As reported by \citet{2014SPPhy.157..191P},  before completing General Relativity (GR), Einstein  investigated the existence of a gravitational analogue of electromagnetic induction \cite{einstein1912there,einstein2009collected} and, in addition, he suggested the presence of a Coriolis-like force inside a rotating spherical mass shell, which provokes \textit{dragging} effects on test masses. Gravitomagnetic effects where subsequently investigated in the framework of GR. In particular, it is relevant to emphasise that the analogy between Einstein's equations in the weak-field and slow-motion approximation and Maxwell's equations was formulated for the first time by H. Thirring \cite{2012GReGr..44.3225T,2012GReGr..44.3217P}. Afterwards, Thirring calculated the dragging effects inside a rotating mass shell and, in collaboration with J. Lense, they solved perturbatively the equation of motion of a test particle in the field of a rotating mass, from which  the so-called Lense-Thirring effect \cite{1984GReGr..16..711M} originated  (see \citet{2014SPPhy.157..191P} for a critical analysis of the content of these papers). However, it was soon clear that these effects are much smaller than the leading Newtonian gravitational ones, which can be called \textit{gravitoelectric} in the spirit of the analogy: indeed, in his letter to Thirring \cite{einstein2009_thirring}, Einstein affirmed that these effects ``remain far below any observable quantity''. This is a consequence of the fact that both the speeds of the sources and that of the test particles (except for massless ones) must be compared to the speed of light, so that the overall effect is very small when we are in linearised GR; on the other hand, more favourable conditions can be met when the gravitational field is strong, which happens for instance for pulsars, black holes and other extreme astrophysical situations.

Nonetheless, since the very beginning of the space exploration era,  proposals have been made to exploit space technologies to test these effects: in fact, the conceptual design of the Stanford Gravity Probe B (GPB) mission was proposed to NASA in 1961, but its launch took place only in 2004 and the final results were published in 2011 \cite{Everitt:2011hp}. 
In particular, the gravitomagnetic precession of the gyroscope axis was measured  by GPB with 19\% precision. More recently, I. Ciufolini \cite{1986PhRvL..56..278C} suggested to use laser-ranging to measure the modification of the satellites orbits determined by the Earth rotation and obtained initial  confirmation of the prevision of Einstein's theory with 10\% precision \cite{2004Natur.431..958C}, with subsequent improvements \cite{Ciufolini:2019ezb}. A comprehensive analysis of the attempt to measure the Lense-Thirring and related effects in the Solar System can be found in the review written by \citet{2011Ap&SS.331..351I}.

The purpose of this paper is to examine the recent developments in the study of the gravitomagnetic effects and in the search of  observational tests. Actually, in the literature there are already several reviews and monographs on this subject,  which we refer to for a comprehensive discussion (see e.g. \citet{ciufolini1995gravitation,Ruggiero:2002hz,mashhoon03,iorio2007measurement,2014SPPhy.157..191P}); as for us, we will focus on the findings and proposals published during the last twenty years, with the aim to provide an up-to-date reference for all researchers working in this field. We specify that we have limited ourselves to considering gravitomagnetic effects in GR: there are, however, investigations  also in gravitational theories that can be considered  as generalizations of Einstein's theory. Just to mention few of them, we refer to  the
Lorentz-violating Standard-Model Extension \cite{Bailey:2010af,Tasson:2012nx}, $f(R)$ and scalar-tensor theories \cite{Dass:2019kon}, non local gravity \cite{Mashhoon:2019jkq}, {Chern-Simons gravitational theory \cite{ref2.5}}.

The paper is organised as follows: in Section \ref{sec:theo} we review the basic theoretical foundations of the gravitomagnetic analogy, and its different meanings, then in Section \ref{sec:other} we resume recent theoretical progresses. Afterwords, we focus on the various proposals aimed at the measurements of gravitomagnetic effects:  historically, the first attempts were made in the space around the Earth or in the Solar System, which is still a lively scenario for these purposes, as we discuss in Section \ref{sec:solar}. The continuous technological improvements made it possibile to consider the feasibility of Earth-based experiments which are examined in Section \ref{sec:lab}. The exploration of deep space lends itself to analyse various astrophysical events that can be used to verify the predictions of GR and, in particular, for testing gravitomagnetic effects: this is the topic of Section \ref{sec:astro}. Another possibility is offered by analogue models, which arise in different contexts in physics: some of them, which are relevant for gravitomagnetic effects, are discussed in Section \ref{sec:analogue}.  The interplay between quantum phenomena and GR  is explored in Section \ref{sec:quantum}, while in Section \ref{sec:mach} we review recent interpretations of gravitmagnetic effects in view of the Mach principle.

\section{Basic Theoretical Framework}\label{sec:theo}

The term gravitomagnetism is probably due to Thorne \cite{membrane}, even though the analogy with electromagnetism had already been used by \citet{forward1961general} to refer to a formalism useful to deal with GR ``in experimentally realisable conditions'', that is to say when the gravitational field is weak and the speeds are much smaller than speed of light: actually, this is the well known \textit{gravitoelectromagnetic} (GEM) analogy between the linearised version of GR and electromagnetism, which is ultimately rooted in the ``space plus time'' splitting of electromagnetism in flat spacetime. 

Actually, as keenly observed by \citet{Jantzen:1992rg}, gravitoelectromagnetism has ``many faces'', since the description of GR effects in analogy with electromagnetism is somewhat ubiquitous. Here we briefly summarise the relevant features of these analogies, and refer to the works by \citet{Jantzen:1992rg,Jantzen:1996au,lynden,Bini:2002mh,de2010classical} for a thorough description of the approaches to spacetime splitting and the related gravitoelectromagnetic formalisms, and to the papers by \citet{natario,Costa:2021atq} for a comprehensive account of the various gravitelectromagnetic analogies. In addition, the role of rotating observers  is thoroughly  discussed in the monograph edited by \citet{rizzi2004relativity}.

In particular, we broadly follow the presentation of \citet{Costa:2021atq} to emphasize the different ``levels'' of the analogy which, starting from the potentials appearing in the spacetime metric, involves the fields, i.e. derivatives of the metric elements and, in the end, the tidal tensors, which are made of the derivatives of the fields.

Before starting, we believe that it is important to answer two important questions: why and to what extent gravitoelectromagnetism is useful?  The power of analogies in science has been known since the time of Kepler, who wrote:
\begin{quote}
 \textit{And I cherish more than anything else the Analogies, my most trustworthy masters. They know all the secrets of Nature, and they ought to be least neglected in Geometry.}
\end{quote}
In our case, analogies are useful and fruitful because they allow to get a better intuition of what happens in the 4-dimensional geometry of GR, with the possibility to understand new effects in terms of known ones: for instance, the celebrated Lense-Thirring precession is analogous to the precession of a magnetic dipole in the magnetic field \cite{ciufolini1995gravitation}. On the other hand,  it should be clear that there are no motivations to believe that gravitational interactions can be \textit{completely} described in analogy with electromagnetism, since there are deep differences between the two theories: indeed, as we are going to discuss, there are limitations in the analogies that should be taken into account.

The plan of this short outlook on gravitoelectromagnetic analogies is as follows: in Section \ref{ssec:split} we resume some basic ideas of spacetime splitting, which enables to show that relativistic dynamics, once that a class of observers has been defined, can be naturally described using a gravitoelectromagnetic analogy in full GR, without approximation; to introduce the analogy, we use the Sagnac effect, which can be understood in terms of the action of a gravitomagnetic potential on the motion of test particles (Section \ref{ssec:sagnac}). The well known gravitoelectromagnetic analogy arising in linearized GR (Section \ref{ssec:lin}) can be seen as a particular case of the full theory approach , when  a particular class of observers (those at rest at infinity) is considered. Eventually, in Section \ref{ssec:fermi} we focus on the gravitoelectromagnetic analogy that can be build using the components of the curvature tensor.

\subsection{Splitting spacetime}\label{ssec:split}

The idea to split spacetime into \textit{space plus time}  is fundamental in GR to get a better understanding, on the basis of our 3-dimensional experience, of what happens in the 4-dimensional geometry. To do this, in a spacetime manifold $\mathcal{V}_4$ we consider
a congruence $\Gamma $ of timelike (future-oriented) worldlines  that can be thought of as the four-velocity field of a set of test observers filling the spacetime and performing measurements: in this sense, it is possible to say that $\Gamma$ constitute the physical reference frame. {It seems therefore worthwhile examining the propertiers of the pair $(\mathcal{V}_4,\Gamma)$ on geometrical grounds.}

There are different approaches to spacetime splitting in the literature (see  \citet{Jantzen:1992rg,Jantzen:1996au,Bini:2002mh,de2010classical}); we use here {Cattaneo's approach \cite{ref2.1}}, which we appreciate for its clarity and physical-mathematical insight. {In particular, our introduction to Cattaneo's splitting is largely taken from the paper written by  \citet{rizzi2004relativistic} (see also references therein).}

Given a time-like congruence $\Gamma $ it is always
possible to choose a system of admissible
coordinates so that the lines $x^{0}=var$ coincide with the lines of $\Gamma $%
; in this case, such coordinates are said to be adapted
to the physical frame defined by the congruence $\Gamma $. In physical terms, the observers are at rest in this system of coordinates. This splitting approach allows to describe kinematics and dynamics in GR in ``relative'' terms, in the very spirit of Einsteinian approach, since
 they are expressed relatively to the reference frame $\Gamma$. { Let us remark that the entire approach can be applied as well to flat spacetime, when we consider a metric adapted to a congruence of non inertial observers.}

Let $\gamma^\mu$ be the components of the field of unit vectors
 tangent to the world-lines of the congruence
$\Gamma $ in adapted coordinates, parameterized by $\lambda$.  The spatial coordinates\footnote{Latin indices run from 1 to 3, and refer to space components, while Greek indices run from 0 to 3, and label spacetime components. The signature of spacetime is $(-1,1,1,1)$.} $x^i$ are constant
along the lines of $\Gamma$ and  $dx^i=0$ along any
line of $\Gamma$; the same holds for the components
$\displaystyle \gamma^{i}=\frac{dx^i}{d\lambda}$. The
component $\gamma^0$ directly comes from the condition $g_{\mu \nu
}\gamma ^{\mu }\gamma ^{\nu }=-1$:
\begin{equation}
\gamma ^{{0}}=\frac{1}{\sqrt{-g_{{00}}}}. \label{eq:gammazeroup}
\end{equation}
The covariant components can be found in the usual way
\begin{equation}
\gamma_0=g_{0\mu}\gamma^\mu=g_{00}\gamma^{0}=-\sqrt{-g_{00}},
\label{eq:gammazerodown}
\end{equation}
\begin{equation}
\gamma_{i}=g_{i\mu}\gamma^\mu=g_{i0}\gamma^0. \label{eq:gammai}
\end{equation}
Summarizing, we have
\begin{equation}
\left\{
\begin{array}{l}
\gamma ^{{0}}=\frac{1}{\sqrt{-g_{{00}}}}, \\
\gamma ^{i}=0,
\end{array}
\right. \;\;\;\;\;\;\;\;\;\;\;\;\left\{
\begin{array}{l}
\gamma _{0}=-\sqrt{-g_{{00}}}, \\
\gamma _{i}=g_{i{0}}\gamma ^{{0}}.
\end{array}
\right.  \label{eq:gammas11}
\end{equation}

The most general coordinates
transformation which does not change the physical frame, i.e. the
congruence $\Gamma $, has the form:
\begin{equation}
\left\{
\begin{array}{c}
x^{\prime }{}^{{0}}=x^{\prime }{}^{{0}}(x^{{0}},x^{1},x^{2},x^{3}), \\
x^{\prime }{}^{i}=x^{\prime }{}^{i}(x^{1},x^{2},x^{3}),
\end{array}
\right.  \label{eq:gauge_trans}
\end{equation}
with the additional condition $\partial x^{\prime 0}/\partial
x^{{0}}>0$, which  ensures that the  change of time
parameterization does not modify the arrow of time. The
coordinates transformation (\ref
{eq:gauge_trans}) is said to be {internal to the physical frame} $%
\Gamma $.\\

At each point $p$ in spacetime, the tangent space $T_{p}$ can
be split into the direct sum of two subspaces: $\Theta _{p}$,
spanned by $\gamma ^{\alpha }$, the {local
time direction}, and $\Sigma _{p}$, the
3-dimensional subspace which is orthogonal  to $\Theta_{p}$; $\Sigma _{p}$ is called {local
space platform}. Accordingly, the tangent space can be
written as the direct sum
\begin{equation}
T_{p}=\Theta _{p}\oplus \Sigma _{p}.  \label{eq:tangsum}
\end{equation}

Let $\{{e}_\mu\}$ be a basis of $T_p$. A vector ${v}=v^\mu {e}_\mu    \in T_{p}$ can be projected onto $\Theta _{p}$ and $%
\Sigma _{p}$ using the {time projector}
\begin{equation}
P_\Theta \doteq -\gamma _{\mu }\gamma _{\nu } \label{eq:ptheta1}
\end{equation}
and the {space projector}
\begin{equation}
P_\Sigma \doteq \gamma _{\mu \nu }\doteq g_{\mu \nu }+\gamma _{\mu
}\gamma _{\nu } \label{eq:psigma1}
\end{equation}
in the following way:
\begin{equation}
\left\{
\begin{array}{rclll}
\bar{v}_{\mu } & \equiv & P_{\Theta }\left( \,v_\mu\right) &
\doteq &
-\gamma _{\mu }\gamma _{\nu }v^{\nu }, \\
\widetilde{v}_{\mu } & \equiv & P_{\Sigma }\left( \,v_\mu\right) &
\doteq & \gamma _{\mu \nu }v^{\nu }=\left( g_{\mu \nu }+\gamma
_{\mu }\gamma _{\nu }\right) v^{\nu }= v_{\mu }+v^{\nu }\gamma
_{\nu }\gamma _{\mu }.
\end{array}
\right.  \label{eq:DefSplitComp}
\end{equation}

The subspaces $\Theta_p$ and $\Sigma_p$ are
defined by
\begin{equation}
\left\{
\begin{array}{ccl}
\Theta_p & \equiv & \left\{v^{\mu} \in T_p \ | \ \ \  v^{\mu}=\lambda \gamma^{\mu} \ \ \forall \lambda \in \mathbb{R}  \right\},    \\
\Sigma_p & \equiv & \left\{v^{\mu} \in T_p \ | \ \ \ g_{\mu\nu}v^\mu\gamma^\nu=0  \right\},
\end{array}
\right.  \label{eq:projecspacetime1}
\end{equation}
The projectors $P_\Theta,P_\Sigma$ define the mappings
\begin{equation}
P_\Theta: \ \ T_p \longrightarrow \Theta_p \ \ \ \ \ \ P_\Sigma: \
\ T_p \longrightarrow \Sigma_p. \label{eq:projectspacetime2}
\end{equation}

From Eq. (\ref{eq:DefSplitComp}), $\forall v^{\mu} \in T_p$ we have
\begin{equation}
v_\mu=\bar{v}_\mu+\widetilde{v}_\mu=
P_\Theta(v_\mu)+P_\Sigma(v_\mu). \label{eq:projectspacetime3}
\end{equation}
This defines the {natural splitting} of a vector
$v^{\mu}$. The superscripts $^{-},^{\sim }$ denote
respectively a {time vector} and a {space vector},
or more generally, a {time tensor }and {\ a space
tensor}, since the above described procedure can be applied to each tensor index.


To formulate the physical equations relative to the frame $%
\Gamma $, we need the  {transverse partial derivative} defined by
\begin{equation}
\tilde{\partial}_{\mu }\doteq \partial _{\mu }+\gamma _{\mu }\gamma ^{{0}%
}\partial _{{0}}.  \label{eq:dertras}
\end{equation}
By definition, it is a space vector: $\displaystyle \tilde{\partial}_0=\partial_0+\gamma_0 \gamma^0 \partial_0=0$, since $\gamma_0 \gamma^0=-1$. Accordingly, this operator can be used to define the transverse gradient:  in fact for a
generic scalar field $\varphi (x)$ we obtain:
\begin{equation}
P_{\Sigma }(\partial _{\mu }\varphi )=\tilde{\partial}_{\mu
}\varphi. \label{eq:protrans}
\end{equation}
{The metric $g$ and the fields
\begin{align}
    \gamma=\gamma^\mu \partial_{{\mu}}, \quad \omega^0=\gamma_\mu dx^\mu,
\end{align}
 are basic geometrical quantities associated to the pair $(\mathcal{V}_4,\Gamma)$. Using differential operators we can generate first order geometrical objects from these fields
\begin{align}
    C= -\mathcal{L}_\gamma(\omega^0),\quad \Omega=-2 d\omega^0, \quad K=\mathcal{L}_\gamma(g),
\end{align}
where $\mathcal{L}_V$ is the Lie derivative with respect to the field $V$. These new objects are  respectively called the curvature vector, the vortex tensor and the Killing tensor. Using the splitting procedure defined above we get 
\begin{align}
    C_\mu&= \Tilde{C}_\mu,\\
\Omega_{\mu\nu}&=\Tilde{\Omega}_{\mu\nu}+C_\mu \gamma_\nu-C_\nu \gamma_\mu, \\
K_{\mu\nu}&=\Tilde{K}_{\mu\nu}-\gamma_\mu C_\nu-\gamma_\nu C_\mu.
\end{align}
Using the adapted coordinates as before, defined by the condition $\gamma(x^i)=0$  where $i=1,2,3$, we find
\begin{eqnarray}
\widetilde{C}_\mu &=&\gamma ^{\nu }\nabla _{\nu }\gamma _{\mu },
\label{eq:cmu} \\
\widetilde{\Omega }_{\mu \nu } &=&\gamma _{0}\left[ \widetilde{\partial }%
_{\mu }\left( \frac{\gamma _{\nu }}{\gamma _{0}}\right)
-\widetilde{\partial }_{\nu }\left( \frac{\gamma _{\mu }}{\gamma
_{0}}\right) \right],
\label{eq:vortex} \\
\widetilde{K}_{\mu \nu } &=&\gamma ^{{0}}\partial _{{0}}\gamma
_{\mu \nu }. \label{eq:born}
\end{eqnarray}
$\widetilde{C }_\mu$ is the {curvature vector\footnote{Where
\[
\widetilde{C }_\mu= \gamma^\nu \nabla_\nu
\gamma_\mu=\frac{dx^\nu}{ds} \frac{D \gamma_\mu}{dx^\nu}=\frac{D
\gamma_\mu}{ds}.
\]}} of the curve $x^0=var$ of the
congruence. When this vector is null, the curve is geodesic. If
this is true for all curves of the congruence $\Gamma$, the frame
is {freely falling} (and the congruence
is said to be {geodesic}). $\widetilde{\Omega }%
_{\mu \nu }$ is the {space vortex tensor}
 which gives the
local angular velocity of the reference fluid.  When this tensor is null,
the frame is said to be {non rotating} or
\textit{time-orthogonal}; for our purposes, it is relevant to point out that this tensor is not null when $g_{0i} \neq 0$. Actually, it is possible to show \cite{rindlerperlick} that this tensor is simply related to velocity of rotation of the particle of the congruence relative to a Fermi-Walker frame, which is the standard for a non rotating frame in GR: consequently, when the vortex tensor is null, the coordinates are adapted to a non rotating frame.
Eventually, 
$\widetilde{K}_{\mu \nu }$ is the {Born space tensor},
which gives the deformation rate of the reference fluid. {This tensor provides the rate of deformation for the reference fluid: When this tensor equals zero, the frame is described as ``rigid''. Specifically, the Born tensor is zero when the metric components do not exhibit any time dependence.}
 {The natural decomposition of the covariant vector field $\gamma$ becomes
\begin{align}
    \nabla_\mu\gamma_\nu=\frac{1}{2} \left(\Tilde{\Omega}_{\mu\nu}+\tilde{K}_{\mu\nu}\right)-\gamma_\mu C_\nu.
\end{align}
}

{We need some definitions to explain  the relative formulation of kinematics
and dynamics.} To begin with, let us
consider two infinitesimally close events in spacetime, whose
coordinates are $x^\alpha$ and $x^\alpha+dx^\alpha$. We can
introduce the  definitions of the {standard relative time
interval}
\begin{equation}
dT=-\frac{1}{c}\gamma_\mu dx^\mu, \label{eq:dTdef}
\end{equation}
and the {standard relative space element}
\begin{equation}
d\sigma^2=\gamma_{\alpha\beta}dx^\alpha dx^\beta \equiv
\gamma_{ij}dx^idx^j.\label{eq:dsigmadef}
\end{equation}

{These quantities are clearly reliant on the physical frame established by the vector field ${\gamma}^{\mu}(x)$ and establish the temporal and spatial intervals as perceived by an observer within the congruence. As we will demonstrate, they have a pivotal role in the standard relative formulation of a particle's kinematics and dynamics in either an inertial or gravitational field. By its very definition as indicated in Eq. (\ref{eq:dTdef}), the standard relative time interval is found to be an invariant within spacetime.} It represents the projection of the  vector $dx^\mu$
along the vector of the congruence $\gamma^{\mu}$. By using  Eqs. (\ref{eq:dTdef}) and (\ref{eq:dsigmadef}) the spacetime invariant $ds^2$ can be written in the
form
\begin{equation}
ds^2=d\sigma^2-c^2dT^2. \label{eq:dsdsgimadT}
\end{equation}

Now, we are in position to define and describe the motion as seen by the observers in the congruence: {as before, we follow the approach outlined by \citet{rizzi2004relativistic}}. A point particle
$P$ is at rest in $\Gamma$ if its
world-line coincides with one of the lines of the congruence.  On the contrary, when the world-line of $P$ does not
coincide with any of the lines of $\Gamma$, the particle is said to
be in motion: in this case, since  $dx^i \neq 0$, we
can write the parametric equation of the world-line of $P$ in
terms of a parameter $\lambda$, $x^\alpha=x^\alpha(\lambda)$. Let $dP$ denote the infinitesimal displacement of the particle $P$. It
is either time-like or light-like and in both cases $dT\neq 0$, so
we can express the world-coordinates of the moving particle using
the standard relative time as a parameter:
$x^\alpha=x^\alpha(T)$. 

{The expression
\[
v^\alpha=\frac{dx^\alpha}{dT}
\]
defines the relative 4-velocity of
a particle in motion with respect to the physical frame $\Gamma$. Accordingly,  we call {standard relative velocity} its space
projection}
\begin{equation}
\widetilde{v}_\beta \doteq P_\Sigma
(v_\beta)=\gamma_{\beta\alpha}\frac{dx^\alpha}{dT}=\gamma_{\beta
i}\frac{dx^i}{dT}. \label{eq:relvel1}
\end{equation}
Since $\widetilde{v}_\beta \in \Sigma_p$, then
$\widetilde{v}_0=0$. The contravariant components of the standard
relative velocity are
\begin{equation}
\widetilde{v}^i=\frac{dx^i}{dT} \ \ \ \widetilde{v}^0=-\gamma_i
\frac{\widetilde{v}^i}{\gamma_0}, \label{eq:relvel2}
\end{equation}
(because $\gamma_\alpha \widetilde{v}^\alpha=0$). As a consequence, Eq. (\ref{eq:relvel1}) can be written as
\begin{equation}
\widetilde{v}_i=\gamma_{ij}\widetilde{v}^j. \label{eq:relvel3}
\end{equation}
The (space) norm of the standard relative velocity is
\begin{equation}
\left\| \,v^{\mu}\,\right\| _{\Sigma }\doteq
\widetilde{v}^2=\gamma_{ij}\widetilde{v}^i{}\widetilde{v}^j=\frac{d\sigma^2}{dT^2}.
\label{eq:relvel4}
\end{equation}
For a photon, since $ds^2=0$, we get from eq.
(\ref{eq:dsdsgimadT}) $\widetilde{v}^2=c^2$, which is the same
result obtained SR: this is not surprising, since locally GR reduces to SR.  {When we consider material particles}, we can introduce the proper time
\begin{equation}
d\tau^2=-\frac{1}{c^2}ds^2, \label{eq:tproptime1}
\end{equation}
and, using Eqs. (\ref{eq:dTdef}) and  (\ref{eq:relvel4}) we may write
\begin{equation}
\frac{dT}{d\tau}=\frac{1}{\sqrt{1-\frac{\widetilde{v}^2}{c^2}}}
\label{eq:gammal2}
\end{equation}
{Again, we notice that we obtain a result that is formally identical to the special relativistic one} in terms of the Lorentz factor $\displaystyle \frac{1}{\sqrt{1-\frac{\widetilde{v}^2}{c^2}}}$,  in which the relative velocity appears.

After introducing the quantities that enable to describe relative kinematics, we can pass to relative dynamics: in other words, we need to formulate the geodesic equations in relative terms. To this end, let us consider the
connection coefficients, in the coordinates $\{x^\mu\}$ adapted to
the physical frame, $\Gamma^{\alpha}_{\ \beta\gamma}$.
The geodesic equations are written as
\begin{equation}
\frac{dU^\alpha}{d\tau}+\Gamma^\alpha_{\ \beta\gamma}U^\beta
U^\gamma=0, \label{eq:geo1}
\end{equation}
in terms of the 4-velocity
\begin{displaymath}
U^\alpha=\frac{dx^\alpha}{d\tau}
\end{displaymath}
and the proper time $\tau$. Let $m_0$ be the proper mass of the particle: then the
energy-momentum 4-vector is
\begin{displaymath}
P^{\alpha}=m_0U^{\alpha}=m_0 \frac{dx^\alpha}{d\tau}
\end{displaymath}


Now we want to re-formulate the geodesic equations in their relative
form, i.e. by means of the standard relative quantities that we
have introduced so far. To this end, let us define the
{standard relative momentum}
\begin{equation}
\widetilde{p}_\alpha \doteq P_\Sigma \left( P_\alpha
\right)=\gamma_{\alpha\beta}P^\beta=m_0\gamma_{\alpha
 i }\frac{dx^i}{dT}\frac{dT}{d\tau}=m\widetilde{v}_\alpha,
\label{eq:relmom1}
\end{equation}
where we introduced the {standard relative mass}
\begin{equation}
m\doteq\frac{m_0}{\sqrt{1-\frac{\widetilde{v}^2}{c^2}}}
\label{eq:relmass}
\end{equation}
in formal analogy with  SR. Since
$\widetilde{p}_\alpha \in \Sigma_p$, then $\widetilde{p}_0=0$. If we deal with massless particles, we can proceed as follows to
formally define the momentum 4-vector: consider, for instance, a
"classical", i.e. perfectly localized monochromatic photon, we can
set
\begin{equation}
\lim_{m_0 \rightarrow 0} \left(m_0 \frac{dT}{d\tau}
\right)=\frac{h\nu}{c^2} \label{eq:4mom2}
\end{equation}
and obtain
\begin{equation}
P^\alpha=\frac{h\nu}{c^2}\frac{dx^\alpha}{dT}
\label{eq:palphafotone}
\end{equation}
where $h$ is the Planck constant and, in terms of relative
quantities, the relation that links the  wavelength and the
frequency of the photon to the speed of light is
$\lambda\nu=\frac{d\sigma}{dT}=c$.

It is possible to show (see again, \citet{rizzi2004relativistic} and references therein) that  the space projection of the geodesic equations can be written as
\begin{equation}
\frac{\hat{D}\widetilde{p}_i}{dT}=m\widetilde{G}_i
\label{eq:geospace2}
\end{equation}
where $\frac{\hat{D}}{dT}$ is a suitably defined derivative operator. In other words, the variation of the space momentum
vector is determined by the field $\widetilde{G}_i$:
\begin{equation}
\widetilde{G}_i=-c^2\widetilde{C}_i+c\widetilde{\Omega}_{ij}\widetilde{v}^j
\label{eq:gi}
\end{equation}
It is often useful to split the field $\widetilde{G}_i$ into the
sum of two fields $\widetilde{G'_i},\widetilde{G''_i}$, defined as
follows:
\begin{eqnarray}
\widetilde{G'}_i    & \doteq &
-c^2\widetilde{C}_i=-c^2\left[\gamma_0 \tilde{\partial}_i
\gamma^0-\partial_0\left(\frac{\gamma_i}{\gamma_0}\right) \right]  \nonumber \\
\widetilde{G''}_i   & \doteq & c \widetilde{\Omega}_{ij}
\widetilde{v}^j \label{eq:gprimogsecondo}
\end{eqnarray}
The field $\widetilde{G'}_i$ can be interpreted as a dragging
field ($c^2C_\alpha$ is the 4-acceleration $a_\alpha$ of the
particles of the reference fluid) and the field
$\widetilde{G''}_i$ can be interpreted as a Coriolis-like field.

The contravariant form of eq.
(\ref{eq:geospace2}) (see also (\ref{eq:gi})) turns out to
be:
\begin{equation}
\frac{\hat{D}\widetilde{p}^i}{dT}=-mc^2 \widetilde{C}^i+mc\left(
\widetilde{\Omega}^{i}_{\ j} -\widetilde{K}^i_{\
j}\right)\widetilde{v}^j. \label{eq:geospacecontro}
\end{equation}

The above description of relative dynamics holds true in full GR and it can be explained in terms of  $\widetilde{G}_i$ which, in turn, can be seen as the  sum of the dragging field  $\widetilde{G'}_i$, related to the  curvature of the congruence, and the velocity dependent field  $\widetilde{G''}_i$, related to  vorticity of the congruence. It is relevant to point out that only the field $\widetilde{G}_i$ is invariant with respect to the internal transformations (\ref{eq:gauge_trans}).  This formalism lends itself to introduce an analogy with the electromagnetic dynamics: in other words, the velocity dependent field can be interpreted as a magnetic-like force, while the dragging term can be interpreted as an electric-like force, whence the gravitoelectromagnetic analogy. In order to see how this can be done, in next Section we discuss the Sagnac effect \cite{post}  which, as we are going to show, can be interpreted as the consequence of the action of a gravitomagnetic potential, in analogy with the well known Aharonov-Bohm effect \cite{PhysRev.115.485}.

\subsection{The Sagnac effect and the emerging gravitomagnetic analogy} \label{ssec:sagnac}

We are in position to see how it is possible to describe some known relativistic effects in spacetime, on the basis of a gravitoelectromagnetic analogy deriving from the splitting approach described above. To this end, we consider the line-element in the form
\beq
ds^{2}=g_{00}c^{2}dt^{2}+2g_{0i}cdtdx^{i}+g_{ij}dx^{i}dx^{j}, \label{eq:metricastazionaria}
\eeq
and we suppose that the metric elements do not depend on time. The above metric is quite general in its form and, in particular,  it said to be  {non time-orthogonal}, because $g_{0i} \neq 0$.  We focus on the asymmetry in the propagation times of two signals  in the spacetime described by the line-element (\ref{eq:metricastazionaria}); this asymmetry is the so-called {Sagnac time delay}, {which we describe here  with extensive reference to the papers by \citet{2003GReGr..35.2129R,rizzi2004relativistic,ruggiero2015note,tartaglia_entropy}.}

Let us consider two massive or massless particles simultaneously emitted at a given location:  they propagate in opposite directions along the same path and reach the emission point at different times. {It is possible to demonstrate that for both massless particles and massive particles moving at equal speeds in opposite directions, the variation of their coordinate propagation time can be expressed as follows:}
\beq
\Delta t = \frac 2 c \oint_{\ell} \frac{g_{0i}dx^{i}}{|g_{00}|}=-\frac 2 c \oint_{\ell} \frac{g_{0i}dx^{i}}{g_{00}}. \label{eq:formulafond}
\eeq
{For instance, let us  suppose that spacetime is axially symmetric: accordingly, the metric (\ref{eq:metricastazionaria}) can be written in adapted (cylindrical) coordinates in the form}
\beq
ds^{2}=g_{00}c^{2}dt^{2}+2g_{0\varphi}cdt d\varphi+g_{rr}dr^{2}+g_{\varphi\varphi}d\varphi^2+g_{zz}dz^{2} \label{eq:metricaxis}
\eeq
In particular, the  line element in a uniformly rotating frame of reference in flat Minkowski spacetime is written  in this form
\beq
ds^{2}=-\left(1-\frac{\omega^{2}r^{2}}{c^{2}} \right)c^{2}dt^{2}+2\frac{\omega r^{2}}{c}cdtd\varphi +dr^{2}+r^{2}d\varphi^{2}+dz^{2}, \label{eq:rot0}
\eeq
where $\omega$ is the constant rotation rate. From the above metric, it is possible to derive the classical effect that was first pointed out by Sagnac and is now currently used in laser gyros \cite{lasergyro}: in particular for a circular path of radius $R$ we get the following proper time difference
\begin{equation}
\Delta \tau =
\frac{4\pi R^2 \omega}{c^2}\left( 1-\frac{\omega ^{2}R^{2}}{c^{2}}
\right)^{-1/2} \label{eq:deltatau}
\end{equation} 
(see e.g. the monograph edited by \citet{rizzi2004relativity}).  

If we start from the metric (\ref{eq:metricaxis}), perform the (local) coordinates transformation $\varphi'=\varphi-\Omega t$, and set, for given $r_{0},z_{0}$, $\displaystyle \Omega=-\frac{c g_{0\varphi}}{g_{\varphi\varphi}}$, then in the new metric $g'_{\varphi t}=0$  and the observers at $r=r_{0},z=z_{0}$ do not experience any Sagnac effect. These observers are called Zero Angular Momentum Observers (ZAMO), so we see that they are not rotating with respect to the local spacetime geometry: differently speaking, their vorticity (\ref{eq:vortex}) is zero. It is easy to understand why they are called ZAMO: in fact, from the metric (\ref{eq:metricaxis}) we can write the Lagrangian 
\beq
\mathcal L = \frac 1 2 \left(g_{00}c^{2}\dot t^{2}+2g_{0\varphi}c \dot t \dot \varphi+ g_{rr}\dot r^{2}+g_{\varphi\varphi}\dot \varphi^{2}+g_{zz}\dot z^{2} \right), \label{eq:lagr1}
\eeq
where dot means derivation with respect to an affine parameter $\lambda$.  The angular momentum is given by
\beq
p_{\varphi}=\frac{\partial \mathcal L}{\partial \dot \varphi}=cg_{0\varphi}\dot t+g_{\varphi\varphi} \dot \varphi. \label{eq:pphi}
\eeq
On setting  $\Omega=\frac{d\varphi}{dt}$, we see that when $\displaystyle \Omega=-\frac{cg_{0\varphi}}{g_{\varphi\varphi}}$ the angular momentum is null.

If the metric (\ref{eq:metricaxis}) is  inertial (i.e. Minkowski) at infinity, we see that the observers at rest  have a non zero angular momentum and measure a Sagnac effect; on the other hand, the ZAMO are moving in the metric (\ref{eq:metricaxis}), but measure no Sagnac effect. This the so-called frame-dragging, which means that the ZAMO are dragged by the spacetime metric, and this effect depends on the non diagonal elements of the metric, hence on its vorticity (see also the discussion in Section \ref{sec:mach}).

The spacetime metric (\ref{eq:metricastazionaria}) can be written, using the standard relative time element and the standard relative space element in the form (\ref{eq:dsdsgimadT}); in particular, we may write  
\beq
dT=-\frac 1 c \gamma_{\alpha}dx^{\alpha}=-\frac 1 c \gamma_{0}\left(dx^{0}+\frac{\gamma_{i}}{\gamma_{0}}dx^{i} \right)=\frac 1  c \sqrt{-g_{00}}\left(dx^{0}+\frac{g_{0i}}{g_{00}}dx^{i} \right). \label{eq:dTrev}
\eeq
So, if we set
\beq
A_{i}^{G}=c^{2}\frac{\gamma_{i}}{\gamma_{0}}=c^{2}\frac{g_{0i}}{g_{00}}, \label{eq:defAG0}
\eeq
the Sagnac effect (\ref{eq:formulafond}) can be written in the form
\beq
\Delta t =-\frac{2}{c^{3}}  \oint_{\ell} A_{i}^{G}dx^{i}. \label{eq:formulafond1}
\eeq
As a consequence, the Sagnac effect can be interpreted as a gravitomagnetic Aharonov-Bohm \cite{Ruggiero:2005nd,Ruggiero:2004iaq} effect, determined by the gravitomagnetic potential $A_{i}^{G}$. In addition, we may formally introduce the gravitoelectric potential 
\beq
\phi^{G}=-c^{2}\gamma^{0}. \label{eq:defphiG}
\eeq
The use of this terminology, can be justified considering the equation of motion (\ref{eq:geospace2}). In fact, it is possible to define the \textit{gravitomagnetic field}\footnote{Here and henceforth boldface symbols  refer to space vectors.}
\begin{equation}
\widetilde{B}_{G}^{i}\doteq \left( \widetilde{{\bf \nabla }}\times
\bm{\tilde{A}}_{G}\right) ^{i}  \label{eq:gengravmag}
\end{equation}
and  the \textit{gravitoelectric field}:
\begin{equation}
\widetilde{E}^{G}_i\doteq -\left( -\widetilde{\partial} _{i}\phi
_{G}-\partial _{0}\widetilde{A}^{G}_i\right).  \label{eq:egen}
\end{equation}
Then, the equation of motion (\ref{eq:geospace2}) can be written
in the form
\begin{equation}
\frac{\hat{D}\widetilde{p}_{i}}{dT}=m
\widetilde{E}^{G}_i+m\gamma _{0}\left(
\frac{\bm{\tilde{v}}}{c}\times \bm{\tilde{B}}_{G}\right) _{i}
\label{eq:motogen1}
\end{equation}
which looks like the equation of motion of a particle acted upon by a generalized Lorentz force.

Consequently, a gravitoelectromagnetic analogy naturally emerges in  full GR when we are dealing with a non time-orthogonal metric like (\ref{eq:metricastazionaria}). However, the great majority of experimental studies on GR is done in weak-field conditions, i.e. when the magnitude of the gravitational field allows a linearization of the relevant equations: this will be discussed in the next Section.

\subsection{Gravitoelectromagnetic analogy in linearized General Relativity} \label{ssec:lin}

Einstein equations
\beq
G_{\mu\nu}=\frac{8\pi G}{c^{4}}T_{\mu\nu}. \label{eq:Einstein0}
\eeq
can be solved in the weak-field and slow-motion approximation: in this case the gravitational field can be considered as a perturbation of flat spacetime, described by the Minkowski tensor $\eta_{\mu\nu}$. {Here and henceforth,  we closely follow the approach given by \citet{Ruggiero:2021uag} (see also references therein) to the solution of Einstein equations in this approximation.}

As a consequence, the metric tensor can be written in the form $g_{\mu\nu}=\eta_{\mu\nu}+h_{\mu\nu}$, where $h_{\mu\nu}$ is a perturbation:  {$|h_{\mu\nu}|\ll |\eta_{\mu\nu}|$}. If we introduce $\bar h_{\mu\nu}=h_{\mu\nu}-\frac 1 2 \eta_{\mu\nu}h$, where $h=h_{\mu}^{\ \mu}$ and perform a linear approximation, Einstein equations (\ref{eq:Einstein0}) become (see e.g. \citet{straumann2013applications})
\beq
-\square{ \bar h}_{\mu\nu}-\eta_{\mu\nu}\bar h_{\alpha\beta}^{\ \ \ ,\alpha\beta}+\bar h_{\mu\alpha, \nu}^{\ \ \ \ \ \alpha}+\bar h_{\nu\alpha, \mu}^{\ \ \ \ \ \alpha}=\frac{16\pi G}{c^{4}}T_{\mu\nu}.  \label{eq:Einstein1}
\eeq
The gauge freedom can be exploited setting   the \textit{Hilbert gauge condition}
\beq
\bar h^{\mu\nu}_{\ \ ,\nu}=0. \label{eq:Hilibert}
\eeq
 Then,  from (\ref{eq:Einstein1}) we get
\beq
\square{ \bar h}_{\mu\nu}=-\frac{16\pi G}{c^{4}}T_{\mu\nu}.  \label{eq:Einsteinweak}
\eeq
{We remark that condition (58) can always be met through a gauge transformation. In fact, the Einstein equations remain invariant under infinitesimal transformations of this kind:
\beq
h_{\mu\nu} \rightarrow h_{\mu\nu}+\xi_{\mu,\nu}+\xi_{\nu,\mu}, \label{eq:gauge1}
\eeq
which, in terms of $\bar h_{\mu\nu}$ becomes
\beq
\bar h_{\mu\nu} \rightarrow \bar h_{\mu\nu}+\xi_{\mu,\nu}+\xi_{\nu,\mu}-\eta_{\mu\nu}\xi^{\alpha}_{\ ,\alpha}. \label{eq:gauge2}
\eeq
As a consequence, when  $\bar h^{\mu\nu}_{\ \ ,\nu}\neq0$, we may choose $\xi^{\mu}$ to be a solution of $\square \xi^{\mu}=-\bar h^{\mu\nu}_{ \ \ ,\nu}$.}

{Eqs. (\ref{eq:Einsteinweak})  exhibit a  manifest resemblance to Maxwell's equations for the electromagnetic four-potential. Consequently, they can be approached in a similar manner. Specifically, by disregarding the solution to the homogeneous wave equation, the general solution can be expressed using the concept of retarded potentials:
\beq
 {\bar h}_{\mu\nu}=\frac{4G}{c^4}\int_{V} \frac{T_{\mu\nu}(ct-|{\mathbf x}-{\mathbf x}'|, {\mathbf x}')}{|{\mathbf x}-{\mathbf x}'|}\rmd^3 x'\ , \label{eq:solgem1}
\eeq
In the above equation  integration is extended to the volume $V$, containing the source. The components of the energy-momentum tensors are defined as  $T^{00}=\rho_{  } c^2$ and $T^{0i}=cj_{  }^i$, in terms of the mass density $\rho_{  }$ and mass current $j_{  }^{i}$ of the source; as a consequence,   $j_{  }^\mu=\left(c\rho_{  },j_{g}^{i} \right)=\left(c\rho_{  },{\mathbf j_{  }}\right)$ is the mass-current four vector  of the source. In linear approximation, $\displaystyle T^{\mu\nu}_{\ \ ,\nu}=0$, so we obtain the continuity equation 
\beq
\ppar{\rho}{t}+\bm \nabla \cdot \mb j=0. \label{eq:continuity}
\eeq
To fix ideas, let us assume that the source consists of a finite distribution of slowly moving matter, with velocity $\mb v$ such that  $|\mb v|\ll c$; consequently, we obtain that  $T_{ij} \simeq \rho v_{i}v_{j}+p\delta_{ij}$, where $p$ is the pressure. In particular, from Eq. (\ref{eq:solgem1}) we see that $\bar h_{ij} =O(c^{-4})$:  as a result, in this linear  approach, we may neglect in the metric tensor terms that are $O(c^{-4})$.\\
In summary, the solution  of Eq. (\ref{eq:solgem1}) can be written in the form
\beq
{\bar h}_{00}=\frac{4G}{c^{2}}\int_{V} \frac{\rho(ct-|{\mathbf x}-{\mathbf x}'|, {\mathbf x}')}{|{\mathbf x}-{\mathbf x}'|}\rmd^3 x'\ , \label{eq:solgemh00}
\eeq
\beq
{\bar h}_{0i}=-\frac{4G}{c^{3}}\int_{V} \frac{j^{i}(ct-|{\mathbf x}-{\mathbf x}'|, {\mathbf x}')}{|{\mathbf x}-{\mathbf x}'|}\rmd^3 x'\ . \label{eq:solgemh0i}
\eeq
The other components of $\bar h_{\mu\nu}$ are zero at the given approximation level.\\
If we exploit the already mentioned analogy with electromagnetism,  we may  introduce  the \textit{gravitoelectromagnetic potentials}: namely, the gravitoelectric  $\Phi_{  }$  and gravitomagnetic $A^{i}_{  }$ {potentials},  defined by
\beq
\bar h_{00} \doteq 4\frac{\Phi_{  }}{c^{2}}, \quad \bar h_{0i}=-2 \frac{A_{i}}{c^{2}}, \label{eq:defphiAigem}
\eeq
Taking into account Eqs. (\ref{eq:solgemh00}) and (\ref{eq:solgemh0i}), their expressions turn out to be
\beq
\Phi ={G}\int_{V} \frac{\rho(ct-|{\mathbf x}-{\mathbf x}'|, {\mathbf x}')}{|{\mathbf x}-{\mathbf x}'|}\rmd^3 x'\ , \label{eq:solgemphi1}
\eeq
\beq
A_{i}=\frac{2G}{c}\int_{V} \frac{j^{i}(ct-|{\mathbf x}-{\mathbf x}'|, {\mathbf x}')}{|{\mathbf x}-{\mathbf x}'|}\rmd^3 x'\ . \label{eq:solgemAi1}
\eeq
In the end, the spacetime metric that characterizes the solutions of Einstein's equation in the weak-field approximation takes on the following form:
\beq
\mathrm{d} s^2= -c^2 \left(1-2\frac{\Phi}{c^2}\right)\rmd t^2 -\frac4c A_{i}\rmd x^{i}\rmd t +
 \left(1+2\frac{\Phi}{c^2}\right)\delta_{ij}\rmd x^i \rmd x^j. \  \label{eq:weakfieldmetric1}
\eeq 
\vspace{0.2cm}}

It is useful to see the link between this \textit{linear} gravitoelectromagnetic analogy and the one described before and valid in full theory. In the above metric, the weak-field approximation means that $\displaystyle \left|\frac{\Phi}{c^{2}}\right| \ll 1$, $\displaystyle \left| \frac{A_{i}}{c^{2}}\right|\ll 1$. In particular, we have $\displaystyle \frac{\Phi}{c^{2}}=\frac{1+g_{00}}{2}$ and $\displaystyle \frac{A_{i}}{c^{2}}=-\frac{g_{0i}}{2}$. Now, if we consider the definition (\ref{eq:defphiG}),  in the weak-field approximation, we have
\beq
\phi^{G}=-c^{2}\gamma^{0}=-c^{2}\frac{1}{\sqrt{-g_{00}}}=-c^{2}\frac{1}{\sqrt{1-2\frac{\Phi}{c^{2}}}} \simeq \Phi  \label{eq:wfPhi}
\eeq
Similarly, we have
\beq
A_{i}^{G}=c^{2}\frac{\gamma_{i}}{\gamma_{0}}=c^{2}\frac{g_{0i}}{g_{00}} \simeq -c^{2}g_{0i} = 2 A_{i} \label{eq:wfA}
\eeq
As a consequence, apart from a 2 factor in the definition of gravitomagnetic effects, we obtain a correspondence between the two definitions.

Let us focus on how this analogy translates into the expression of the geodesic equations.
{Let us start from the  line element (\ref{eq:weakfieldmetric1}) and calculate the geodesic equations up to linear order in $\bfbeta={\mathbf v}/c$. From
\beq
\dttau x^{\mu}+\Gamma^{\mu}_{\alpha\beta} \dtau{x^{\alpha}} \dtau{x^{\beta}}=0,  \label{eq:geott1}
\eeq
we obtain for the space components \cite{natario,Bini:2008cy}:
\beq
\frac{\rmd v^i}{\rmd t}=\frac{\partial \Phi}{\partial x^{i}}-2({\pmb \beta}\times {\mathbf B})_i+2\frac{\partial A_{i}}{c\partial t}-3\beta^i \frac{\partial \Phi }{c\partial t}.\ \label{eq:geonew}
\eeq
Then, if we define the gravitoelectromagnetic fields as
\beq
\mb B= \bm \nabla \wedge \mb A, \quad \mb E= -\bm \nabla \Phi-\frac{2}{c} \ppar{\mb A}{t}, \label{eq:defEtime}
\eeq
the above equation (\ref{eq:geonew}) becomes
 {\beq
\frac{\rmd v^i}{\rmd t}=-E^{i}-2({\pmb \beta}\times {\mathbf B})_i-3\beta^i \frac{\partial \Phi }{c\partial t}. \label{eq:lor001}
\eeq}
Accordingly, it is not warranted that the geodesic equations take a Lorentz-like form,   due to the presence of the last term in Eq. (\ref{eq:lor001}),  if the metric elements are time-depending \cite{Ruggiero:2021uag}: this can be also seen  from Eq. (\ref{eq:geospacecontro}), where an additional term appears, depending on the Born tensor, which is not null for time-depending metrics.
If no time-dependency is present, we have a Lorentz-like equation in the form
{\beq
\frac{\rmd v^i}{\rmd t}=-E^{i}-2({\pmb \beta}\times {\mathbf B})_i. \label{eq:lor0012}
\eeq}
 
In addition, starting from Eq. (\ref{eq:Einsteinweak}), it is possible to show \cite{Ruggiero:2002hz,mashhoon03} that Einstein's equation can be formally written in analogy to Maxwell's equations for the gravitoelectromagnetic fields. However, as discussed by \citet{Bini:2008cy,natario,Ruggiero:2021uag}, when the metric is time-depending, it is not possible to obtain a one-to-one gravitoelectromagnetic analogy both for the geodesic equations and the field equations, since, in any case, non-Maxwellian terms appear (see also \citet{Bakopoulos:2014exa,Williams:2020fgi}).

Another limitation of this linear  analogy is that there are no gravitoelectromagnetic waves: differently speaking, there is not a propagation of the metric components $h_{0i}$ and $h_{00}$; in fact, the gauge condition (\ref{eq:Hilibert}) implies that $h^{0\alpha}=0$ in the transverse and traceless (TT) frame (see e.g. \citet{ref3.2}, Chapter 18);  {hence, it becomes impossible to represent gravitoelectromagnetic waves within the TT frame, implying their nonexistence in any frame. This is because we can consistently cancel a stationary, uniform gravitational potential $\Phi$ and a gravitomagnetic vector potential $A_{i}$ through a suitable coordinate transformation to a locally inertial frame.}

The metric (\ref{eq:weakfieldmetric1}) is expressed in coordinates that are adapted to observers at rest at infinity: as we have seen, the gravitomagnetic effects derive from mass currents. To fix ideas, in the case of a localised source whose center of mass is at rest with respect to these observers, the gravitomagnetic potential is related to the angular momentum of the source. In particular, we get
\beq
\Phi=\frac{GM}{r}, \quad  A_{i} =  \frac{G}{c} \frac{\left(\mb J \wedge \mb x \right)_{i}}{r^{3}}
\label{Angularmomentum}
\eeq
where $r=|{\bf x}|$, in terms of the mass  $M$ and angular momentum ${\bf J}$ of the source \cite{Ruggiero:2002hz,mashhoon03}.
On the other hand, as we have seen in Section \ref{ssec:sagnac}, gravitomagnetic effects arise also because of the rotation of the observers. To summarize, both the motion of the observers and that of the sources might contribute to the definition of the gravitomagnetic effects: a non null vortex tensor, in fact, is generally related to the rotation of the reference frame (see Section \ref{ssec:split}).

\subsection{Gravitoelectromagnetic analogy from the curvature tensor}\label{ssec:fermi}

We showed in the previous Sections how to build a gravitoelectromagnetic analogy: in full GR this is possible using a splitting approach for a given congruence of observers; this approach reduces to the gravitoelectromagnetic formalism of linerized GR, when we consider  inertial observers around a rotating localized source. Here, we focus on the gravitoelectromagnetic analogy that can be build using the components of the curvature tensor: in particular, we will show that, under suitable hypotheses, the geodesic equations take the Lorentz-like form when we use Fermi coordinates.  {The formulation of the spacetime element in Fermi coordinates depends on both the characteristics of the reference frame, such as the acceleration and rotation of the congruence, and the spacetime curvature, as influenced by the Riemann curvature tensor. Our focus is on the impacts of the curvature tensor, leading us to examine a geodesic and non-rotating frame. Nevertheless, it is important to note that, in general, there will be contributions stemming from the world-line acceleration and the tetrad rotation.}  (see e.g. \citet{Ruggiero_2020}).  {The approach to gravitoelectromagnetism in Fermi coordinates which are describing here is largely taken from \citet{Ruggiero:2022gzl} (see also references therein).}

{If we use  Fermi coordinates $(cT,X,Y,Z)$, up to quadratic displacements $|X^{i}|$ from the reference world-line, the line element turns out to be (see e.g.  \citet{manasse1963fermi,MTW})
\beq
ds^{2}=-\left(1+R_{0i0j}X^iX^j \right)c^{2}dT^{2}-\frac 4 3 R_{0jik}X^jX^k cdT dX^{i}+\left(\delta_{ij}-\frac{1}{3}R_{ikjl}X^kX^l \right)dX^{i}dX^{j}. \label{eq:mmmetric}
\eeq
Here, $R_{\alpha \beta \gamma \delta}(T)$ is the projection of the 
Riemann curvature tensor on the orthonormal tetrad $e^{\mu}_{(\alpha)}(\tau)$ of the
reference observer, parameterized by the proper time\footnote{In $e^{\mu}_{(\alpha)}$ tetrad indices like $(\alpha)$ are within parentheses, while  $\mu$ is a  background spacetime index; however, for the sake of simplicity, we drop here and henceforth parentheses to refer to tetrad indices, which are the only ones used.} $\tau$: $\displaystyle R_{\alpha \beta \gamma \delta}(T) = R_{\alpha \beta \gamma \delta}(\tau)=R_{\mu\nu \rho
\sigma}e^\mu_{(\alpha)}(\tau)e^\nu_{(\beta)}(\tau)e^\rho_{(\gamma)}(\tau)e^\sigma 
_{(\delta)}(\tau)$ and it is evaluated along the reference geodesic, where $T=\tau$ and $\mb X=0$.
If we set
\[
\frac{\Phi}{c^{2}}=\frac{g_{00}+1}{2} \quad \frac{\Psi_{ij}}{c^{2}}=\frac{g_{ij}-\delta_{ij}}{2} \quad \frac{A_{i}}{c^{2}}=-\frac{g_{0i}}{2},
\]
the spacetime metric (\ref{eq:mmmetric}) turns out to be
\beq
\mathrm{d} s^2= -c^2 \left(1-2\frac{\Phi}{c^2}\right)\rmd T^2 -\frac4c A_{i}\rmd X^{i}\rmd T  +
 \left(\delta_{ij}+2\frac{\Psi_{ij}}{c^2}\right)\rmd X^i \rmd X^j\ , \label{eq:weakfieldmetric11}
\eeq
with the following definitions
\begin{eqnarray}
\Phi (T, { X^{i}})&=&-\frac{c^{2}}{2}R_{0i0j}(T )X^iX^j, \label{eq:defPhiG}\\
A^{}_{i}(T ,{X^{i}})&=&\frac{c^{2}}{3}R_{0jik}(T )X^jX^k, \label{eq:defAG}\\
\Psi_{ij} (T, {X^{i}}) & = & -\frac{c^{2}}{6}R_{ikjl}(T)X^{k}X^{l}, \label{eq:defPsiG}
\end{eqnarray}
where $\Phi$ and $A_{i}$ are, respectively, the {gravitoelectric} and {gravitomagnetic} potential, and $\Psi_{ij}$ is the perturbation of the spatial metric. We point out that the line element (\ref{eq:weakfieldmetric11}) is a perturbation of flat Minkowski spacetime, that is to say $\displaystyle |\frac{\Phi}{c^{2}}| \ll 1$, $\displaystyle |\frac{\Psi_{ij}}{c^{2}}| \ll 1$, $\displaystyle |\frac{A_{i}}{c^{2}}| \ll 1$.}  

We see that even though the metric elements have a different meaning, the form of the spacetime interval in Eq. (\ref{eq:weakfieldmetric11}) is quite similar to Eq. (\ref{eq:weakfieldmetric1}): accordingly,  the same consequences can be drawn for the geodesic equations, which we write up to linear order  in ${\bm\beta}={\mathbf V}/c$, where $\displaystyle V^{i}=\frac{\rmd  X^i}{\rmd T}$. We define the {gravitomagnetic field}
\beq
\mb B= \bm \nabla \wedge \mb A, \label{eq:defB}
\eeq
or, in terms of the curvature tensor
\beq
B^{}_i(T ,{\mb R})=-\frac{c^{2}}{2}\epsilon_{ijk}R^{jk}_{\;\;\;\; 0l}(T )X^l. \label{eq:defB000}
\eeq
Accordingly, the space components of the geodesic equations are
\beq
\frac{\rmd^{2} X^i}{\rmd T^{2}}=\frac{\partial \Phi}{\partial X^{i}}-2({\bm \beta}\times {\mathbf B})_i+2\frac{\partial A_{i}}{c\partial T}-2\beta^j \frac{\partial \Psi_{ij} }{c\partial T}-\beta^i \frac{\partial \Phi }{c\partial T}. \label{eq:geonew1}
\eeq
In addition, exploiting once again the analogy with electromagnetism, we define the \textit{gravitoelectric field}
\beq
 \quad \mb E= -\bm \nabla \Phi-\frac{2}{c} \ppar{\mb A}{T}, \label{eq:defEtime1}
\eeq
where, in terms of the curvature tensor, we have
\beq
E^{}_i=c^{2}R_{0i0j}(T) X^j, \label{eq:defEIEG}
\eeq
In summary,  Eq. (\ref{eq:geonew1}) becomes
\beq
\frac{\rmd^{2} X^i}{\rmd T^{2}}=-{ E}^{i}-2 \left(\frac{{\mathbf V}}{c}\times {\mathbf B}\right)^{i}-2\frac{V^{j}}{c} \frac{\partial \Psi_{ij} }{c\partial T}-\frac{V^{i}}{c} \frac{\partial \Phi }{c\partial T}.  \label{eq:lor2}
\eeq
Let us examine the meaning of  Eq. (\ref{eq:lor2}) and  relevance of the various terms.  First of all, it is important to stress that this equation defines the motion of a test mass with respect to the reference observer. Consequently, all quantities involved are \textit{relative} to the reference observer at the origin of the frame. 
In addition, the geodesic equations do not take a Lorentz-like form if the fields are not static, due to the presence of the last terms which contain time-derivatives \cite{Ruggiero:2021uag}. However,  both terms - according to the definitions (\ref{eq:defPhiG}) and (\ref{eq:defPsiG}) - are quadratic in the displacements from the reference world-line.  So, even if the fields are time-depending (such as in the case of a gravitational wave) we obtain the Lorentz-like force
\beq
\frac{\rmd^{2} \mb X}{\rmd T^{2}}=-{ \mb E}-2 \left(\frac{{\mathbf V}}{c}\times {\mathbf B}\right),  \label{eq:lor2lin}
\eeq
if we confine ourselves to linear displacements from the reference world-line. In particular, in this case the gravitoelectric field  turns out to be
\beq
\mb E= -\bm \nabla \Phi. \label{eq:defEzero}
\eeq
We see that a gravitoelectromagnetic analogy for the force equation holds true only if suitable hypotheses are assumed.

\subsection{Summary} \label{ssec:summ} 

What we have shown in the previous Sections is the possibility to describe gravitational dynamics in analogy with electromagnetism: in all cases, the dynamics of free particles is formally described by  Lorentz-like equations (\ref{eq:motogen1}),(\ref{eq:lor0012}),(\ref{eq:lor2lin}), when suitable hypotheses are taken into account.  The similarity with Maxwell's theory, then, allows to explain and investigate gravitational effects in terms of known electromagnetic ones, according to the reasonable principle for which similar equations lead to similar solutions. The presence of gravitomagnetic effects has no counterparts in Newtonian gravity and it is relevant not only from an experimental or observational point of view but, also, for fundamental reasons. For instance, the gravitational Larmor theorem \cite{larmor} completes Einstein's formulation of the principle of equivalence, which can be rephrased in terms of the equivalence between the translational acceleration of the Einstein elevator and the Newtonian (i.e. gravitoelctric) field: however, a rotation of the elevator is needed  to take into account the existence of the gravitomagnetic field. 

The  framework described so far is sufficient to encompass different approaches to the study of gravitomagnetic effects, both from a theoretical and experimental viewpoints. We refer to previous review works for a more detailed description of the mathematical and physical aspects of this formalism  \cite{ciufolini1995gravitation,Jantzen:1992rg,Jantzen:1996au,Bini:2002mh,Ruggiero:2002hz,mashhoon03,natario,Costa:2021atq}.

\section{Other Theoretical Developments} \label{sec:other} 
\

In the previous Sections we have described  the basic features of the gravitoelectromagnetic analogies; these theoretical foundations are the bases  on which further progresses have been made, in different contexts, for the purpose to verify  gravitomagnetic effects in experiments or observations. 

The discovery of the first double pulsar \cite{Lyne:2004cj} was an unprecedented opportunity to test  GR effects; more generally speaking,  pulsar astrophysics is an exciting laboratory to test relativistic gravity and, also,  gravitomagnetic effects  \cite{Shao:2014wja,2020Sci...367..577V}. In these systems we may evaluate  the effect of the interaction of the spins of the sources and the orbital angular momentum: in particular \citet{OConnell:2004btu} focused on the possibility to measure spin-orbit effects in binary systems, which may produce not only a correction to the advance of periastron, but also a precession of the orbit about the spin direction. Surveys on general-relativistic spin  effects in binary systems can be found in \cite{OConnell:2008wyk,OC}. Binary systems are relevant for gravitational waves physics: in this context \citet{Kaplan:2008dh} showed that the Maxwell-like formalism can be useful to get physical insights into the numerical-relativity simulations of compact objects. 

The analogy with electromagnetism can be applied to time-varying gravitomagnetic fields, as discussed by \citet{Mashhoon:2008kq}, who considered a linear temporal variation of the vector potential and analyzed its possible impact on some experimental tests, such as the gyroscope precession. Actually, it is possible to develop the analogy in order to introduce a gravitational induction  law \cite{Bini:2008cy} which, for instance, can be used to understand higher order corrections in the interaction of a plane gravitational wave with a detector \cite{Ruggiero:2022gzl}. 

\citet{Costa:2019loe} showed that the gravitomagnetic features can be used to distinguish (in the Weyl class of solutions describing the gravitational field of infinite cylinders) between the case of a static solution (the so-called Levi-Civita solution) and the one accounting for a rotating cylinder: in particular, in the latter case, a gravitomagnetic vector potential that cannot be eliminated by a global coordinate transformation is present.  \citet{Herrera:2007au,Herrera:2021aqv} analysed the vorticity of the congruence of the observers world lines which, as we have seen above, is responsible for the dragging of inertial frames; their findings show that vorticity is related to the presence of a circular flow of super energy in the plane orthogonal to the vorticity vector, and this happens in stationary vacuum spacetimes and also in general Bondi-Sachs radiative ones.  

The clock effect is another gravitomagnetic effect which refers to the difference in the proper time measurements of two clocks (freely) orbiting in opposite directions around a rotating source, and it was initially studied by \citet{clock}. \citet{Lichtenegger:2002af} explained it in analogy with electromagnetism. Generalisations of this effect were discussed by \citet{Hackmann:2014aga}, with possible applications to  the GPS and geostationary satellites. 

As we have seen in Section \ref{ssec:fermi}, it is possible to obtain a gravitoelectromagnetic analogy based on the curvature tensor:  a visualization  tecnique of the electric-like and magnetic-like properties of the curvature tensor was developed by \citet{Owen:2010fa,Nichols:2011pu,Zhang:2012jj,Nichols:2012jn}, with the aim to better understand the dynamics of compact objects merging.

\citet{Mashhoon:2021qtc} investigated the Stern-Gerlach force in a gravitomagnetic framework, in connection with spin-gravity coupling, and showed that it reduces to Mathisson's spin-curvature force. The dynamics of spinning particles in spacetime and the analogies with electromagnetism were investigated by \citet{PhysRevD.93.104006}, and the case of the spacetime of a gravitational wave was considered by \citet{biniortolan2017}.
 A Stern-Gerlach force, together with a Faraday rotation, naturally arises in magnetized Kerr and Reissner-Nordstr\"om spacetimes \cite{Chakraborty:2021bsb}, and this might have consequences in understanding the impact of  both magnetic and gravitomagnetic fields on the propagation of electromagnetic signals in the strong field of compact objects. Magnetic helicity is related to the twist and braiding of the magnetic field lines: \citet{Bini:2021gdb} discussed gravitomagnetic helicity, both in the linear and spacetime curvature approaches to gravitoelectromagnetism.

The interplay between gravitational and electromagnetic field and its  implications on gravitomagnetic effects was investigated by different authors. For instance, \citet{Ahmedov:2003fd} considered the possibility that the interaction of the gravitomagnetic field with the electric field could lead to  new measurements strategies. In \cite{Abdujabbarov:2008mz} the interaction of the Kerr-Taub-NUT spacetime with a magnetic field is investigated. The relation between the vorticity tensor  and the electromagnetic Poynting vector was studied by \citet{Herrera:2006cw}, to understand its role in producing gravitomagnetic effects, while in  \cite{Herrera:2012ke} the production of vorticity by electromagnetic radiation is focused on. The possibility that the Lense-Thirring effect could be produced by the simultaneous presence of electric and magnetic fields was studied by \citet{Gutierrez-Ruiz:2015cnp}. Eventually, gravitoelectromagnetic resonances were investigated by \citet{2011PhRvD..84d3524T} and the interaction between magnetic fields and gravitational waves, with emphasis on the gravitomagnetic effects, was studied by \citet{Tsagas:2009cr}.

There are analyses of gravitomagnetic effects in cosmology: they naturally arise in the so-called post-Friedmann formalism, which is an approach to the perturbation of the Friedmann-Lemaitre-Robertson-Walker  (FLRW) spacetime \cite{Bruni:2013mua,Thomas:2015kua,Barrera-Hinojosa:2020gnx,Barrera-Hinojosa:2021msx}. Instead, FLRW universes were studied with a gravitoelectromagnetic formalism in regions that are small if compared with the Hubble scale by \citet{Faraoni:2022coe}. {From a historical perspective, it is important to mention the rotating cosmological models of Kurt G\"odel \cite{ref2.6,ref2.7}.}

A gravitomagnetic origin of friction for black-hole dynamics was investigated by \citet{Cashen:2016neh}, while \citet{Gutierrez-Ruiz:2018tre} focused on the possible impact of frame-dragging on chaotic dynamics of test particles around a family of stationary axially-symmetric solutions of Einstein's equation coupled with electromagnetic fields.  The Aharonov-Bohm effect for the gravitational field of rotating cosmic string was studied by \citet{Barros:2003hq}.

\section{Solar System Tests}\label{sec:solar}

From a historical point of view, the first attempts were aimed at the measurements of the effects of the  terrestrial gravitomagnetic field;   in this case, a major difficulty is the coexistence, in Earth-based laboratories,  of the Coriolis  field due to the diurnal rotation, which is much greater than the terrestrial gravitomagnetic field but has a quite similar behaviour. Consequently, researchers turned their interest towards the space around the Earth first (where the Coriolis field is not present) and, then, in the Solar System. 

A prototypical case is the already mentioned GPB mission, whose basic concept stemmed from Schiff's seminal paper \cite{schiff} (even though the idea was independently considered by \citet{pugh}), where it was shown that a gyroscope orbiting around the Earth undergoes a geodetic precession, due to its motion in curved spacetime, and a gravitomagnetic precession, entirely determined by the rotation of the Earth. 
The space mission was proposed at the beginning of the 60's, and launched in 2004 with the aim, in particular, to measure the gravitomagnetic precession to a precision of 1\%. Actually, the gyroscope undergoes a precession
\begin{equation}
\boldsymbol{\Omega}_P=\frac{G}{c^2}\left[\frac{3({{\bf J 
}}\cdot {{\bf r}}){{\bf r}}}{r^{5}}-\frac{{{\bf J
}}}{r^{3}} \right],\ \label{eq:GPB}
\end{equation}
as determined by the Earth's angular momentum $\bf J$, where $\mb r$ is its position vector with respect to the center of the Earth; in the case of the GPB mission orbit, the integrated effect from (\ref{eq:GPB}) corresponds to 39 mas/year. However, due to experimental problems (electrostatic patches caused by the non-uniform coating of the gyroscopes), the effective accuracy was of 19\% only (0.28 \%  for the geodetic precession), in any case in agreement with the predictions of GR \cite{Everitt:2011hp}.

Another approach to the measurement of the gravitomagnetic field of the Earth is the analysis of the orbits of satellites  performed by laser-ranging. Actually, the line of the nodes of a test particle orbiting around a source of gravitational field is dragged by the angular momentum of the central body; however, for satellites around the Earth, this effect is much smaller than the one deriving from the non sphericity of the Earth.  \citet{1986PhRvL..56..278C} originally suggested to use a couple of satellites, with supplementary inclinations, in order to get rid of the leading non relativistic effects. A detailed account account of the first approaches to this kind of measurements was given by \citet{ciufolini}.  Later results claimed for a confirmation of the GR predictions for the gravitomagnetic effect of the Earth within a 10\% uncertainty \cite{2004Natur.431..958C}, which were followed by a discussion on the {error budget \cite{2011EL.....9630001I,Ries:2011qho,Iorio:2013zha,Iorio:2015wza,Iorio:2018ntb,ref2.2}}. Eventually, the latest findings report for a confirmation of the gravitomagnetic effect within 2\% uncertainty \cite{Ciufolini:2019ezb}. {Furthermore, the LARES 2 satellite was recently successfully launched for improving the accuracy of previous tests of gravitomagnetism \cite{ref2.3}.}

Lunar laser ranging (LLR) provided, over the years, several precision tests of GR  \cite{LLR}. \citet{Murphy:2007nt} suggested that LLR provided accurate test of gravitomagnetic effects on the lunar orbit relative to Earth: in this case it is not the angular momentum of the Earth that determines this effect, rather it is due to the  orbital motion of the  Earth and the Moon  in the Solar System, which can be seen as mass currents. According to \citet{Ciufolini:2008ma}, this kind of gravitomagnetic effect should thought of as \textit{extrinsic} and opposed to the \textit{intrinsic} gravitomagnetic effect determined by the spin angular momentum of a rotating object (see also \citet{Costa:2016iwu} for an analysis, based on curvature invariants, of the notions of intrinsic and exstrinsic gravitomagnetic effects)  A subsequent debate \cite{Kopeikin:2007sq,Murphy:2007qu,Kopeikin:2008xa,Xie:2009pv}  focused on the possibility that the extrinsic gravitomagnetic effect in the Earth-Moon system could be within the range of measurability with LLR; in addition, there are doubts about  the possibility to measure the intrinsic gravitomagnetic effect of the Earth by LLR \cite{Iorio:2008gr}.

Proposals and ideas to measure the gravitomagnetic effects in the Solar System by accurate determination of the orbits of planets and satellites were considered in various works: for instance, \citet{2009JCAP...03..024I} focused on Kerr-de Sitter solution, also to put constraint on the cosmological constant. Starting from the work by \citet{Bini:2008cy}, subsequent studies \cite{Ruggiero:2010yln,Bini:2015pta} focused on the possibility to measure induction effects (due to a time-dependence of the gravitomagnetic field) on the motion of test particles. The analysis of the orbital plane of the Mars Global Surveyor (MGS) spacecraft suggested a possible evidence of the gravitomagnetic field of Mars \cite{Iorio:2006wr}, {which raised a subsequent debate \cite{ref2.4}}; other gravitomagnetic effects modelled as perturbation of the dynamics of binary system  were also considered by \citet{Iorio:2018wwe,Iorio:2018jys}. Corrections to relativistic orbits due to higher order gravitomagnetic effects were studied by \citet{Capozziello:2008dv}.

Additional proposals to investigate gravitomagnetic effects in the space around the Earth were made by \citet{Ruggiero:2018jab}, who suggested to use geostationary satellites to broadcast electromagnetic signals and measure their propagation times, in order to evidence the asymmetry - determined by the Earth rotation - for signals propagating in opposite directions (a sort of generalized Sagnac effect). \citet{Mirza:2019swf} analysed anomalies in  Earth flybys of several missions, and suggested that the interplay between the magnetic field and gravitational field can enhance the gravitomagnetic effects, which might have some relevance in explaining what is observed. Recently, \citet{Tartaglia:2021idn} suggested to exploit the future LISA mission, which is designed as a detector of gravitational waves in space, to measure the gravitomagnetic field of the Milky Way,  using the propagation times of electromagnetic signals. A model of accurate satellites tracking by means of electromagnetic signals that can be used to measure gravitomagnetic effects was considered by \citet{Scharer:2017zol}. Eventually, \citet{esposito} studied the quantum corrections to the time delay in the gravitational field of a rotating object, and showed that these corrections are in any case too small to be detected in the Solar System.

\section{Laboratory Tests}\label{sec:lab}
Gravitomagnetic effects are in general much smaller than Newtonian ones, which makes it difficult to measure them,  as we have already discussed. Nonetheless, several experimental concepts were proposed over the years. In 1977,  in a seminal paper on the possibility to test relativistic gravity in terrestrial laboratory, \citet{PhysRevD.15.2047}, asserted that ``Advancing technology will soon make possible a new class of gravitation experiments: pure laboratory experiments with laboratory sources of non-Newtonian gravity and laboratory
detectors''. Even though none of the proposed experiments (which heavily exploited the gravitoelectromagnetic analogy) were performed up today, those proposals are still topical, and other ones were suggested. 

For instance, \citet{Pascual-Sanchez:2002rro}, on the basis of a previous proposal by \citet{PhysRevLett.53.863}, made a preliminary study on the possibility to use a Foucalult pendulum to measure the Lense-Thirring effect; the idea was to setup the experimental apparatus at the South pole, to get rid of the larger Coriolis effect due to Earth's rotation.

Other experimental proposals refer to the gravitational properties of coherent quantum systems, such as superconductors or superfluids; more generally speaking, also the quantum features of gravitation  {could be relevant \cite{Modanese:1995tx,Modanese:1996zm,PhysRevLett.34.1472}} (see also Section \ref{sec:quantum} below).  {Some recent reviews by \citet{Gallerati:2022nwm,Gallerati:2022pgh} carefully analyse} these topics. For instance,  {rotating superconductors \cite{PhysRevLett.62.845,pod,Tajmar:2006gh}} showed apparently singular properties and it was suggested that they can be explained by gravitomagnetic fields much larger than those predicted by GR \cite{deMatos:2006us,deMatos:2005tbu}. However, the presence of such ``strange'' gravitomagnetic fields would be at odds with known properties of compact objects, such as neutron stars \cite{Bambi:2007vq}.  {In addition, a recent work by \citet{ref3.1} imposes narrower constraints on the impact of gravitomagnetic effects on the explanation of anomalous Cooper pair mass excess.}

The results presented in that publication are imposing narrower constraints on a possible explanation of anomalous Cooper pair mass excess from gravitoelectromagnetism.

More recently, \citet{Ummarino:2017bvz}, worked out the Maxwell and London equations taking into account the gravitational corrections in linear approximations, expressed in terms of gravitoelectric and gravitomagnetic fields, and investigated the modification of the gravitational field in a superconductor; the same formalism was used to suggest that a Josephson AC effect between two superconductors could be determined by the Earth's gravitational field \cite{Ummarino:2020loo} and to investigate the effects in a physical setup where the external electric and magnetic fields determine the presence of a vortex lattice \cite{Ummarino:2021tpz}.

The effects of rotations described as gravitomagnetic effects on topological superconductors and superfluids were studied by \citet{2012PhRvL.108b6802N,Sekine:2015bqa}, while the impact on Bose-Einstein condensates was considered by \citet{Camacho:2012ce}, with possible implications on the origin of dark matter \cite{Sarkar:2017aje}.

The development of atom interferometry suggested to use this technique to perform precision tests of GR (see e.g. \citet{Dimopoulos:2006nk,Dimopoulos:2008hx} and references therein). In principle, also the Lense-Thirring effect could be tested but,
as we already stated, a major obstacle is the need to isolate this GR effect from the much larger Coriolis effect due to the rotation of the Earth; an alternative proposal was made by \citet{Angonin-Willaime:2003rbe} who suggested to use a satellite around the Earth to host an atomic interferometer. 

Ring Laser Gyroscopes \cite{lasergyro} (which are very precise rotation sensors whose operation is based on the Sagnac effect)  are considered as very promising candidates to measure GR  effects, such as the Lense-Thirring and de Sitter precessions, in a terrestrial laboratory: this is currently under investigation by the GINGER (Gyroscopes IN GEneral Relativity) collaboration \cite{PhysRevD.84.122002,Tartaglia:2016jfo,DiVirgilio:2020ior,Altucci:2022rxr,DiVirgilio:2023nrc}.

\section{Astrophysics}\label{sec:astro}

Astrophysics events are a natural arena to observe gravitomagnetic effects. After discussing in general terms their impact, we will focus on some specific phenomena, such as the propagation of electromagnetic signals in Section \ref{ssec:lensing}, galactic dynamics in Section \ref{ssec:galaxies} and gravitational waves in Section \ref{ssec:GW}.

The impact of gravitomagnetic effects is investigated for different astrophysical phenomena, such as the analysis of the stars motion in the Milky way.  For instance, \citet{Kannan:2008nm} studied the effects of a Kerr field in weak-field approximation on the stars orbiting near the center of the Milky way; \citet{Iorio:2010hw} evaluated several relativistic effects on the radial velocity of a star orbiting the supermassive black hole in the galactic center. 
Various gravitomagnetic effects are evaluated in black holes environments, such as their role in the acceleration in accretion disks \cite{Poirier:2015hga};  \citet{Rueda:2022fgz} considered the interaction of Kerr black hole with the magnetic field in order to understand the relation with  gamma ray bursts; \citet{Ricarte:2022wpd} pointed out the effect of frame dragging on infalling gas. Moreover, the gravitomagnetic back reaction of a heavy object (not a test particle) on a black hole was discussed by \citet{Herdeiro:2009qy}, while the gravitomagnetic field of rotating rings was studied by \citet{Ruggiero:2015pva,Ruggiero:2015ima}.

Pulsars are a natural laboratory to test relativistic gravity and, in particular, gravitomagnetic effects: \citet{Ruggiero:2005dg} studied the impact of the gravitomagnetic correction to the Shapiro time delay and its relevance in pulsar timing. The interplay between the gravitomagnetic field of a black hole and the spin of the pulsar companion can modify the rate at which pulses are received and this can give information on the black hole \cite{Kocherlakota:2017hkn}. The observation of the orbital inclination of the pulsar PSR J1141-6545 in the binary system that contains a white dwarf can be explained as a combination of the effect of the Newtonian quadrupole and the Lense-Thirring effect, as suggested by \citet{VenkatramanKrishnan:2020pbi}, even if \citet{Iorio:2020xos} pointed out that care must be paid in the interpretation of these results. Gravitomagnetic tidal resonances \cite{flanagan,Favata:2005da,Poisson:2020eki} are important in the motion of a binary system, and they can impact on the emission of gravitational waves. Eventually, gravitomagnetic effects on circumbinary (i.e. orbiting two stars) exoplanets are studied by \citet{Iorio:2022raj}.

\subsection{Effects on electromagnetic signals propagation} \label{ssec:lensing}

The bending effect  on the propagation of light rays is one of the classical tests of GR; initially, only the effect of the mass was considered but it is clear that, at higher order, also mass currents contribute to this effect. The same was true for the so-called Shapiro effect, which is the time delay on the propagation of light signals in a gravitational field. 

\citet{Kopeikin:2001dz}, in  weak-field approximation, focused on light propagation in the gravitational field determined by self-gravitating spinning bodies, that are moving with arbitrary velocity. Several effects are considered, such as the correction to  the Shapiro time delay, the modification of the bending angle due to the spin of the gravitating bodies, the rotation of the plane of polarisation of electromagnetic waves. 

The gravitomagnetic correction to the time delay of electromagnetic signals  in gravitational lensing were calculated by \citet{Ciufolini:2002iq,Ciufolini:2003yy}, who considered both the case of propagation in a rotating mass shell and in the field of a rotating source, again in weak-field approximation. 

The modification of  the deflection angle and the time delay function due to gravitomagnetic effects were considered by  \citet{Sereno:2003tk}, in a more general framework which can be applied to a post-Newtonian spacetime, and this formalism was used for spinning stars \cite{Sereno:2003kx}, spiral galaxies \cite{Capozziello:2003hh}; the time delay for extended rotating sources, modelled as   isothermal spheres, was focused on by \citet{Sereno:2004nc}, and the effect due to moving lenses was considered by \citet{Sereno:2005wr}. In addition, the Faraday rotation due to the gravitomagnetic field was  considered by \citet{Sereno:2004jx}, and a possible application to binary pulsar systems by \citet{Ruggiero:2006rh}.

In the above papers, gravitomagnetic effects on light rays propagation were studied in  the weak-field approximation, which is undoubtedly sufficient in the Solar System.  On the other hand, \citet{Kraniotis:2005zm} exactly solved the geodesic equations for test particles and photons in Kerr spacetime and, also,   in the spacetime of a rotating electrically charged black hole (Kerr-Newman) \cite{Kraniotis:2014paa}; in addition,   \citet{Kraniotis:2019ked} calculated the frequency shift of light emitted by geodesics test particles in  Kerr-Newman-de Sitter and Kerr-Newman spacetimes; in the latter paper also the pericentre shifts were calculated,  which were already been considered by the same author \cite{Kraniotis:2007zz}.

Always in Kerr spacetime, the bending angle which takes into accounts also of the motion of the observer was considered by \citet{Arakida:2018szn}: in particular, the case is considered where the observer is not located in a flat asymptotic region, so the effect of a finite distance from the lens is taken into account. 

\citet{Iyer:2018omj} obtained an exact analytical expression for the bending angle of light on the equatorial plane of a Kerr black-hole, and then expanded it in power series. In particular, the asymmetry between the direct and the retrograde propagation, due to the rotation of the sources, was evidenced (see also \citet{PhysRevD.80.124023}).


\subsection{Galactic Dynamics} \label{ssec:galaxies}

The dark matter (DM) hypothesis, i.e., the hypothesis of the existence of a non-baryonic component of mass dominating the matter density budget in the Universe, is one of the foundations of the widely accepted $\Lambda$CDM cosmological model \citep[e.g.][]{planck2018b} and, at the same time, one of the greatest mysteries in physics. The DM hypothesis has been incredibly successful in interpreting different astrophysical observables, such as the velocity distribution of galaxies in galaxy clusters \citep[GCs, since][]{zwicky}, the rotation curves (RCs) of disc galaxies \citep[e.g.][]{rotation_curves}, the thermodynamical properties of X-ray emitting gas in GCs \citep{gas}, the gravitational lensing produced by their mass distributions \citep{lensing} and the observations of the two Bullet Clusters \citep{bullet1, bullet2}; as well as cosmological ones, e.g., the anisotropies observed in the cosmic microwave background \citep{CMB1, CMB2} and the growth of cosmic structures from such anisotropies \citep{perturb1, perturb2}. 

For what concerns disc galaxies, the main evidence supporting the existence of DM are based on the observed rotation curves, whose flatness at large radii cannot be explained by the distribution of visible (baryonic) matter, if interpreted on the basis of Newtonian dynamics \cite{strigari2013galactic}.  {In this scenario, Newtonian gravity is employed instead of GR. This choice is made because, when we are far from the galactic center (where the flat behavior is observed), we can reasonably assume that the gravitational field is weak, and the stars within the galaxy are not moving at speeds close to the speed of light. However, there has been a suggestion that GR might still have relevance in this context. Specifically, researchers have explored the issue of galactic rotation curves by examining it from two angles: one involving exact solutions of the equations of GR and the other using weak-field approximations.}

{It was proposed that considering gravitomagnetic effects might lead to a different interpretation of the role of dark matter in explaining the observed phenomena:} even though there is no doubt that in galaxies the Newtonian approximation correctly applies locally, globally it may fail due to the overall rotation of the system.  This may indicate the need of general relativistic corrections to the Newtonian approach. 

This point of view was originally advocated in two pioneering works by \citet{Cooperstock:2005qw, Cooperstock:2006dt} and subsequently further developed by \citet{Carrick:2011ac}. Since some of these works were criticised by \citet{Cross:2006rx,Menzies:2007dm}, subsequently \citet{Balasin:2006cg} proposed another solution of Einstein's equations which resolved some problematic features. The velocity profile derived from the BG model was first used by \citet{crosta2020testing}
as a basis for studing the Milky Way rotation curve: after recasting the BG model to make it consistent 
with the Gaia stellar data \cite{GAIA1,GAIA2}, they  showed
that the GR rotation curve was in quite good agreement with the Gaia data, at a level statistically comparable to the state of the art 
CDM models they compared to in their article.  It is relevant to point out that to obtain the astrometric accuracies  needed in a mission like Gaia, it is important to correctly take into account relativistic effects, such as the gravitomagnetic ones (see the recent review by \citet{crosta_astrometry}).

 { A recent paper by \citet{Costa:2023lrn} analyses the solution obtained by Balasin and Grumiller (BG), and shows 
that it is not appropriate as a galactic model. In particular, the BG model is a rigid solution. The study of the generalization to non rigid rotation was done in \citet{Astesiano:2021ren,astesiano}.}

Also \citet{2020arXiv200914553G} investigated the possibility of accounting for the observed anomalous velocities of stars in galaxies to be a result of a dynamic overall rotation of an inertial frame dragged by a galaxy.
A somewhat different but related approach, which takes into account the gravitomagnetic  effects originating from mass currents into the solution of Einstein equations in  weak-field and slow-motion approximation was put forward by \citet{Ludwig}, \citet{Ruggiero:2021lpf}, \citet{Srivastava:2022eza} and  analyzed with detailed criticism by \citet{Ciotti_2022}. In \citet{Astesiano:2022ghr}, the GR results of these papers are shown to be related to the existence of the homogeneous solutions of the gravitomagnetic field and not directly related to the mass currents. These homogeneous solutions can produce a ``strong gravitomagnetic limit'' where these effects are of the same order as Newtonian ones. Criticisms toward the application of gravitomagnetic effects to study galactic dynamics were made by \citet{Lasenby:2023zdo}. 
Another different approach was proposed using Post-Newtonian corrections in galactic dynamics by \citet{Ramos-Caro:2012ren} and  \citet{2014arXiv1406.6082L} who also proposed an explanation for spiral arms, without the presence of an exotic form of matter. The formalism is based on Boltzmann transport equation for the collisional matter and on the very-low-velocity gravitomagnetism. 

The off-diagonal terms responsible for the gravitomagnetic effects contribute to the modified virial theorem as shown in \cite{Astesiano:2022gph}. They can also give rise to a consistent definition of a “gravitational mass" in this very specific set up \cite{2018arXiv181110602S}.

\subsection{Gravitational Waves} \label{ssec:GW}

The first direct detection of Gravitational Waves (GW), in 2015 \cite{abbott2016observation}, marked the beginning of gravitational waves astronomy and cosmology: {in fact, apart from serving as a test for the theory, gravitational waves have become a potent instrument for exploring the Universe in the era of multi-messenger astronomy;} technological developments and dedicated missions will help to greatly improve the information that can be obtained within this window. Accordingly, it is very important to properly model the measurement process\footnote{Indeed, we focus here on the measurement process only, and neglect the impact of gravitomagnetic effects in the emission of gravitational waves, which is a very interesting topic, fully described in the literature on this field. } and, in this context, it is relevant to emphasise that the interaction of GW with a detector, modelled as a set of test masses, can be described in terms of gravitoelectromagnetic analogy  \cite{Ruggiero_2020,Ruggiero:2021qnu}.  {By adopting this approach, one can readily comprehend that while current devices detect the interaction between test masses and the gravitoelectric components of the wave, there are also gravitomagnetic interactions that offer the potential to observe the influence of gravitational waves on moving masses and spinning particles \cite{biniortolan2017}.}

{Notably,  \citet{Ruggiero_2020b,Ruggiero:2021dri}  demonstrated the potential occurrence of a gravitomagnetic resonance phenomenon involving spinning particles within the influence of a gravitational wave. This observation opens up possibilities for devising innovative detectors capable of measuring collective spin excitations, such as spin waves within magnetized materials.}  In addition, \citet{PhysRevD.73.084003}, using the gravitoelectromagnetic formalism for the components of the curvature tensor, studied the coupling of the helicity of the gravitational wave with the possible rotation of the detector.

{Even though we consider spinless test masses, there are gravitomagnetic effects that need to be taken into account: \citet{baskaran} 
demonstrated that gravitomagnetic terms play a significant role in describing the displacements of interferometer test masses, particularly up to the second order in the distance parameter. As a result, they are important for precise measurements and the determination of gravitational waves source parameters. Specifically, while at the first order a detector's interaction with gravitational waves can be attributed to the influence of a gravitoelectric and gravitomagnetic field perpendicular to the propagation direction, at the second order, the gravitoelectric field exhibits a non-zero component along the propagation direction. These phenomena can be elucidated through the concept of gravitational induction  \cite{Ruggiero:2022gzl}.\\ \citet{Iorio:2009gn} conducted a thorough examination of the interferometric response to these phenomena. Furthermore, it's noteworthy to underscore that investigating these types of effects could hold significance in testing alternative gravity theories beyond GR. Indeed, these theories often introduce longitudinal effects in gravitational radiation, which can arise from various sources such as massive modes, scalar fields, or a more complex geometric structure (see e.g. \citet{capozziello1,capozziello2,corda1,corda2}).}

\section{Analogue Models}\label{sec:analogue}

Analogue gravity (see e.g. \citet{Barcelo:2005fc} and references therein) investigates the possibility to describe physical systems in analogy with the formalism of curved spacetime, and this is typically fruitful for systems that can be experimentally tested in a laboratory  and can give back new insights on the physics of general or special relativity. For example, it is possible to consider  sound waves in a moving fluid in analogy with light waves in a curved spacetime: in particular, if the fluid flow is supersonic, it is possible to get a ``dumb hole'', i.e.  the acoustic analogue of a black hole \cite{PhysRevLett.46.1351,Visser:1998qn}. Accordingly, it is possible to mimic gravitomagnetic effects in these analogue models. 
For instance, \citet{2008arXiv0808.3404P} studied the analogy between the equation of linearized turbulent fluid and those of General Relativity, in linear approximation, that leads to the linear analogy described in Section \ref{ssec:lin}. On the contrary, \citet{Kivotides:2020nue} studies gravitomagnetic effects \textit{on} turbulent fluids.

In addition, \citet{Chakraborty:2015ioa} focused on  a model of the Lense-Thirring effect for a rotating acoustic analogue black hole, and suggested that Bose-Einstein condensate systems (BEC)  could provide an important setup to test these effects, while \citet{Banerjee:2018wev} addressed frame-dragging studying the hydrodynamics of nematic active fluids.

Another related interesting field of investigation pertains to the study of the gravitational field produced by light \cite{Ratzel_2016,Spengler_2022}. In particular, it is possible to calculate in linear approximation of GR the gravitational field of electromagnetic beams called optical vortices, carrying orbital angular momentum \cite{Padgett:17}: the point is that every photon in a laser beam could carry angular momentum in addition to the angular momentum associated with its spin.  Accordingly, these beams can generate a gravitational field that produces gravitomagnetic effects \cite{Strohaber:2011aa}  which  are very small to be detected, even if in principle  present and important from the theoretical point of view. Similarly, it is possible to study the gravitomagnetic effects determined  by the spin angular momentum of the light beams \cite{2018CQGra..35s5007S,2019CQGra..36t5007S,2019CQGra..36k9501S}, which remains, however,  too small to be detected with current technology.

\section{Quantum Effects}\label{sec:quantum}
 Despite the progress made in these fields, there are still many unanswered questions about the relationship between gravity and quantum mechanics. Additionally, there is still much to learn about the behavior of matter and energy at the Planck length, which is currently beyond the scope of experimental observation.

In the past years much attention was given to the theory of quantum fields in classical background gravitational
fields, in particular regarding Hawking radiation by black holes (see \citet{Page:2004xp} for a review on the subject). In this regard, the phenomenological thermodynamic properties of black holes are  well understood, at least for quasi-stationary semiclassical black holes.\\
Parallel to this, some investigations have been done on quantum systems in classical background gravitational fields, for example
for atomic beam interferometry \cite{Dimopoulos:2006nk} and on neutrons in the Earth’s field \cite{PhysRevLett.34.1472}. A review on different aspects of the interaction of mesoscopic quantum systems with gravitational fields was written by \citet{Kiefer:2004hv} (see also the discussion in Section \ref{sec:lab}). In particular  two proposed interaction mechanisms are considered:
\begin{itemize}
    \item the use of quantum fluids as generator and/or detector of gravitational waves in the laboratory;
    \item  the inclusion of gravitomagnetic fields in the study of the properties of rotating superconductors.
\end{itemize}
Experiments to detect such effects are expected to be quite difficult, but they would be of fundamental interest. In particular, in the gravitomagnetic limit, there are experimental setup for probing the “diagonal part", or in other words, the so called “inverse-square law" of gravity using quantum interference \cite{Hammad:2019udg} and there are also quantum detection tests to observe the “off-diagonal" terms, which defines what we call frame dragging \cite{Cong:2020xlt}.
\\

A systematic treatment of the behaviour of a quantum system under the effects of a small and slowly rotating gravitational field was proposed by \citet{Adler:2009du} for spin zero particles and by \citet{Adler:2011bg} for spin $1/2$ particles. These works also take into account the possibile presence of an electromagnetic field, to which  the Klein-Gordon and Dirac equations are minimally coupled.\\
A different approach is what in the literature is referred to as the Schr\"{o}dinger-Newton model \cite{Bahrami:2014gwa}, which describes non-relativistic quantum objects under self-gravitation. In this model, in the Schr\"{o}dinger equation also the Newtonian gravitational potential term appears and the source of this term is given by the square of the module of the wave function. The Schr\"{o}dinger-Newton equation was first proposed to study self-gravitating bosonic stars by \citet{PhysRev.187.1767}. Since this model was completely non-relativistic,  some authors recently considered a modification of the Schr\"{o}dinger-Newton equation by taking into account certain relativistic corrections \cite{Brizuela:2022fcb},  with a particular focus on gravitomagnetic corrections \cite{Manfredi:2014hna}. Under the above approximation, \citet{Zhao:2017xjj}  considered a short distance modification of the Schr\"{o}dinger-Newton equation which also results in a short distance modification of the quantum mechanical virial theorem.\\
For what concerns quantum information, the dynamics of an
orbiting qubits under the effect of gravitational frame dragging was studied by \citet{Lanzagorta:2012tn}. In particular the author considered the Kerr spacetime geometry and a spin $1/2$ qubits. Subsequently, another possible test for frame dragging effect using the same setup was proposed by \citet{Lanzagorta:2016mhh}.\\
We also note that the effect of 
a scattering process with gravitons as an intermediate state was investigated by \citet{Jesus:2022ywg}. In this setup, the gravitomagnetic limit is considered. This allows a  Lagrangian formulation which includes interactions of gravitons with fermions and photons. On the other hand, the presence of an external gravitational field and how the frame dragging effects affect the scattering process was focused on by \citet{Kim:2022iub}.

\section{Connections with the Mach's Principle}\label{sec:mach}

The idea underlying the Mach's principle is that there is a relationshipe between the local and the global Universe and the nature of this link is mechanical. We are already familiar with connections of a different kind, for example the known “Olber's paradox" \cite{olbers}, where the link is optical. Actually, according to Mach, the local inertia of a body is a consequence of the global distribution of matter in the Universe.

This view is a contrast to  the old Newtonian paradigm, which states that inertia is an intrinsic property of matter and therefore it is completely independent from the rest of the Universe. Of course, until now, no such change in the value of inertia has been measured experimentally. From the Machian perspective, the fact that this change has not been detected is not a big deal, since nearby matter would make a little contribution to the inertia of a test body compared to the vastness of the observed Universe.\\
The Newtonian paradigm is based on  absolute space and time: consequently, the notion of absolute motion comes naturally, as relative to the absolute space, and the absolute acceleration is the one appearing in the Newton second law. To prove the existence of absolute motion, in the \textit{Principia} Newton describes a  thought experiment to detect absolute rotation, which is the famous ``bucket experiment''. Newton's interpretation of this thought experiment is that  if we measure the centrifugal forces
responsible for the concavity of the water surface we are in fact measuring the absolute rotation of a body. In his interpretation, the centrifugal forces arise as a result of the rotation of water with respect to absolute space, since all possible inertial frames are tied to this fundamental ``entity''. The Focault's pendulum is a  concrete realization of the above discussion, and it shows the absolute rotation of the Earth.

The main criticism toward the Newtonian interpretation were pointed out by Berkeley and, after many years, by Mach \cite{mach}. They claimed that all we can observe is that the centrifugal forces are due to the motion of the water with respect to the rest of the matter in the Universe, not only respect to the walls of the bucket. In particular Mach used the sentence ``respect to the fixed stars''. According to Mach, we do not know whether the result of the experiment would be the same if all the matter of the universe were removed,
nor whether such centrifugal forces could be produced by the rotation of the rest of the Universe while the walls of the bucket stay  ``fixed''. \\
The theory of General Relativity presents Machian features, such as the dragging effects  ({see the monograph by \citet{ciufolini1995gravitation} and the paper by
\citet{2019arXiv190110766V} for a recent review}). To see how these effects arise in GR,  let us consider the weak-field limit of a spacetime outside a slowly rotating stationary body  in adapted coordinates; in particular, the line element (\ref{eq:weakfieldmetric1}) can be written as
\begin{align}
    ds^2 =- c^2\left(1-\frac{2GM}{c^2 r}\right) dt^2+\left(1+\frac{2GM}{c^2 r}\right) dr^2+ r^2 \left(d\theta^2+r^2 \sin^2\theta  d\phi^2\right)- 4\frac{J}{r} \sin\theta dt d\phi, \label{OutsideKerr}
\end{align}
where the non diagonal terms are given by Eqs. (\ref{eq:solgemAi1}) and (\ref{Angularmomentum}). It is the field $\mathbf B$ in Eq. (\ref{eq:defEtime}) which gives the Coriolis forces acting on test particles outside the body and the precession of gyroscopes. A similar argument applies for the forces arising inside a rotating shell of matter \cite{1984GReGr..16..711M}. Let us consider a gyroscope at rest respect to the static observers. The static observers are defined by the four-velocity $Z$
\begin{align}
    Z= \frac{1}{\sqrt{-g_{tt}}} \partial_t\simeq \partial_t,
\end{align}
and are at rest in the coordinate system of (\ref{OutsideKerr}). Due to the assumptions on the asymptotic flatness of the metric, the reference frame
defined by the  congruence of worldlines $\partial_t$ corresponds to a rigid frame anchored to the
asymptotic inertial frame at infinity. The key feature is that this reference frame is at rest compared to the distant stars, but it has non-zero angular momentum. The spin axis $ \Xi$ of the gyroscope is Fermi-Walker transported along its worldline \cite{MTW}
\begin{align}
    \frac{D\Xi^\mu}{d\tau}=\Xi^\nu a_\nu V^\mu,
\end{align}
where $\tau$ is the proper time, $V=Z$ is  the four velocity of the gyroscope and $a$ is the four acceleration. We can read off the Christoffel symbols from Eq. (\ref{eq:lor001}) and after imposing the orthogonality condition $\Xi^\mu V_\mu=0$ we find
\begin{align}
    \frac{D\Xi^i}{d\tau}= \frac{1}{2} (\Xi \times \mathbf{B})^i.
\end{align}

After using Eq. (\ref{Angularmomentum}), which we report here for easier reference 
\begin{align}
    A_{i} =  \frac{G}{c} \frac{\left(\mb J \wedge \mb x \right)_{i}}{r^{3}},
\end{align} 
we get the result in Eq. (\ref{eq:GPB}). This discussion shows, as stated before, that General Relativity possess some Machian properties: in fact, the gyroscope determines the axes of a local inertial frame and they are affected by the mass distribution and its motion.\\
For a more detailed application of gravitomagnetism to geodetic precession and frame dragging see \citet{2012arXiv1206.4593C}, where another quantitative confirmation of Mach's arguments can be found.\\

Such effects can arise also in setups which are different from the one that has been discussed. For example, in the context of cosmology,  \citet{Schmid:2008np} showed that there is  dragging of local inertial frames by a weighted
average of the cosmological energy currents via gravitomagnetism, for all linear perturbations of all
Friedmann-Robertson-Walker (FRW) universes.

\section{Conclusions}\label{sec:conc}

Gravitoelectromagnetic analogies arise in different context in General Relativity and are often used to provide a better insight into complicated gravitational phenomena, even though it is clear that gravitational and electromagnetic interactions are essentially different. In this work, we reviewed these analogies and emphasized also the  hypotheses on which they are based. 

First,  we showed that a splitting approach in full theory leads to a non-linear analogy with electromagnetism: to this end, we introduced the spatial tensor algebra formalism using Cattaneo's  projection techniques. This led  us to  the form of the force equation for a test particle, as seen by an observer described by a timelike congruence $\Gamma$ in the spacetime $\mathcal{V}_4$. This force contains a term which is proportional to the curvature of $\Gamma$ and a term proportional to the spatial velocity of the test particle, which makes it possible to introduce an analogy with the electromagnetic dynamics: in particular, a gravitoelectric field is associated to the local linear acceleration, while a gravitomagnetic field is associated to the local angular acceleration. 

Then, we pointed out that the well known gravitoelectromagnetic analogy that arises in linearized General Relativity can be seen as a limiting case of the previously discussed exact analogy. In particular, the linear formalism is very useful to deal with experiments and observations which are often performed in conditions where the gravitational field is weak and the speeds are small compared to the speed of light. 
Accordingly, the linear gravitoelectromagnetic analogy is a powerful tool to explain new gravitational effects in terms of known electromagnetic ones: we remark that gravitomagnetic effects are peculiar to General Relativity, since in Newtonian gravity there are no gravitational effects arising from mass currents. {In addition, we pointed out some limitations of this linear analogy that arises when we are dealing with time-depending gravitational fields; in particular, the radiative regime is one of the limits of applicability of the gravitoelectromagnetic approximation to General Relativity} 

Eventually, we briefly sketched the gravitoelectromagnetic analogy that is based on the magnetic-like and electric-like parts of the Riemann  tensor, using Fermi coordinates. This formal approach is useful, for instance, when we are directly dealing with tidal effects, such as in the case of gravitational waves physics.

After this basic  introduction,  we reviewed the recent theoretical developments which are aimed to suggest new possible tests of gravitomagnetic effects. The continuous improvements in technology and measurement techniques made it possible to imagine a pletora of possible consequences of these effects. Accordingly, we made a survey of the proposals to test gravitomagnetic effects emerged during the last twenty years, which refer to Solar System and Earth-based experiments, astrophysical observations, analogue models. In addition we
reviewed the interplay between gravitomagnetic effects and other areas of physics, such as quantum effects and the Mach principle.

The presence of ``gravitomagnetic effects'' in a somewhat wider sense is a natural consequence of the general relativistic approach to the description of the gravitational interaction and, consequently, even if these effects are generally very small and difficult to
distinguish from other competing ones, there should be no doubts on their existence. On the other hand, their study is important to correctly model and understand complex gravitational phenomena on different scales, which range from the near space in the Solar System, to astrophysical events and, eventually, to galactic and cosmological dynamics. 

We conjecture a further  evolution of this  useful formalism, and we hope that this review will be a helpful reference for researchers involved in the study of gravitational physics.

\section*{Acknowledgments}
MLR acknowledges the contribution of  the local research  project Modelli gravitazionali per lo studio dell'universo (2022) - Dipartimento di Matematica ``G.Peano'', Universit\`a degli Studi di Torino, that supported the visit of DA  to Torino, during which this work was started. This  work is done within the activity of the Gruppo Nazionale per la Fisica Matematica (GNFM). 
 {The authors thank Filipe Costa, Mariateresa Crosta,  Antonio Gallerati, Jos\'e Nat\'ario for useful comments.}



\begin{thebibliography}{272}%
\makeatletter
\providecommand \@ifxundefined [1]{%
 \@ifx{#1\undefined}
}%
\providecommand \@ifnum [1]{%
 \ifnum #1\expandafter \@firstoftwo
 \else \expandafter \@secondoftwo
 \fi
}%
\providecommand \@ifx [1]{%
 \ifx #1\expandafter \@firstoftwo
 \else \expandafter \@secondoftwo
 \fi
}%
\providecommand \natexlab [1]{#1}%
\providecommand \enquote  [1]{``#1''}%
\providecommand \bibnamefont  [1]{#1}%
\providecommand \bibfnamefont [1]{#1}%
\providecommand \citenamefont [1]{#1}%
\providecommand \href@noop [0]{\@secondoftwo}%
\providecommand \href [0]{\begingroup \@sanitize@url \@href}%
\providecommand \@href[1]{\@@startlink{#1}\@@href}%
\providecommand \@@href[1]{\endgroup#1\@@endlink}%
\providecommand \@sanitize@url [0]{\catcode `\\12\catcode `\$12\catcode
  `\&12\catcode `\#12\catcode `\^12\catcode `\_12\catcode `\%12\relax}%
\providecommand \@@startlink[1]{}%
\providecommand \@@endlink[0]{}%
\providecommand \url  [0]{\begingroup\@sanitize@url \@url }%
\providecommand \@url [1]{\endgroup\@href {#1}{\urlprefix }}%
\providecommand \urlprefix  [0]{URL }%
\providecommand \Eprint [0]{\href }%
\providecommand \doibase [0]{http://dx.doi.org/}%
\providecommand \selectlanguage [0]{\@gobble}%
\providecommand \bibinfo  [0]{\@secondoftwo}%
\providecommand \bibfield  [0]{\@secondoftwo}%
\providecommand \translation [1]{[#1]}%
\providecommand \BibitemOpen [0]{}%
\providecommand \bibitemStop [0]{}%
\providecommand \bibitemNoStop [0]{.\EOS\space}%
\providecommand \EOS [0]{\spacefactor3000\relax}%
\providecommand \BibitemShut  [1]{\csname bibitem#1\endcsname}%
\let\auto@bib@innerbib\@empty
\bibitem [{\citenamefont {McDonald}(1997)}]{McDonald:1997fd}%
  \BibitemOpen
  \bibfield  {author} {\bibinfo {author} {\bibfnamefont {K.~T.}\ \bibnamefont
  {McDonald}},\ }\href@noop {} {\bibfield  {journal} {\bibinfo  {journal}
  {American Journal of Physics}\ }\textbf {\bibinfo {volume} {65}},\ \bibinfo
  {pages} {591} (\bibinfo {year} {1997})}\BibitemShut {NoStop}%
\bibitem [{\citenamefont {Heaviside}(1894)}]{heaviside1894electromagnetic}%
  \BibitemOpen
  \bibfield  {author} {\bibinfo {author} {\bibfnamefont {O.}~\bibnamefont
  {Heaviside}},\ }\href@noop {} {\emph {\bibinfo {title} {Electromagnetic
  Theory}}}\ (\bibinfo  {publisher} {The Electrician Printing and Publishing
  Co, London},\ \bibinfo {year} {1894})\BibitemShut {NoStop}%
\bibitem [{\citenamefont {{Herrmann}}\ and\ \citenamefont
  {{Pohlig}}(2022)}]{hermann}%
  \BibitemOpen
  \bibfield  {author} {\bibinfo {author} {\bibfnamefont {F.}~\bibnamefont
  {{Herrmann}}}\ and\ \bibinfo {author} {\bibfnamefont {M.}~\bibnamefont
  {{Pohlig}}},\ }\href {\doibase 10.1119/10.0009888} {\bibfield  {journal}
  {\bibinfo  {journal} {American Journal of Physics}\ }\textbf {\bibinfo
  {volume} {90}},\ \bibinfo {pages} {410} (\bibinfo {year} {2022})}\BibitemShut
  {NoStop}%
\bibitem [{\citenamefont {{Pfister}}(2014)}]{2014SPPhy.157..191P}%
  \BibitemOpen
  \bibfield  {author} {\bibinfo {author} {\bibfnamefont {H.}~\bibnamefont
  {{Pfister}}},\ }in\ \href {\doibase 10.1007/978-3-319-06761-2_24} {\emph
  {\bibinfo {booktitle} {Springer Proceedings in Physics}}},\ \bibinfo {series}
  {Springer Proceedings in Physics}, Vol.\ \bibinfo {volume} {157}\ (\bibinfo
  {year} {2014})\ p.\ \bibinfo {pages} {191}\BibitemShut {NoStop}%
\bibitem [{\citenamefont {Einstein}(1912)}]{einstein1912there}%
  \BibitemOpen
  \bibfield  {author} {\bibinfo {author} {\bibfnamefont {A.}~\bibnamefont
  {Einstein}},\ }\href@noop {} {\bibfield  {journal} {\bibinfo  {journal}
  {Vierteljahrsschrift fur gerichtliche Medizin und offentliches
  Sanitatswesen}\ }\textbf {\bibinfo {volume} {44}},\ \bibinfo {pages} {37}
  (\bibinfo {year} {1912})}\BibitemShut {NoStop}%
\bibitem [{\citenamefont
  {Einstein}(2009{\natexlab{a}})}]{einstein2009collected}%
  \BibitemOpen
  \bibfield  {author} {\bibinfo {author} {\bibfnamefont {A.}~\bibnamefont
  {Einstein}},\ }\href@noop {} {\emph {\bibinfo {title} {The Collected Papers
  of Albert Einstein, Volume 4 (English): The Swiss Years: Writings, 1912-1914.
  (English translation supplement)}}}\ (\bibinfo  {publisher} {Princeton
  University Press},\ \bibinfo {year} {2009})\ p.\ \bibinfo {pages}
  {126}\BibitemShut {NoStop}%
\bibitem [{\citenamefont {{Thirring}}(2012)}]{2012GReGr..44.3225T}%
  \BibitemOpen
  \bibfield  {author} {\bibinfo {author} {\bibfnamefont {H.}~\bibnamefont
  {{Thirring}}},\ }\href {\doibase 10.1007/s10714-012-1451-3} {\bibfield
  {journal} {\bibinfo  {journal} {General Relativity and Gravitation}\ }\textbf
  {\bibinfo {volume} {44}},\ \bibinfo {pages} {3225} (\bibinfo {year}
  {2012})}\BibitemShut {NoStop}%
\bibitem [{\citenamefont {{Pfister}}(2012)}]{2012GReGr..44.3217P}%
  \BibitemOpen
  \bibfield  {author} {\bibinfo {author} {\bibfnamefont {H.}~\bibnamefont
  {{Pfister}}},\ }\href {\doibase 10.1007/s10714-012-1450-4} {\bibfield
  {journal} {\bibinfo  {journal} {General Relativity and Gravitation}\ }\textbf
  {\bibinfo {volume} {44}},\ \bibinfo {pages} {3217} (\bibinfo {year}
  {2012})}\BibitemShut {NoStop}%
\bibitem [{\citenamefont {{Mashhoon}}\ \emph {et~al.}(1984)\citenamefont
  {{Mashhoon}}, \citenamefont {{Hehl}},\ and\ \citenamefont
  {{Theiss}}}]{1984GReGr..16..711M}%
  \BibitemOpen
  \bibfield  {author} {\bibinfo {author} {\bibfnamefont {B.}~\bibnamefont
  {{Mashhoon}}}, \bibinfo {author} {\bibfnamefont {F.~W.}\ \bibnamefont
  {{Hehl}}}, \ and\ \bibinfo {author} {\bibfnamefont {D.~S.}\ \bibnamefont
  {{Theiss}}},\ }\href {\doibase 10.1007/BF00762913} {\bibfield  {journal}
  {\bibinfo  {journal} {General Relativity and Gravitation}\ }\textbf {\bibinfo
  {volume} {16}},\ \bibinfo {pages} {711} (\bibinfo {year} {1984})}\BibitemShut
  {NoStop}%
\bibitem [{\citenamefont
  {Einstein}(2009{\natexlab{b}})}]{einstein2009_thirring}%
  \BibitemOpen
  \bibfield  {author} {\bibinfo {author} {\bibfnamefont {A.}~\bibnamefont
  {Einstein}},\ }\href@noop {} {\emph {\bibinfo {title} {The Collected Papers
  of Albert Einstein, Volume 8 (English): Schulman, R., et al. "The collected
  papers of Albert Einstein.-Vol. 8: The Berlin years: correspondence
  1914-1918."}}}\ (\bibinfo  {publisher} {Princeton University Press},\
  \bibinfo {year} {2009})\ p.\ \bibinfo {pages} {365}\BibitemShut {NoStop}%
\bibitem [{\citenamefont {Everitt}\ \emph {et~al.}(2011)\citenamefont {Everitt}
  \emph {et~al.}}]{Everitt:2011hp}%
  \BibitemOpen
  \bibfield  {author} {\bibinfo {author} {\bibfnamefont {C.~W.~F.}\
  \bibnamefont {Everitt}} \emph {et~al.},\ }\href {\doibase
  10.1103/PhysRevLett.106.221101} {\bibfield  {journal} {\bibinfo  {journal}
  {Phys. Rev. Lett.}\ }\textbf {\bibinfo {volume} {106}},\ \bibinfo {pages}
  {221101} (\bibinfo {year} {2011})},\ \Eprint {http://arxiv.org/abs/1105.3456}
  {arXiv:1105.3456 [gr-qc]} \BibitemShut {NoStop}%
\bibitem [{\citenamefont {{Ciufolini}}(1986)}]{1986PhRvL..56..278C}%
  \BibitemOpen
  \bibfield  {author} {\bibinfo {author} {\bibfnamefont {I.}~\bibnamefont
  {{Ciufolini}}},\ }\href {\doibase 10.1103/PhysRevLett.56.278} {\bibfield
  {journal} {\bibinfo  {journal} {\prl}\ }\textbf {\bibinfo {volume} {56}},\
  \bibinfo {pages} {278} (\bibinfo {year} {1986})}\BibitemShut {NoStop}%
\bibitem [{\citenamefont {{Ciufolini}}\ and\ \citenamefont
  {{Pavlis}}(2004)}]{2004Natur.431..958C}%
  \BibitemOpen
  \bibfield  {author} {\bibinfo {author} {\bibfnamefont {I.}~\bibnamefont
  {{Ciufolini}}}\ and\ \bibinfo {author} {\bibfnamefont {E.~C.}\ \bibnamefont
  {{Pavlis}}},\ }\href {\doibase 10.1038/nature03007} {\bibfield  {journal}
  {\bibinfo  {journal} {\nat}\ }\textbf {\bibinfo {volume} {431}},\ \bibinfo
  {pages} {958} (\bibinfo {year} {2004})}\BibitemShut {NoStop}%
\bibitem [{\citenamefont {Ciufolini}\ \emph {et~al.}(2019)\citenamefont
  {Ciufolini}, \citenamefont {Paolozzi}, \citenamefont {Pavlis}, \citenamefont
  {Sindoni}, \citenamefont {Ries}, \citenamefont {Matzner}, \citenamefont
  {Koenig}, \citenamefont {Paris}, \citenamefont {Gurzadyan},\ and\
  \citenamefont {Penrose}}]{Ciufolini:2019ezb}%
  \BibitemOpen
  \bibfield  {author} {\bibinfo {author} {\bibfnamefont {I.}~\bibnamefont
  {Ciufolini}}, \bibinfo {author} {\bibfnamefont {A.}~\bibnamefont {Paolozzi}},
  \bibinfo {author} {\bibfnamefont {E.~C.}\ \bibnamefont {Pavlis}}, \bibinfo
  {author} {\bibfnamefont {G.}~\bibnamefont {Sindoni}}, \bibinfo {author}
  {\bibfnamefont {J.}~\bibnamefont {Ries}}, \bibinfo {author} {\bibfnamefont
  {R.}~\bibnamefont {Matzner}}, \bibinfo {author} {\bibfnamefont
  {R.}~\bibnamefont {Koenig}}, \bibinfo {author} {\bibfnamefont
  {C.}~\bibnamefont {Paris}}, \bibinfo {author} {\bibfnamefont
  {V.}~\bibnamefont {Gurzadyan}}, \ and\ \bibinfo {author} {\bibfnamefont
  {R.}~\bibnamefont {Penrose}},\ }\href {\doibase
  10.1140/epjc/s10052-019-7386-z} {\bibfield  {journal} {\bibinfo  {journal}
  {Eur. Phys. J. C}\ }\textbf {\bibinfo {volume} {79}},\ \bibinfo {pages} {872}
  (\bibinfo {year} {2019})},\ \Eprint {http://arxiv.org/abs/1910.09908}
  {arXiv:1910.09908 [gr-qc]} \BibitemShut {NoStop}%
\bibitem [{\citenamefont {{Iorio}}\ \emph {et~al.}(2011)\citenamefont
  {{Iorio}}, \citenamefont {{Lichtenegger}}, \citenamefont {{Ruggiero}},\ and\
  \citenamefont {{Corda}}}]{2011Ap&SS.331..351I}%
  \BibitemOpen
  \bibfield  {author} {\bibinfo {author} {\bibfnamefont {L.}~\bibnamefont
  {{Iorio}}}, \bibinfo {author} {\bibfnamefont {H.~I.~M.}\ \bibnamefont
  {{Lichtenegger}}}, \bibinfo {author} {\bibfnamefont {M.~L.}\ \bibnamefont
  {{Ruggiero}}}, \ and\ \bibinfo {author} {\bibfnamefont {C.}~\bibnamefont
  {{Corda}}},\ }\href {\doibase 10.1007/s10509-010-0489-5} {\bibfield
  {journal} {\bibinfo  {journal} {Astrophysics and Space Science volume}\
  }\textbf {\bibinfo {volume} {331}},\ \bibinfo {pages} {351} (\bibinfo {year}
  {2011})},\ \Eprint {http://arxiv.org/abs/1009.3225} {arXiv:1009.3225 [gr-qc]}
  \BibitemShut {NoStop}%
\bibitem [{\citenamefont {Ciufolini}\ and\ \citenamefont
  {Wheeler}(1995)}]{ciufolini1995gravitation}%
  \BibitemOpen
  \bibfield  {author} {\bibinfo {author} {\bibfnamefont {I.}~\bibnamefont
  {Ciufolini}}\ and\ \bibinfo {author} {\bibfnamefont {J.~A.}\ \bibnamefont
  {Wheeler}},\ }\href@noop {} {\emph {\bibinfo {title} {Gravitation and
  inertia}}}\ (\bibinfo  {publisher} {Princeton University Press},\ \bibinfo
  {year} {1995})\BibitemShut {NoStop}%
\bibitem [{\citenamefont {Ruggiero}\ and\ \citenamefont
  {Tartaglia}(2002)}]{Ruggiero:2002hz}%
  \BibitemOpen
  \bibfield  {author} {\bibinfo {author} {\bibfnamefont {M.~L.}\ \bibnamefont
  {Ruggiero}}\ and\ \bibinfo {author} {\bibfnamefont {A.}~\bibnamefont
  {Tartaglia}},\ }\href@noop {} {\bibfield  {journal} {\bibinfo  {journal}
  {Nuovo Cim.}\ }\textbf {\bibinfo {volume} {B117}},\ \bibinfo {pages} {743}
  (\bibinfo {year} {2002})},\ \Eprint {http://arxiv.org/abs/gr-qc/0207065}
  {arXiv:gr-qc/0207065 [gr-qc]} \BibitemShut {NoStop}%
\bibitem [{\citenamefont {{Mashhoon}}(2007)}]{mashhoon03}%
  \BibitemOpen
  \bibfield  {author} {\bibinfo {author} {\bibfnamefont {B.}~\bibnamefont
  {{Mashhoon}}},\ }in\ \href@noop {} {\emph {\bibinfo {booktitle} {The
  measurement of gravitomagnetism: a challenging enterprise}}},\ \bibinfo
  {editor} {edited by\ \bibinfo {editor} {\bibfnamefont {L.}~\bibnamefont
  {Iorio}}}\ (\bibinfo  {publisher} {NOVA publishers},\ \bibinfo {year}
  {2007})\ \Eprint {http://arxiv.org/abs/gr-qc/0311030} {arXiv:gr-qc/0311030}
  \BibitemShut {NoStop}%
\bibitem [{\citenamefont {Iorio}(2007)}]{iorio2007measurement}%
  \BibitemOpen
  \bibinfo {editor} {\bibfnamefont {L.}~\bibnamefont {Iorio}},\ ed.,\
  \href@noop {} {\emph {\bibinfo {title} {The measurement of gravitomagnetism:
  a challenging enterprise}}}\ (\bibinfo  {publisher} {NOVA publishers},\
  \bibinfo {year} {2007})\BibitemShut {NoStop}%
\bibitem [{\citenamefont {Bailey}(2010)}]{Bailey:2010af}%
  \BibitemOpen
  \bibfield  {author} {\bibinfo {author} {\bibfnamefont {Q.~G.}\ \bibnamefont
  {Bailey}},\ }\href {\doibase 10.1103/PhysRevD.82.065012} {\bibfield
  {journal} {\bibinfo  {journal} {Phys. Rev. D}\ }\textbf {\bibinfo {volume}
  {82}},\ \bibinfo {pages} {065012} (\bibinfo {year} {2010})},\ \Eprint
  {http://arxiv.org/abs/1005.1435} {arXiv:1005.1435 [gr-qc]} \BibitemShut
  {NoStop}%
\bibitem [{\citenamefont {Tasson}(2012)}]{Tasson:2012nx}%
  \BibitemOpen
  \bibfield  {author} {\bibinfo {author} {\bibfnamefont {J.~D.}\ \bibnamefont
  {Tasson}},\ }\href {\doibase 10.1103/PhysRevD.86.124021} {\bibfield
  {journal} {\bibinfo  {journal} {Phys. Rev. D}\ }\textbf {\bibinfo {volume}
  {86}},\ \bibinfo {pages} {124021} (\bibinfo {year} {2012})},\ \Eprint
  {http://arxiv.org/abs/1211.4850} {arXiv:1211.4850 [hep-ph]} \BibitemShut
  {NoStop}%
\bibitem [{\citenamefont {Dass}\ and\ \citenamefont
  {Liberati}(2019)}]{Dass:2019kon}%
  \BibitemOpen
  \bibfield  {author} {\bibinfo {author} {\bibfnamefont {A.}~\bibnamefont
  {Dass}}\ and\ \bibinfo {author} {\bibfnamefont {S.}~\bibnamefont
  {Liberati}},\ }\href {\doibase 10.1007/s10714-019-2568-4} {\bibfield
  {journal} {\bibinfo  {journal} {Gen. Rel. Grav.}\ }\textbf {\bibinfo {volume}
  {51}},\ \bibinfo {pages} {84} (\bibinfo {year} {2019})},\ \Eprint
  {http://arxiv.org/abs/1903.10059} {arXiv:1903.10059 [gr-qc]} \BibitemShut
  {NoStop}%
\bibitem [{\citenamefont {Mashhoon}\ and\ \citenamefont
  {Hehl}(2019)}]{Mashhoon:2019jkq}%
  \BibitemOpen
  \bibfield  {author} {\bibinfo {author} {\bibfnamefont {B.}~\bibnamefont
  {Mashhoon}}\ and\ \bibinfo {author} {\bibfnamefont {F.~W.}\ \bibnamefont
  {Hehl}},\ }\href {\doibase 10.3390/universe5090195} {\bibfield  {journal}
  {\bibinfo  {journal} {Universe}\ }\textbf {\bibinfo {volume} {5}},\ \bibinfo
  {pages} {195} (\bibinfo {year} {2019})},\ \Eprint
  {http://arxiv.org/abs/1908.05431} {arXiv:1908.05431 [gr-qc]} \BibitemShut
  {NoStop}%
\bibitem [{\citenamefont {{Smith}}\ \emph {et~al.}(2008)\citenamefont
  {{Smith}}, \citenamefont {{Erickcek}}, \citenamefont {{Caldwell}},\ and\
  \citenamefont {{Kamionkowski}}}]{ref2.5}%
  \BibitemOpen
  \bibfield  {author} {\bibinfo {author} {\bibfnamefont {T.~L.}\ \bibnamefont
  {{Smith}}}, \bibinfo {author} {\bibfnamefont {A.~L.}\ \bibnamefont
  {{Erickcek}}}, \bibinfo {author} {\bibfnamefont {R.~R.}\ \bibnamefont
  {{Caldwell}}}, \ and\ \bibinfo {author} {\bibfnamefont {M.}~\bibnamefont
  {{Kamionkowski}}},\ }\href {\doibase 10.1103/PhysRevD.77.024015} {\bibfield
  {journal} {\bibinfo  {journal} {\prd}\ }\textbf {\bibinfo {volume} {77}},\
  \bibinfo {eid} {024015} (\bibinfo {year} {2008})},\ \Eprint
  {http://arxiv.org/abs/0708.0001} {arXiv:0708.0001 [astro-ph]} \BibitemShut
  {NoStop}%
\bibitem [{\citenamefont {Thorne}\ \emph {et~al.}(1986)\citenamefont {Thorne},
  \citenamefont {Thorne}, \citenamefont {Price},\ and\ \citenamefont
  {MacDonald}}]{membrane}%
  \BibitemOpen
  \bibfield  {author} {\bibinfo {author} {\bibfnamefont {K.~S.}\ \bibnamefont
  {Thorne}}, \bibinfo {author} {\bibfnamefont {K.~S.}\ \bibnamefont {Thorne}},
  \bibinfo {author} {\bibfnamefont {R.~H.}\ \bibnamefont {Price}}, \ and\
  \bibinfo {author} {\bibfnamefont {D.~A.}\ \bibnamefont {MacDonald}},\
  }\href@noop {} {\emph {\bibinfo {title} {Black holes: the membrane
  paradigm}}}\ (\bibinfo  {publisher} {Yale university press},\ \bibinfo {year}
  {1986})\BibitemShut {NoStop}%
\bibitem [{\citenamefont {Forward}(1961)}]{forward1961general}%
  \BibitemOpen
  \bibfield  {author} {\bibinfo {author} {\bibfnamefont {R.~L.}\ \bibnamefont
  {Forward}},\ }\href@noop {} {\bibfield  {journal} {\bibinfo  {journal}
  {Proceedings of the IRE}\ }\textbf {\bibinfo {volume} {49}},\ \bibinfo
  {pages} {892} (\bibinfo {year} {1961})}\BibitemShut {NoStop}%
\bibitem [{\citenamefont {Jantzen}\ \emph {et~al.}(1992)\citenamefont
  {Jantzen}, \citenamefont {Carini},\ and\ \citenamefont
  {Bini}}]{Jantzen:1992rg}%
  \BibitemOpen
  \bibfield  {author} {\bibinfo {author} {\bibfnamefont {R.~T.}\ \bibnamefont
  {Jantzen}}, \bibinfo {author} {\bibfnamefont {P.}~\bibnamefont {Carini}}, \
  and\ \bibinfo {author} {\bibfnamefont {D.}~\bibnamefont {Bini}},\ }\href
  {\doibase 10.1016/0003-4916(92)90297-Y} {\bibfield  {journal} {\bibinfo
  {journal} {Annals Phys.}\ }\textbf {\bibinfo {volume} {215}},\ \bibinfo
  {pages} {1} (\bibinfo {year} {1992})},\ \Eprint
  {http://arxiv.org/abs/gr-qc/0106043} {arXiv:gr-qc/0106043} \BibitemShut
  {NoStop}%
\bibitem [{\citenamefont {Jantzen}\ \emph {et~al.}(1996)\citenamefont
  {Jantzen}, \citenamefont {Carini},\ and\ \citenamefont
  {Bini}}]{Jantzen:1996au}%
  \BibitemOpen
  \bibfield  {author} {\bibinfo {author} {\bibfnamefont {R.~T.}\ \bibnamefont
  {Jantzen}}, \bibinfo {author} {\bibfnamefont {P.}~\bibnamefont {Carini}}, \
  and\ \bibinfo {author} {\bibfnamefont {D.}~\bibnamefont {Bini}},\ }in\
  \href@noop {} {\emph {\bibinfo {booktitle} {{7th Marcel Grossmann Meeting on
  General Relativity (MG 7)}}}}\ (\bibinfo {year} {1996})\ pp.\ \bibinfo
  {pages} {133--152},\ \Eprint {http://arxiv.org/abs/gr-qc/0105096}
  {arXiv:gr-qc/0105096} \BibitemShut {NoStop}%
\bibitem [{\citenamefont {{Lynden-Bell}}\ and\ \citenamefont
  {{Nouri-Zonoz}}(1998)}]{lynden}%
  \BibitemOpen
  \bibfield  {author} {\bibinfo {author} {\bibfnamefont {D.}~\bibnamefont
  {{Lynden-Bell}}}\ and\ \bibinfo {author} {\bibfnamefont {M.}~\bibnamefont
  {{Nouri-Zonoz}}},\ }\href {\doibase 10.1103/RevModPhys.70.427} {\bibfield
  {journal} {\bibinfo  {journal} {Reviews of Modern Physics}\ }\textbf
  {\bibinfo {volume} {70}},\ \bibinfo {pages} {427} (\bibinfo {year} {1998})},\
  \Eprint {http://arxiv.org/abs/gr-qc/9612049} {arXiv:gr-qc/9612049 [gr-qc]}
  \BibitemShut {NoStop}%
\bibitem [{\citenamefont {Bini}\ and\ \citenamefont
  {Jantzen}(2003)}]{Bini:2002mh}%
  \BibitemOpen
  \bibfield  {author} {\bibinfo {author} {\bibfnamefont {D.}~\bibnamefont
  {Bini}}\ and\ \bibinfo {author} {\bibfnamefont {R.~T.}\ \bibnamefont
  {Jantzen}},\ }\href@noop {} {\bibfield  {journal} {\bibinfo  {journal} {Nuovo
  Cim. B}\ }\textbf {\bibinfo {volume} {117}},\ \bibinfo {pages} {983}
  (\bibinfo {year} {2003})},\ \Eprint {http://arxiv.org/abs/gr-qc/0202085}
  {arXiv:gr-qc/0202085} \BibitemShut {NoStop}%
\bibitem [{\citenamefont {De~Felice}\ and\ \citenamefont
  {Bini}(2010)}]{de2010classical}%
  \BibitemOpen
  \bibfield  {author} {\bibinfo {author} {\bibfnamefont {F.}~\bibnamefont
  {De~Felice}}\ and\ \bibinfo {author} {\bibfnamefont {D.}~\bibnamefont
  {Bini}},\ }\href@noop {} {\emph {\bibinfo {title} {Classical measurements in
  curved space-times}}}\ (\bibinfo  {publisher} {Cambridge University Press},\
  \bibinfo {year} {2010})\BibitemShut {NoStop}%
\bibitem [{\citenamefont {{Costa}}\ and\ \citenamefont
  {{Nat{\'a}rio}}(2014)}]{natario}%
  \BibitemOpen
  \bibfield  {author} {\bibinfo {author} {\bibfnamefont {L.~F.~O.}\
  \bibnamefont {{Costa}}}\ and\ \bibinfo {author} {\bibfnamefont
  {J.}~\bibnamefont {{Nat{\'a}rio}}},\ }\href {\doibase
  10.1007/s10714-014-1792-1} {\bibfield  {journal} {\bibinfo  {journal}
  {General Relativity and Gravitation}\ }\textbf {\bibinfo {volume} {46}},\
  \bibinfo {eid} {1792} (\bibinfo {year} {2014})},\ \Eprint
  {http://arxiv.org/abs/1207.0465} {arXiv:1207.0465 [gr-qc]} \BibitemShut
  {NoStop}%
\bibitem [{\citenamefont {Costa}\ and\ \citenamefont
  {Nat\'ario}(2021)}]{Costa:2021atq}%
  \BibitemOpen
  \bibfield  {author} {\bibinfo {author} {\bibfnamefont {L.~F.~O.}\
  \bibnamefont {Costa}}\ and\ \bibinfo {author} {\bibfnamefont
  {J.}~\bibnamefont {Nat\'ario}},\ }\href {\doibase 10.3390/universe7100388}
  {\bibfield  {journal} {\bibinfo  {journal} {Universe}\ }\textbf {\bibinfo
  {volume} {7}},\ \bibinfo {pages} {388} (\bibinfo {year} {2021})},\ \Eprint
  {http://arxiv.org/abs/2109.14641} {arXiv:2109.14641 [gr-qc]} \BibitemShut
  {NoStop}%
\bibitem [{\citenamefont {Rizzi}\ and\ \citenamefont
  {Ruggiero}(2004{\natexlab{a}})}]{rizzi2004relativity}%
  \BibitemOpen
  \bibfield  {author} {\bibinfo {author} {\bibfnamefont {G.}~\bibnamefont
  {Rizzi}}\ and\ \bibinfo {author} {\bibfnamefont {M.~L.}\ \bibnamefont
  {Ruggiero}},\ }\href@noop {} {\emph {\bibinfo {title} {Relativity in rotating
  frames: relativistic physics in rotating reference frames}}}\ (\bibinfo
  {publisher} {Springer},\ \bibinfo {year} {2004})\BibitemShut {NoStop}%
\bibitem [{\citenamefont {Cattaneo}(1961)}]{ref2.1}%
  \BibitemOpen
  \bibfield  {author} {\bibinfo {author} {\bibfnamefont {C.}~\bibnamefont
  {Cattaneo}},\ }\href@noop {} {\emph {\bibinfo {title} {Introduzione alla
  teoria einsteiniana della gravitazione}}}\ (\bibinfo  {publisher} {Libreria
  Eredi Virgilio Veschi, Roma},\ \bibinfo {year} {1961})\BibitemShut {NoStop}%
\bibitem [{\citenamefont {Rizzi}\ and\ \citenamefont
  {Ruggiero}(2004{\natexlab{b}})}]{rizzi2004relativistic}%
  \BibitemOpen
  \bibfield  {author} {\bibinfo {author} {\bibfnamefont {G.}~\bibnamefont
  {Rizzi}}\ and\ \bibinfo {author} {\bibfnamefont {M.~L.}\ \bibnamefont
  {Ruggiero}},\ }\href@noop {} {\bibfield  {journal} {\bibinfo  {journal}
  {Relativity in Rotating Frames: Relativistic Physics in Rotating Reference
  Frames}\ ,\ \bibinfo {pages} {179}} (\bibinfo {year}
  {2004}{\natexlab{b}})}\BibitemShut {NoStop}%
\bibitem [{\citenamefont {{Rindler}}\ and\ \citenamefont
  {{Perlick}}(1990)}]{rindlerperlick}%
  \BibitemOpen
  \bibfield  {author} {\bibinfo {author} {\bibfnamefont {W.}~\bibnamefont
  {{Rindler}}}\ and\ \bibinfo {author} {\bibfnamefont {V.}~\bibnamefont
  {{Perlick}}},\ }\href {\doibase 10.1007/BF00757816} {\bibfield  {journal}
  {\bibinfo  {journal} {General Relativity and Gravitation}\ }\textbf {\bibinfo
  {volume} {22}},\ \bibinfo {pages} {1067} (\bibinfo {year}
  {1990})}\BibitemShut {NoStop}%
\bibitem [{\citenamefont {{Post}}(1967)}]{post}%
  \BibitemOpen
  \bibfield  {author} {\bibinfo {author} {\bibfnamefont {E.~J.}\ \bibnamefont
  {{Post}}},\ }\href {\doibase 10.1103/RevModPhys.39.475} {\bibfield  {journal}
  {\bibinfo  {journal} {Reviews of Modern Physics}\ }\textbf {\bibinfo {volume}
  {39}},\ \bibinfo {pages} {475} (\bibinfo {year} {1967})}\BibitemShut
  {NoStop}%
\bibitem [{\citenamefont {Aharonov}\ and\ \citenamefont
  {Bohm}(1959)}]{PhysRev.115.485}%
  \BibitemOpen
  \bibfield  {author} {\bibinfo {author} {\bibfnamefont {Y.}~\bibnamefont
  {Aharonov}}\ and\ \bibinfo {author} {\bibfnamefont {D.}~\bibnamefont
  {Bohm}},\ }\href {\doibase 10.1103/PhysRev.115.485} {\bibfield  {journal}
  {\bibinfo  {journal} {Phys. Rev.}\ }\textbf {\bibinfo {volume} {115}},\
  \bibinfo {pages} {485} (\bibinfo {year} {1959})}\BibitemShut {NoStop}%
\bibitem [{\citenamefont {{Rizzi}}\ and\ \citenamefont
  {{Ruggiero}}(2003)}]{2003GReGr..35.2129R}%
  \BibitemOpen
  \bibfield  {author} {\bibinfo {author} {\bibfnamefont {G.}~\bibnamefont
  {{Rizzi}}}\ and\ \bibinfo {author} {\bibfnamefont {M.~L.}\ \bibnamefont
  {{Ruggiero}}},\ }\href {\doibase 10.1023/A:1027345505786} {\bibfield
  {journal} {\bibinfo  {journal} {General Relativity and Gravitation}\ }\textbf
  {\bibinfo {volume} {35}},\ \bibinfo {pages} {2129} (\bibinfo {year}
  {2003})},\ \Eprint {http://arxiv.org/abs/gr-qc/0306128} {arXiv:gr-qc/0306128
  [gr-qc]} \BibitemShut {NoStop}%
\bibitem [{\citenamefont {Ruggiero}\ and\ \citenamefont
  {Tartaglia}(2015)}]{ruggiero2015note}%
  \BibitemOpen
  \bibfield  {author} {\bibinfo {author} {\bibfnamefont {M.~L.}\ \bibnamefont
  {Ruggiero}}\ and\ \bibinfo {author} {\bibfnamefont {A.}~\bibnamefont
  {Tartaglia}},\ }\href@noop {} {\bibfield  {journal} {\bibinfo  {journal} {The
  European Physical Journal Plus}\ }\textbf {\bibinfo {volume} {130}},\
  \bibinfo {pages} {1} (\bibinfo {year} {2015})}\BibitemShut {NoStop}%
\bibitem [{\citenamefont {{Tartaglia}}\ and\ \citenamefont
  {{Ruggiero}}(2021)}]{tartaglia_entropy}%
  \BibitemOpen
  \bibfield  {author} {\bibinfo {author} {\bibfnamefont {A.}~\bibnamefont
  {{Tartaglia}}}\ and\ \bibinfo {author} {\bibfnamefont {M.~L.}\ \bibnamefont
  {{Ruggiero}}},\ }\href {\doibase 10.48550/arXiv.2103.03763} {\bibfield
  {journal} {\bibinfo  {journal} {arXiv e-prints}\ ,\ \bibinfo {eid}
  {arXiv:2103.03763}} (\bibinfo {year} {2021})},\ \Eprint
  {http://arxiv.org/abs/2103.03763} {arXiv:2103.03763 [gr-qc]} \BibitemShut
  {NoStop}%
\bibitem [{\citenamefont {{Chow}}\ \emph {et~al.}(1985)\citenamefont {{Chow}},
  \citenamefont {{Gea-Banacloche}}, \citenamefont {{Pedrotti}}, \citenamefont
  {{Sanders}}, \citenamefont {{Schleich}},\ and\ \citenamefont
  {{Scully}}}]{lasergyro}%
  \BibitemOpen
  \bibfield  {author} {\bibinfo {author} {\bibfnamefont {W.~W.}\ \bibnamefont
  {{Chow}}}, \bibinfo {author} {\bibfnamefont {J.}~\bibnamefont
  {{Gea-Banacloche}}}, \bibinfo {author} {\bibfnamefont {L.~M.}\ \bibnamefont
  {{Pedrotti}}}, \bibinfo {author} {\bibfnamefont {V.~E.}\ \bibnamefont
  {{Sanders}}}, \bibinfo {author} {\bibfnamefont {W.}~\bibnamefont
  {{Schleich}}}, \ and\ \bibinfo {author} {\bibfnamefont {M.~O.}\ \bibnamefont
  {{Scully}}},\ }\href {\doibase 10.1103/RevModPhys.57.61} {\bibfield
  {journal} {\bibinfo  {journal} {Reviews of Modern Physics}\ }\textbf
  {\bibinfo {volume} {57}},\ \bibinfo {pages} {61} (\bibinfo {year}
  {1985})}\BibitemShut {NoStop}%
\bibitem [{\citenamefont {Ruggiero}(2005)}]{Ruggiero:2005nd}%
  \BibitemOpen
  \bibfield  {author} {\bibinfo {author} {\bibfnamefont {M.~L.}\ \bibnamefont
  {Ruggiero}},\ }\href {\doibase 10.1007/s10714-005-0190-0} {\bibfield
  {journal} {\bibinfo  {journal} {Gen. Rel. Grav.}\ }\textbf {\bibinfo {volume}
  {37}},\ \bibinfo {pages} {1845} (\bibinfo {year} {2005})},\ \Eprint
  {http://arxiv.org/abs/gr-qc/0510047} {arXiv:gr-qc/0510047} \BibitemShut
  {NoStop}%
\bibitem [{\citenamefont {Ruggiero}(2004)}]{Ruggiero:2004iaq}%
  \BibitemOpen
  \bibfield  {author} {\bibinfo {author} {\bibfnamefont {M.~L.}\ \bibnamefont
  {Ruggiero}},\ }\href {\doibase 10.1393/ncb/i2004-10185-7} {\bibfield
  {journal} {\bibinfo  {journal} {Nuovo Cim. B}\ }\textbf {\bibinfo {volume}
  {119}},\ \bibinfo {pages} {893} (\bibinfo {year} {2004})},\ \Eprint
  {http://arxiv.org/abs/1007.3857} {arXiv:1007.3857 [gr-qc]} \BibitemShut
  {NoStop}%
\bibitem [{\citenamefont {Ruggiero}(2021{\natexlab{a}})}]{Ruggiero:2021uag}%
  \BibitemOpen
  \bibfield  {author} {\bibinfo {author} {\bibfnamefont {M.~L.}\ \bibnamefont
  {Ruggiero}},\ }\href {\doibase 10.3390/universe7110451} {\bibfield  {journal}
  {\bibinfo  {journal} {Universe}\ }\textbf {\bibinfo {volume} {7}},\ \bibinfo
  {pages} {451} (\bibinfo {year} {2021}{\natexlab{a}})},\ \Eprint
  {http://arxiv.org/abs/2111.09008} {arXiv:2111.09008 [gr-qc]} \BibitemShut
  {NoStop}%
\bibitem [{\citenamefont {Straumann}(2013)}]{straumann2013applications}%
  \BibitemOpen
  \bibfield  {author} {\bibinfo {author} {\bibfnamefont {N.}~\bibnamefont
  {Straumann}},\ }\href@noop {} {\emph {\bibinfo {title} {General Relativity,
  With Applications to Astrophysics}}}\ (\bibinfo  {publisher} {Springer},\
  \bibinfo {year} {2013})\BibitemShut {NoStop}%
\bibitem [{\citenamefont {Bini}\ \emph {et~al.}(2008)\citenamefont {Bini},
  \citenamefont {Cherubini}, \citenamefont {Chicone},\ and\ \citenamefont
  {Mashhoon}}]{Bini:2008cy}%
  \BibitemOpen
  \bibfield  {author} {\bibinfo {author} {\bibfnamefont {D.}~\bibnamefont
  {Bini}}, \bibinfo {author} {\bibfnamefont {C.}~\bibnamefont {Cherubini}},
  \bibinfo {author} {\bibfnamefont {C.}~\bibnamefont {Chicone}}, \ and\
  \bibinfo {author} {\bibfnamefont {B.}~\bibnamefont {Mashhoon}},\ }\href
  {\doibase 10.1088/0264-9381/25/22/225014} {\bibfield  {journal} {\bibinfo
  {journal} {Class. Quant. Grav.}\ }\textbf {\bibinfo {volume} {25}},\ \bibinfo
  {pages} {225014} (\bibinfo {year} {2008})},\ \Eprint
  {http://arxiv.org/abs/0803.0390} {arXiv:0803.0390 [gr-qc]} \BibitemShut
  {NoStop}%
\bibitem [{\citenamefont {Bakopoulos}\ and\ \citenamefont
  {Kanti}(2014)}]{Bakopoulos:2014exa}%
  \BibitemOpen
  \bibfield  {author} {\bibinfo {author} {\bibfnamefont {A.}~\bibnamefont
  {Bakopoulos}}\ and\ \bibinfo {author} {\bibfnamefont {P.}~\bibnamefont
  {Kanti}},\ }\href {\doibase 10.1007/s10714-014-1742-y} {\bibfield  {journal}
  {\bibinfo  {journal} {Gen. Rel. Grav.}\ }\textbf {\bibinfo {volume} {46}},\
  \bibinfo {pages} {1742} (\bibinfo {year} {2014})},\ \Eprint
  {http://arxiv.org/abs/1405.0265} {arXiv:1405.0265 [gr-qc]} \BibitemShut
  {NoStop}%
\bibitem [{\citenamefont {Williams}\ and\ \citenamefont
  {Inan}(2021)}]{Williams:2020fgi}%
  \BibitemOpen
  \bibfield  {author} {\bibinfo {author} {\bibfnamefont {L.~L.}\ \bibnamefont
  {Williams}}\ and\ \bibinfo {author} {\bibfnamefont {N.}~\bibnamefont
  {Inan}},\ }\href {\doibase 10.1088/1367-2630/abf322} {\bibfield  {journal}
  {\bibinfo  {journal} {New J. Phys.}\ }\textbf {\bibinfo {volume} {23}},\
  \bibinfo {pages} {053019} (\bibinfo {year} {2021})},\ \Eprint
  {http://arxiv.org/abs/2012.08077} {arXiv:2012.08077 [gr-qc]} \BibitemShut
  {NoStop}%
\bibitem [{\citenamefont {{Hobson}}\ \emph {et~al.}(2005)\citenamefont
  {{Hobson}}, \citenamefont {{Efstathiou}},\ and\ \citenamefont
  {{Lasenby}}}]{ref3.2}%
  \BibitemOpen
  \bibfield  {author} {\bibinfo {author} {\bibfnamefont {M.~P.}\ \bibnamefont
  {{Hobson}}}, \bibinfo {author} {\bibfnamefont {G.~P.}\ \bibnamefont
  {{Efstathiou}}}, \ and\ \bibinfo {author} {\bibfnamefont {A.~N.}\
  \bibnamefont {{Lasenby}}},\ }\href@noop {} {\emph {\bibinfo {title} {{General
  Relativity}}}}\ (\bibinfo {year} {2005})\BibitemShut {NoStop}%
\bibitem [{\citenamefont {Ruggiero}\ and\ \citenamefont
  {Ortolan}(2020{\natexlab{a}})}]{Ruggiero_2020}%
  \BibitemOpen
  \bibfield  {author} {\bibinfo {author} {\bibfnamefont {M.~L.}\ \bibnamefont
  {Ruggiero}}\ and\ \bibinfo {author} {\bibfnamefont {A.}~\bibnamefont
  {Ortolan}},\ }\href {\doibase 10.1088/2399-6528/ab9320} {\bibfield  {journal}
  {\bibinfo  {journal} {Journal of Physics Communications}\ }\textbf {\bibinfo
  {volume} {4}},\ \bibinfo {pages} {055013} (\bibinfo {year}
  {2020}{\natexlab{a}})}\BibitemShut {NoStop}%
\bibitem [{\citenamefont {Ruggiero}(2022)}]{Ruggiero:2022gzl}%
  \BibitemOpen
  \bibfield  {author} {\bibinfo {author} {\bibfnamefont {M.~L.}\ \bibnamefont
  {Ruggiero}},\ }\href {\doibase 10.1007/s10714-022-02983-8} {\bibfield
  {journal} {\bibinfo  {journal} {Gen. Rel. Grav.}\ }\textbf {\bibinfo {volume}
  {54}},\ \bibinfo {pages} {97} (\bibinfo {year} {2022})},\ \Eprint
  {http://arxiv.org/abs/2204.08914} {arXiv:2204.08914 [gr-qc]} \BibitemShut
  {NoStop}%
\bibitem [{\citenamefont {Manasse}\ and\ \citenamefont
  {Misner}(1963)}]{manasse1963fermi}%
  \BibitemOpen
  \bibfield  {author} {\bibinfo {author} {\bibfnamefont {F.}~\bibnamefont
  {Manasse}}\ and\ \bibinfo {author} {\bibfnamefont {C.~W.}\ \bibnamefont
  {Misner}},\ }\href@noop {} {\bibfield  {journal} {\bibinfo  {journal}
  {Journal of mathematical physics}\ }\textbf {\bibinfo {volume} {4}},\
  \bibinfo {pages} {735} (\bibinfo {year} {1963})}\BibitemShut {NoStop}%
\bibitem [{\citenamefont {Misner}\ \emph {et~al.}(1973)\citenamefont {Misner},
  \citenamefont {Thorne},\ and\ \citenamefont {Wheeler}}]{MTW}%
  \BibitemOpen
  \bibfield  {author} {\bibinfo {author} {\bibfnamefont {C.~W.}\ \bibnamefont
  {Misner}}, \bibinfo {author} {\bibfnamefont {K.~S.}\ \bibnamefont {Thorne}},
  \ and\ \bibinfo {author} {\bibfnamefont {J.~A.}\ \bibnamefont {Wheeler}},\
  }\href@noop {} {\emph {\bibinfo {title} {Gravitation}}}\ (\bibinfo
  {publisher} {San Francisco: WH Freeman and Co.},\ \bibinfo {year}
  {1973})\BibitemShut {NoStop}%
\bibitem [{\citenamefont {{Mashhoon}}(1993)}]{larmor}%
  \BibitemOpen
  \bibfield  {author} {\bibinfo {author} {\bibfnamefont {B.}~\bibnamefont
  {{Mashhoon}}},\ }\href {\doibase 10.1016/0375-9601(93)90248-X} {\bibfield
  {journal} {\bibinfo  {journal} {Physics Letters A}\ }\textbf {\bibinfo
  {volume} {173}},\ \bibinfo {pages} {347} (\bibinfo {year}
  {1993})}\BibitemShut {NoStop}%
\bibitem [{\citenamefont {Lyne}\ \emph {et~al.}(2004)\citenamefont {Lyne} \emph
  {et~al.}}]{Lyne:2004cj}%
  \BibitemOpen
  \bibfield  {author} {\bibinfo {author} {\bibfnamefont {A.~G.}\ \bibnamefont
  {Lyne}} \emph {et~al.},\ }\href {\doibase 10.1126/science.1094645} {\bibfield
   {journal} {\bibinfo  {journal} {Science}\ }\textbf {\bibinfo {volume}
  {303}},\ \bibinfo {pages} {1153} (\bibinfo {year} {2004})},\ \Eprint
  {http://arxiv.org/abs/astro-ph/0401086} {arXiv:astro-ph/0401086} \BibitemShut
  {NoStop}%
\bibitem [{\citenamefont {Shao}\ \emph {et~al.}(2015)\citenamefont {Shao} \emph
  {et~al.}}]{Shao:2014wja}%
  \BibitemOpen
  \bibfield  {author} {\bibinfo {author} {\bibfnamefont {L.}~\bibnamefont
  {Shao}} \emph {et~al.},\ }\href {\doibase 10.22323/1.215.0042} {\bibfield
  {journal} {\bibinfo  {journal} {PoS}\ }\textbf {\bibinfo {volume}
  {AASKA14}},\ \bibinfo {pages} {042} (\bibinfo {year} {2015})},\ \Eprint
  {http://arxiv.org/abs/1501.00058} {arXiv:1501.00058 [astro-ph.HE]}
  \BibitemShut {NoStop}%
\bibitem [{\citenamefont {{Venkatraman Krishnan}}\ \emph
  {et~al.}(2020)\citenamefont {{Venkatraman Krishnan}}, \citenamefont
  {{Bailes}}, \citenamefont {{van Straten}}, \citenamefont {{Wex}},
  \citenamefont {{Freire}}, \citenamefont {{Keane}}, \citenamefont {{Tauris}},
  \citenamefont {{Rosado}}, \citenamefont {{Bhat}}, \citenamefont {{Flynn}},
  \citenamefont {{Jameson}},\ and\ \citenamefont
  {{Os{\l}owski}}}]{2020Sci...367..577V}%
  \BibitemOpen
  \bibfield  {author} {\bibinfo {author} {\bibfnamefont {V.}~\bibnamefont
  {{Venkatraman Krishnan}}}, \bibinfo {author} {\bibfnamefont {M.}~\bibnamefont
  {{Bailes}}}, \bibinfo {author} {\bibfnamefont {W.}~\bibnamefont {{van
  Straten}}}, \bibinfo {author} {\bibfnamefont {N.}~\bibnamefont {{Wex}}},
  \bibinfo {author} {\bibfnamefont {P.~C.~C.}\ \bibnamefont {{Freire}}},
  \bibinfo {author} {\bibfnamefont {E.~F.}\ \bibnamefont {{Keane}}}, \bibinfo
  {author} {\bibfnamefont {T.~M.}\ \bibnamefont {{Tauris}}}, \bibinfo {author}
  {\bibfnamefont {P.~A.}\ \bibnamefont {{Rosado}}}, \bibinfo {author}
  {\bibfnamefont {N.~D.~R.}\ \bibnamefont {{Bhat}}}, \bibinfo {author}
  {\bibfnamefont {C.}~\bibnamefont {{Flynn}}}, \bibinfo {author} {\bibfnamefont
  {A.}~\bibnamefont {{Jameson}}}, \ and\ \bibinfo {author} {\bibfnamefont
  {S.}~\bibnamefont {{Os{\l}owski}}},\ }\href {\doibase
  10.1126/science.aax7007} {\bibfield  {journal} {\bibinfo  {journal}
  {Science}\ }\textbf {\bibinfo {volume} {367}},\ \bibinfo {pages} {577}
  (\bibinfo {year} {2020})},\ \Eprint {http://arxiv.org/abs/2001.11405}
  {arXiv:2001.11405 [astro-ph.HE]} \BibitemShut {NoStop}%
\bibitem [{\citenamefont {O'Connell}(2004)}]{OConnell:2004btu}%
  \BibitemOpen
  \bibfield  {author} {\bibinfo {author} {\bibfnamefont {R.~F.}\ \bibnamefont
  {O'Connell}},\ }\href {\doibase 10.1103/PhysRevLett.93.081103} {\bibfield
  {journal} {\bibinfo  {journal} {Phys. Rev. Lett.}\ }\textbf {\bibinfo
  {volume} {93}},\ \bibinfo {pages} {081103} (\bibinfo {year} {2004})},\
  \Eprint {http://arxiv.org/abs/gr-qc/0409101} {arXiv:gr-qc/0409101}
  \BibitemShut {NoStop}%
\bibitem [{\citenamefont {O'Connell}(2008)}]{OConnell:2008wyk}%
  \BibitemOpen
  \bibfield  {author} {\bibinfo {author} {\bibfnamefont {R.~F.}\ \bibnamefont
  {O'Connell}}\ }(\bibinfo {year} {2008})\ \Eprint
  {http://arxiv.org/abs/0804.3806} {arXiv:0804.3806 [gr-qc]} \BibitemShut
  {NoStop}%
\bibitem [{\citenamefont {{O'Connell}}(2010)}]{OC}%
  \BibitemOpen
  \bibfield  {author} {\bibinfo {author} {\bibfnamefont {R.~F.}\ \bibnamefont
  {{O'Connell}}},\ }in\ \href {\doibase 10.1007/978-90-481-3735-0_14} {\emph
  {\bibinfo {booktitle} {General Relativity and John Archibald Wheeler}}},\
  \bibinfo {series} {Astrophysics and Space Science Library}, Vol.\ \bibinfo
  {volume} {367},\ \bibinfo {editor} {edited by\ \bibinfo {editor}
  {\bibfnamefont {I.}~\bibnamefont {{Ciufolini}}}\ and\ \bibinfo {editor}
  {\bibfnamefont {R.~A.~A.}\ \bibnamefont {{Matzner}}}}\ (\bibinfo {year}
  {2010})\ p.\ \bibinfo {pages} {325},\ \Eprint
  {http://arxiv.org/abs/1009.4401} {arXiv:1009.4401 [gr-qc]} \BibitemShut
  {NoStop}%
\bibitem [{\citenamefont {Kaplan}\ \emph {et~al.}(2009)\citenamefont {Kaplan},
  \citenamefont {Nichols},\ and\ \citenamefont {Thorne}}]{Kaplan:2008dh}%
  \BibitemOpen
  \bibfield  {author} {\bibinfo {author} {\bibfnamefont {J.~D.}\ \bibnamefont
  {Kaplan}}, \bibinfo {author} {\bibfnamefont {D.~A.}\ \bibnamefont {Nichols}},
  \ and\ \bibinfo {author} {\bibfnamefont {K.~S.}\ \bibnamefont {Thorne}},\
  }\href {\doibase 10.1103/PhysRevD.80.124014} {\bibfield  {journal} {\bibinfo
  {journal} {Phys. Rev. D}\ }\textbf {\bibinfo {volume} {80}},\ \bibinfo
  {pages} {124014} (\bibinfo {year} {2009})},\ \Eprint
  {http://arxiv.org/abs/0808.2510} {arXiv:0808.2510 [gr-qc]} \BibitemShut
  {NoStop}%
\bibitem [{\citenamefont {Mashhoon}(2008)}]{Mashhoon:2008kq}%
  \BibitemOpen
  \bibfield  {author} {\bibinfo {author} {\bibfnamefont {B.}~\bibnamefont
  {Mashhoon}},\ }\href {\doibase 10.1088/0264-9381/25/8/085014} {\bibfield
  {journal} {\bibinfo  {journal} {Class. Quant. Grav.}\ }\textbf {\bibinfo
  {volume} {25}},\ \bibinfo {pages} {085014} (\bibinfo {year} {2008})},\
  \Eprint {http://arxiv.org/abs/0802.1356} {arXiv:0802.1356 [gr-qc]}
  \BibitemShut {NoStop}%
\bibitem [{\citenamefont {Costa}\ \emph
  {et~al.}(2021{\natexlab{a}})\citenamefont {Costa}, \citenamefont
  {Nat\'ario},\ and\ \citenamefont {Santos}}]{Costa:2019loe}%
  \BibitemOpen
  \bibfield  {author} {\bibinfo {author} {\bibfnamefont {L.~F.~O.}\
  \bibnamefont {Costa}}, \bibinfo {author} {\bibfnamefont {J.}~\bibnamefont
  {Nat\'ario}}, \ and\ \bibinfo {author} {\bibfnamefont {N.~O.}\ \bibnamefont
  {Santos}},\ }\href {\doibase 10.1088/1361-6382/abc570} {\bibfield  {journal}
  {\bibinfo  {journal} {Class. Quant. Grav.}\ }\textbf {\bibinfo {volume}
  {38}},\ \bibinfo {pages} {055003} (\bibinfo {year} {2021}{\natexlab{a}})},\
  \Eprint {http://arxiv.org/abs/1912.09407} {arXiv:1912.09407 [gr-qc]}
  \BibitemShut {NoStop}%
\bibitem [{\citenamefont {Herrera}\ \emph {et~al.}(2007)\citenamefont
  {Herrera}, \citenamefont {Carot},\ and\ \citenamefont
  {Di~Prisco}}]{Herrera:2007au}%
  \BibitemOpen
  \bibfield  {author} {\bibinfo {author} {\bibfnamefont {L.}~\bibnamefont
  {Herrera}}, \bibinfo {author} {\bibfnamefont {J.}~\bibnamefont {Carot}}, \
  and\ \bibinfo {author} {\bibfnamefont {A.}~\bibnamefont {Di~Prisco}},\ }\href
  {\doibase 10.1103/PhysRevD.76.044012} {\bibfield  {journal} {\bibinfo
  {journal} {Phys. Rev. D}\ }\textbf {\bibinfo {volume} {76}},\ \bibinfo
  {pages} {044012} (\bibinfo {year} {2007})},\ \Eprint
  {http://arxiv.org/abs/0707.0867} {arXiv:0707.0867 [gr-qc]} \BibitemShut
  {NoStop}%
\bibitem [{\citenamefont {Herrera}(2021)}]{Herrera:2021aqv}%
  \BibitemOpen
  \bibfield  {author} {\bibinfo {author} {\bibfnamefont {L.}~\bibnamefont
  {Herrera}},\ }\href@noop {} {\bibfield  {journal} {\bibinfo  {journal}
  {Universe}\ }\textbf {\bibinfo {volume} {7}},\ \bibinfo {pages} {27}
  (\bibinfo {year} {2021})},\ \Eprint {http://arxiv.org/abs/2102.03552}
  {arXiv:2102.03552 [gr-qc]} \BibitemShut {NoStop}%
\bibitem [{\citenamefont {{Cohen}}\ and\ \citenamefont
  {{Mashhoon}}(1993)}]{clock}%
  \BibitemOpen
  \bibfield  {author} {\bibinfo {author} {\bibfnamefont {J.~M.}\ \bibnamefont
  {{Cohen}}}\ and\ \bibinfo {author} {\bibfnamefont {B.}~\bibnamefont
  {{Mashhoon}}},\ }\href {\doibase 10.1016/0375-9601(93)90387-F} {\bibfield
  {journal} {\bibinfo  {journal} {Physics Letters A}\ }\textbf {\bibinfo
  {volume} {181}},\ \bibinfo {pages} {353} (\bibinfo {year}
  {1993})}\BibitemShut {NoStop}%
\bibitem [{\citenamefont {Lichtenegger}\ \emph {et~al.}(2006)\citenamefont
  {Lichtenegger}, \citenamefont {Iorio},\ and\ \citenamefont
  {Mashhoon}}]{Lichtenegger:2002af}%
  \BibitemOpen
  \bibfield  {author} {\bibinfo {author} {\bibfnamefont {H.~I.~M.}\
  \bibnamefont {Lichtenegger}}, \bibinfo {author} {\bibfnamefont
  {L.}~\bibnamefont {Iorio}}, \ and\ \bibinfo {author} {\bibfnamefont
  {B.}~\bibnamefont {Mashhoon}},\ }\href {\doibase 10.1002/andp.200610214}
  {\bibfield  {journal} {\bibinfo  {journal} {Annalen Phys.}\ }\textbf
  {\bibinfo {volume} {15}},\ \bibinfo {pages} {868} (\bibinfo {year} {2006})},\
  \Eprint {http://arxiv.org/abs/gr-qc/0211108} {arXiv:gr-qc/0211108}
  \BibitemShut {NoStop}%
\bibitem [{\citenamefont {Hackmann}\ and\ \citenamefont
  {L\"ammerzahl}(2014)}]{Hackmann:2014aga}%
  \BibitemOpen
  \bibfield  {author} {\bibinfo {author} {\bibfnamefont {E.}~\bibnamefont
  {Hackmann}}\ and\ \bibinfo {author} {\bibfnamefont {C.}~\bibnamefont
  {L\"ammerzahl}},\ }\href {\doibase 10.1103/PhysRevD.90.044059} {\bibfield
  {journal} {\bibinfo  {journal} {Phys. Rev. D}\ }\textbf {\bibinfo {volume}
  {90}},\ \bibinfo {pages} {044059} (\bibinfo {year} {2014})},\ \Eprint
  {http://arxiv.org/abs/1406.6232} {arXiv:1406.6232 [gr-qc]} \BibitemShut
  {NoStop}%
\bibitem [{\citenamefont {Owen}\ \emph {et~al.}(2011)\citenamefont {Owen} \emph
  {et~al.}}]{Owen:2010fa}%
  \BibitemOpen
  \bibfield  {author} {\bibinfo {author} {\bibfnamefont {R.}~\bibnamefont
  {Owen}} \emph {et~al.},\ }\href {\doibase 10.1103/PhysRevLett.106.151101}
  {\bibfield  {journal} {\bibinfo  {journal} {Phys. Rev. Lett.}\ }\textbf
  {\bibinfo {volume} {106}},\ \bibinfo {pages} {151101} (\bibinfo {year}
  {2011})},\ \Eprint {http://arxiv.org/abs/1012.4869} {arXiv:1012.4869 [gr-qc]}
  \BibitemShut {NoStop}%
\bibitem [{\citenamefont {Nichols}\ \emph {et~al.}(2011)\citenamefont {Nichols}
  \emph {et~al.}}]{Nichols:2011pu}%
  \BibitemOpen
  \bibfield  {author} {\bibinfo {author} {\bibfnamefont {D.~A.}\ \bibnamefont
  {Nichols}} \emph {et~al.},\ }\href {\doibase 10.1103/PhysRevD.84.124014}
  {\bibfield  {journal} {\bibinfo  {journal} {Phys. Rev. D}\ }\textbf {\bibinfo
  {volume} {84}},\ \bibinfo {pages} {124014} (\bibinfo {year} {2011})},\
  \Eprint {http://arxiv.org/abs/1108.5486} {arXiv:1108.5486 [gr-qc]}
  \BibitemShut {NoStop}%
\bibitem [{\citenamefont {Zhang}\ \emph {et~al.}(2012)\citenamefont {Zhang},
  \citenamefont {Zimmerman}, \citenamefont {Nichols}, \citenamefont {Chen},
  \citenamefont {Lovelace}, \citenamefont {Matthews}, \citenamefont {Owen},\
  and\ \citenamefont {Thorne}}]{Zhang:2012jj}%
  \BibitemOpen
  \bibfield  {author} {\bibinfo {author} {\bibfnamefont {F.}~\bibnamefont
  {Zhang}}, \bibinfo {author} {\bibfnamefont {A.}~\bibnamefont {Zimmerman}},
  \bibinfo {author} {\bibfnamefont {D.~A.}\ \bibnamefont {Nichols}}, \bibinfo
  {author} {\bibfnamefont {Y.}~\bibnamefont {Chen}}, \bibinfo {author}
  {\bibfnamefont {G.}~\bibnamefont {Lovelace}}, \bibinfo {author}
  {\bibfnamefont {K.~D.}\ \bibnamefont {Matthews}}, \bibinfo {author}
  {\bibfnamefont {R.}~\bibnamefont {Owen}}, \ and\ \bibinfo {author}
  {\bibfnamefont {K.~S.}\ \bibnamefont {Thorne}},\ }\href {\doibase
  10.1103/PhysRevD.86.084049} {\bibfield  {journal} {\bibinfo  {journal} {Phys.
  Rev. D}\ }\textbf {\bibinfo {volume} {86}},\ \bibinfo {pages} {084049}
  (\bibinfo {year} {2012})},\ \Eprint {http://arxiv.org/abs/1208.3034}
  {arXiv:1208.3034 [gr-qc]} \BibitemShut {NoStop}%
\bibitem [{\citenamefont {Nichols}\ \emph {et~al.}(2012)\citenamefont
  {Nichols}, \citenamefont {Zimmerman}, \citenamefont {Chen}, \citenamefont
  {Lovelace}, \citenamefont {Matthews}, \citenamefont {Owen}, \citenamefont
  {Zhang},\ and\ \citenamefont {Thorne}}]{Nichols:2012jn}%
  \BibitemOpen
  \bibfield  {author} {\bibinfo {author} {\bibfnamefont {D.~A.}\ \bibnamefont
  {Nichols}}, \bibinfo {author} {\bibfnamefont {A.}~\bibnamefont {Zimmerman}},
  \bibinfo {author} {\bibfnamefont {Y.}~\bibnamefont {Chen}}, \bibinfo {author}
  {\bibfnamefont {G.}~\bibnamefont {Lovelace}}, \bibinfo {author}
  {\bibfnamefont {K.~D.}\ \bibnamefont {Matthews}}, \bibinfo {author}
  {\bibfnamefont {R.}~\bibnamefont {Owen}}, \bibinfo {author} {\bibfnamefont
  {F.}~\bibnamefont {Zhang}}, \ and\ \bibinfo {author} {\bibfnamefont {K.~S.}\
  \bibnamefont {Thorne}},\ }\href {\doibase 10.1103/PhysRevD.86.104028}
  {\bibfield  {journal} {\bibinfo  {journal} {Phys. Rev. D}\ }\textbf {\bibinfo
  {volume} {86}},\ \bibinfo {pages} {104028} (\bibinfo {year} {2012})},\
  \Eprint {http://arxiv.org/abs/1208.3038} {arXiv:1208.3038 [gr-qc]}
  \BibitemShut {NoStop}%
\bibitem [{\citenamefont {Mashhoon}(2021)}]{Mashhoon:2021qtc}%
  \BibitemOpen
  \bibfield  {author} {\bibinfo {author} {\bibfnamefont {B.}~\bibnamefont
  {Mashhoon}},\ }\href {\doibase 10.3390/e23040445} {\bibfield  {journal}
  {\bibinfo  {journal} {Entropy}\ }\textbf {\bibinfo {volume} {23}},\ \bibinfo
  {pages} {445} (\bibinfo {year} {2021})},\ \Eprint
  {http://arxiv.org/abs/2102.06433} {arXiv:2102.06433 [gr-qc]} \BibitemShut
  {NoStop}%
\bibitem [{\citenamefont {Costa}\ \emph {et~al.}(2016)\citenamefont {Costa},
  \citenamefont {Nat\'ario},\ and\ \citenamefont
  {Zilh\~ao}}]{PhysRevD.93.104006}%
  \BibitemOpen
  \bibfield  {author} {\bibinfo {author} {\bibfnamefont {L.~F.~O.}\
  \bibnamefont {Costa}}, \bibinfo {author} {\bibfnamefont {J.}~\bibnamefont
  {Nat\'ario}}, \ and\ \bibinfo {author} {\bibfnamefont {M.}~\bibnamefont
  {Zilh\~ao}},\ }\href {\doibase 10.1103/PhysRevD.93.104006} {\bibfield
  {journal} {\bibinfo  {journal} {Phys. Rev. D}\ }\textbf {\bibinfo {volume}
  {93}},\ \bibinfo {pages} {104006} (\bibinfo {year} {2016})}\BibitemShut
  {NoStop}%
\bibitem [{\citenamefont {Bini}\ \emph {et~al.}(2017)\citenamefont {Bini},
  \citenamefont {Geralico},\ and\ \citenamefont {Ortolan}}]{biniortolan2017}%
  \BibitemOpen
  \bibfield  {author} {\bibinfo {author} {\bibfnamefont {D.}~\bibnamefont
  {Bini}}, \bibinfo {author} {\bibfnamefont {A.}~\bibnamefont {Geralico}}, \
  and\ \bibinfo {author} {\bibfnamefont {A.}~\bibnamefont {Ortolan}},\ }\href
  {\doibase 10.1103/PhysRevD.95.104044} {\bibfield  {journal} {\bibinfo
  {journal} {Phys. Rev. D}\ }\textbf {\bibinfo {volume} {95}},\ \bibinfo
  {pages} {104044} (\bibinfo {year} {2017})}\BibitemShut {NoStop}%
\bibitem [{\citenamefont {Chakraborty}(2022)}]{Chakraborty:2021bsb}%
  \BibitemOpen
  \bibfield  {author} {\bibinfo {author} {\bibfnamefont {C.}~\bibnamefont
  {Chakraborty}},\ }\href {\doibase 10.1103/PhysRevD.105.064072} {\bibfield
  {journal} {\bibinfo  {journal} {Phys. Rev. D}\ }\textbf {\bibinfo {volume}
  {105}},\ \bibinfo {pages} {064072} (\bibinfo {year} {2022})},\ \Eprint
  {http://arxiv.org/abs/2106.03520} {arXiv:2106.03520 [gr-qc]} \BibitemShut
  {NoStop}%
\bibitem [{\citenamefont {Bini}\ \emph {et~al.}(2022)\citenamefont {Bini},
  \citenamefont {Mashhoon},\ and\ \citenamefont {Obukhov}}]{Bini:2021gdb}%
  \BibitemOpen
  \bibfield  {author} {\bibinfo {author} {\bibfnamefont {D.}~\bibnamefont
  {Bini}}, \bibinfo {author} {\bibfnamefont {B.}~\bibnamefont {Mashhoon}}, \
  and\ \bibinfo {author} {\bibfnamefont {Y.~N.}\ \bibnamefont {Obukhov}},\
  }\href {\doibase 10.1103/PhysRevD.105.064028} {\bibfield  {journal} {\bibinfo
   {journal} {Phys. Rev. D}\ }\textbf {\bibinfo {volume} {105}},\ \bibinfo
  {pages} {064028} (\bibinfo {year} {2022})},\ \Eprint
  {http://arxiv.org/abs/2112.07550} {arXiv:2112.07550 [gr-qc]} \BibitemShut
  {NoStop}%
\bibitem [{\citenamefont {Ahmedov}\ and\ \citenamefont
  {Rakhmatov}(2003)}]{Ahmedov:2003fd}%
  \BibitemOpen
  \bibfield  {author} {\bibinfo {author} {\bibfnamefont {B.~J.}\ \bibnamefont
  {Ahmedov}}\ and\ \bibinfo {author} {\bibfnamefont {N.~I.}\ \bibnamefont
  {Rakhmatov}},\ }\href {\doibase 10.1023/A:1023722704270} {\bibfield
  {journal} {\bibinfo  {journal} {Found. Phys.}\ }\textbf {\bibinfo {volume}
  {33}},\ \bibinfo {pages} {625} (\bibinfo {year} {2003})},\ \Eprint
  {http://arxiv.org/abs/gr-qc/0608036} {arXiv:gr-qc/0608036} \BibitemShut
  {NoStop}%
\bibitem [{\citenamefont {Abdujabbarov}\ \emph {et~al.}(2008)\citenamefont
  {Abdujabbarov}, \citenamefont {Ahmedov},\ and\ \citenamefont
  {Kagramanova}}]{Abdujabbarov:2008mz}%
  \BibitemOpen
  \bibfield  {author} {\bibinfo {author} {\bibfnamefont {A.~A.}\ \bibnamefont
  {Abdujabbarov}}, \bibinfo {author} {\bibfnamefont {B.~J.}\ \bibnamefont
  {Ahmedov}}, \ and\ \bibinfo {author} {\bibfnamefont {V.~G.}\ \bibnamefont
  {Kagramanova}},\ }\href {\doibase 10.1007/s10714-008-0635-3} {\bibfield
  {journal} {\bibinfo  {journal} {Gen. Rel. Grav.}\ }\textbf {\bibinfo {volume}
  {40}},\ \bibinfo {pages} {2515} (\bibinfo {year} {2008})},\ \Eprint
  {http://arxiv.org/abs/0802.4349} {arXiv:0802.4349 [gr-qc]} \BibitemShut
  {NoStop}%
\bibitem [{\citenamefont {Herrera}\ \emph {et~al.}(2006)\citenamefont
  {Herrera}, \citenamefont {Gonzalez}, \citenamefont {Pachon},\ and\
  \citenamefont {Rueda}}]{Herrera:2006cw}%
  \BibitemOpen
  \bibfield  {author} {\bibinfo {author} {\bibfnamefont {L.}~\bibnamefont
  {Herrera}}, \bibinfo {author} {\bibfnamefont {G.~A.}\ \bibnamefont
  {Gonzalez}}, \bibinfo {author} {\bibfnamefont {L.~A.}\ \bibnamefont
  {Pachon}}, \ and\ \bibinfo {author} {\bibfnamefont {J.~A.}\ \bibnamefont
  {Rueda}},\ }\href {\doibase 10.1088/0264-9381/23/7/011} {\bibfield  {journal}
  {\bibinfo  {journal} {Class. Quant. Grav.}\ }\textbf {\bibinfo {volume}
  {23}},\ \bibinfo {pages} {2395} (\bibinfo {year} {2006})},\ \Eprint
  {http://arxiv.org/abs/gr-qc/0602040} {arXiv:gr-qc/0602040} \BibitemShut
  {NoStop}%
\bibitem [{\citenamefont {Herrera}\ and\ \citenamefont
  {Barreto}(2012)}]{Herrera:2012ke}%
  \BibitemOpen
  \bibfield  {author} {\bibinfo {author} {\bibfnamefont {L.}~\bibnamefont
  {Herrera}}\ and\ \bibinfo {author} {\bibfnamefont {W.}~\bibnamefont
  {Barreto}},\ }\href {\doibase 10.1103/PhysRevD.86.064014} {\bibfield
  {journal} {\bibinfo  {journal} {Phys. Rev. D}\ }\textbf {\bibinfo {volume}
  {86}},\ \bibinfo {pages} {064014} (\bibinfo {year} {2012})},\ \Eprint
  {http://arxiv.org/abs/1206.0413} {arXiv:1206.0413 [gr-qc]} \BibitemShut
  {NoStop}%
\bibitem [{\citenamefont {Guti\'errez-Ruiz}\ and\ \citenamefont
  {Pach\'on}(2015)}]{Gutierrez-Ruiz:2015cnp}%
  \BibitemOpen
  \bibfield  {author} {\bibinfo {author} {\bibfnamefont {A.~F.}\ \bibnamefont
  {Guti\'errez-Ruiz}}\ and\ \bibinfo {author} {\bibfnamefont {L.~A.}\
  \bibnamefont {Pach\'on}},\ }\href {\doibase 10.1103/PhysRevD.91.124047}
  {\bibfield  {journal} {\bibinfo  {journal} {Phys. Rev. D}\ }\textbf {\bibinfo
  {volume} {91}},\ \bibinfo {pages} {124047} (\bibinfo {year} {2015})},\
  \Eprint {http://arxiv.org/abs/1504.01763} {arXiv:1504.01763 [gr-qc]}
  \BibitemShut {NoStop}%
\bibitem [{\citenamefont {{Tsagas}}(2011)}]{2011PhRvD..84d3524T}%
  \BibitemOpen
  \bibfield  {author} {\bibinfo {author} {\bibfnamefont {C.~G.}\ \bibnamefont
  {{Tsagas}}},\ }\href {\doibase 10.1103/PhysRevD.84.043524} {\bibfield
  {journal} {\bibinfo  {journal} {\prd}\ }\textbf {\bibinfo {volume} {84}},\
  \bibinfo {eid} {043524} (\bibinfo {year} {2011})},\ \Eprint
  {http://arxiv.org/abs/1105.5103} {arXiv:1105.5103 [gr-qc]} \BibitemShut
  {NoStop}%
\bibitem [{\citenamefont {Tsagas}(2010)}]{Tsagas:2009cr}%
  \BibitemOpen
  \bibfield  {author} {\bibinfo {author} {\bibfnamefont {C.~G.}\ \bibnamefont
  {Tsagas}},\ }\href {\doibase 10.1103/PhysRevD.81.043501} {\bibfield
  {journal} {\bibinfo  {journal} {Phys. Rev. D}\ }\textbf {\bibinfo {volume}
  {81}},\ \bibinfo {pages} {043501} (\bibinfo {year} {2010})},\ \Eprint
  {http://arxiv.org/abs/0912.2749} {arXiv:0912.2749 [astro-ph.CO]} \BibitemShut
  {NoStop}%
\bibitem [{\citenamefont {Bruni}\ \emph {et~al.}(2014)\citenamefont {Bruni},
  \citenamefont {Thomas},\ and\ \citenamefont {Wands}}]{Bruni:2013mua}%
  \BibitemOpen
  \bibfield  {author} {\bibinfo {author} {\bibfnamefont {M.}~\bibnamefont
  {Bruni}}, \bibinfo {author} {\bibfnamefont {D.~B.}\ \bibnamefont {Thomas}}, \
  and\ \bibinfo {author} {\bibfnamefont {D.}~\bibnamefont {Wands}},\ }\href
  {\doibase 10.1103/PhysRevD.89.044010} {\bibfield  {journal} {\bibinfo
  {journal} {Phys. Rev. D}\ }\textbf {\bibinfo {volume} {89}},\ \bibinfo
  {pages} {044010} (\bibinfo {year} {2014})},\ \Eprint
  {http://arxiv.org/abs/1306.1562} {arXiv:1306.1562 [astro-ph.CO]} \BibitemShut
  {NoStop}%
\bibitem [{\citenamefont {Thomas}\ \emph {et~al.}(2015)\citenamefont {Thomas},
  \citenamefont {Bruni},\ and\ \citenamefont {Wands}}]{Thomas:2015kua}%
  \BibitemOpen
  \bibfield  {author} {\bibinfo {author} {\bibfnamefont {D.~B.}\ \bibnamefont
  {Thomas}}, \bibinfo {author} {\bibfnamefont {M.}~\bibnamefont {Bruni}}, \
  and\ \bibinfo {author} {\bibfnamefont {D.}~\bibnamefont {Wands}},\ }\href
  {\doibase 10.1093/mnras/stv1390} {\bibfield  {journal} {\bibinfo  {journal}
  {Mon. Not. Roy. Astron. Soc.}\ }\textbf {\bibinfo {volume} {452}},\ \bibinfo
  {pages} {1727} (\bibinfo {year} {2015})},\ \Eprint
  {http://arxiv.org/abs/1501.00799} {arXiv:1501.00799 [astro-ph.CO]}
  \BibitemShut {NoStop}%
\bibitem [{\citenamefont {Barrera-Hinojosa}\ \emph {et~al.}(2021)\citenamefont
  {Barrera-Hinojosa}, \citenamefont {Li}, \citenamefont {Bruni},\ and\
  \citenamefont {He}}]{Barrera-Hinojosa:2020gnx}%
  \BibitemOpen
  \bibfield  {author} {\bibinfo {author} {\bibfnamefont {C.}~\bibnamefont
  {Barrera-Hinojosa}}, \bibinfo {author} {\bibfnamefont {B.}~\bibnamefont
  {Li}}, \bibinfo {author} {\bibfnamefont {M.}~\bibnamefont {Bruni}}, \ and\
  \bibinfo {author} {\bibfnamefont {J.-h.}\ \bibnamefont {He}},\ }\href
  {\doibase 10.1093/mnras/staa4025} {\bibfield  {journal} {\bibinfo  {journal}
  {Mon. Not. Roy. Astron. Soc.}\ }\textbf {\bibinfo {volume} {501}},\ \bibinfo
  {pages} {5697} (\bibinfo {year} {2021})},\ \Eprint
  {http://arxiv.org/abs/2010.08257} {arXiv:2010.08257 [astro-ph.CO]}
  \BibitemShut {NoStop}%
\bibitem [{\citenamefont {Barrera-Hinojosa}\ \emph {et~al.}(2022)\citenamefont
  {Barrera-Hinojosa}, \citenamefont {Li},\ and\ \citenamefont
  {Cai}}]{Barrera-Hinojosa:2021msx}%
  \BibitemOpen
  \bibfield  {author} {\bibinfo {author} {\bibfnamefont {C.}~\bibnamefont
  {Barrera-Hinojosa}}, \bibinfo {author} {\bibfnamefont {B.}~\bibnamefont
  {Li}}, \ and\ \bibinfo {author} {\bibfnamefont {Y.-C.}\ \bibnamefont {Cai}},\
  }\href {\doibase 10.1093/mnras/stab3657} {\bibfield  {journal} {\bibinfo
  {journal} {Mon. Not. Roy. Astron. Soc.}\ }\textbf {\bibinfo {volume} {510}},\
  \bibinfo {pages} {3589} (\bibinfo {year} {2022})},\ \Eprint
  {http://arxiv.org/abs/2109.02632} {arXiv:2109.02632 [astro-ph.CO]}
  \BibitemShut {NoStop}%
\bibitem [{\citenamefont {Faraoni}\ \emph {et~al.}(2022)\citenamefont
  {Faraoni}, \citenamefont {Vachon}, \citenamefont {Vanderwee},\ and\
  \citenamefont {Jose}}]{Faraoni:2022coe}%
  \BibitemOpen
  \bibfield  {author} {\bibinfo {author} {\bibfnamefont {V.}~\bibnamefont
  {Faraoni}}, \bibinfo {author} {\bibfnamefont {G.}~\bibnamefont {Vachon}},
  \bibinfo {author} {\bibfnamefont {R.}~\bibnamefont {Vanderwee}}, \ and\
  \bibinfo {author} {\bibfnamefont {S.}~\bibnamefont {Jose}},\ }\href {\doibase
  10.1103/PhysRevD.105.083530} {\bibfield  {journal} {\bibinfo  {journal}
  {Phys. Rev. D}\ }\textbf {\bibinfo {volume} {105}},\ \bibinfo {pages}
  {083530} (\bibinfo {year} {2022})},\ \Eprint
  {http://arxiv.org/abs/2201.05287} {arXiv:2201.05287 [gr-qc]} \BibitemShut
  {NoStop}%
\bibitem [{\citenamefont {{G{\"o}del}}(2000)}]{ref2.6}%
  \BibitemOpen
  \bibfield  {author} {\bibinfo {author} {\bibfnamefont {K.}~\bibnamefont
  {{G{\"o}del}}},\ }\href {\doibase 10.1023/A:1001911308752} {\bibfield
  {journal} {\bibinfo  {journal} {General Relativity and Gravitation}\ }\textbf
  {\bibinfo {volume} {32}},\ \bibinfo {pages} {1419} (\bibinfo {year}
  {2000})}\BibitemShut {NoStop}%
\bibitem [{\citenamefont {{G{\"o}del}}(1949)}]{ref2.7}%
  \BibitemOpen
  \bibfield  {author} {\bibinfo {author} {\bibfnamefont {K.}~\bibnamefont
  {{G{\"o}del}}},\ }\href {\doibase 10.1103/RevModPhys.21.447} {\bibfield
  {journal} {\bibinfo  {journal} {Reviews of Modern Physics}\ }\textbf
  {\bibinfo {volume} {21}},\ \bibinfo {pages} {447} (\bibinfo {year}
  {1949})}\BibitemShut {NoStop}%
\bibitem [{\citenamefont {Cashen}\ \emph {et~al.}(2017)\citenamefont {Cashen},
  \citenamefont {Aker},\ and\ \citenamefont {Kesden}}]{Cashen:2016neh}%
  \BibitemOpen
  \bibfield  {author} {\bibinfo {author} {\bibfnamefont {B.}~\bibnamefont
  {Cashen}}, \bibinfo {author} {\bibfnamefont {A.}~\bibnamefont {Aker}}, \ and\
  \bibinfo {author} {\bibfnamefont {M.}~\bibnamefont {Kesden}},\ }\href
  {\doibase 10.1103/PhysRevD.95.064014} {\bibfield  {journal} {\bibinfo
  {journal} {Phys. Rev. D}\ }\textbf {\bibinfo {volume} {95}},\ \bibinfo
  {pages} {064014} (\bibinfo {year} {2017})},\ \Eprint
  {http://arxiv.org/abs/1610.01590} {arXiv:1610.01590 [gr-qc]} \BibitemShut
  {NoStop}%
\bibitem [{\citenamefont {Guti\'errez-Ruiz}\ \emph {et~al.}(2021)\citenamefont
  {Guti\'errez-Ruiz}, \citenamefont {C\'ardenas-Avenda\~no}, \citenamefont
  {Yunes},\ and\ \citenamefont {Pach\'on}}]{Gutierrez-Ruiz:2018tre}%
  \BibitemOpen
  \bibfield  {author} {\bibinfo {author} {\bibfnamefont {A.~F.}\ \bibnamefont
  {Guti\'errez-Ruiz}}, \bibinfo {author} {\bibfnamefont {A.}~\bibnamefont
  {C\'ardenas-Avenda\~no}}, \bibinfo {author} {\bibfnamefont {N.}~\bibnamefont
  {Yunes}}, \ and\ \bibinfo {author} {\bibfnamefont {L.~A.}\ \bibnamefont
  {Pach\'on}},\ }\href {\doibase 10.1088/1361-6382/abff99} {\bibfield
  {journal} {\bibinfo  {journal} {Class. Quant. Grav.}\ }\textbf {\bibinfo
  {volume} {38}},\ \bibinfo {pages} {145013} (\bibinfo {year} {2021})},\
  \Eprint {http://arxiv.org/abs/1806.06476} {arXiv:1806.06476 [gr-qc]}
  \BibitemShut {NoStop}%
\bibitem [{\citenamefont {Barros}\ \emph {et~al.}(2003)\citenamefont {Barros},
  \citenamefont {Bezerra},\ and\ \citenamefont {Romero}}]{Barros:2003hq}%
  \BibitemOpen
  \bibfield  {author} {\bibinfo {author} {\bibfnamefont {A.}~\bibnamefont
  {Barros}}, \bibinfo {author} {\bibfnamefont {V.~B.}\ \bibnamefont {Bezerra}},
  \ and\ \bibinfo {author} {\bibfnamefont {C.}~\bibnamefont {Romero}},\ }\href
  {\doibase 10.1142/S0217732303012143} {\bibfield  {journal} {\bibinfo
  {journal} {Mod. Phys. Lett. A}\ }\textbf {\bibinfo {volume} {18}},\ \bibinfo
  {pages} {2673} (\bibinfo {year} {2003})},\ \Eprint
  {http://arxiv.org/abs/gr-qc/0307062} {arXiv:gr-qc/0307062} \BibitemShut
  {NoStop}%
\bibitem [{\citenamefont {{Schiff}}(1960)}]{schiff}%
  \BibitemOpen
  \bibfield  {author} {\bibinfo {author} {\bibfnamefont {L.~I.}\ \bibnamefont
  {{Schiff}}},\ }\href {\doibase 10.1103/PhysRevLett.4.215} {\bibfield
  {journal} {\bibinfo  {journal} {\prl}\ }\textbf {\bibinfo {volume} {4}},\
  \bibinfo {pages} {215} (\bibinfo {year} {1960})}\BibitemShut {NoStop}%
\bibitem [{\citenamefont {{Pugh}}(2003)}]{pugh}%
  \BibitemOpen
  \bibfield  {author} {\bibinfo {author} {\bibfnamefont {G.~E.}\ \bibnamefont
  {{Pugh}}},\ }in\ \href {\doibase 10.1142/9789812564818_0034} {\emph {\bibinfo
  {booktitle} {Nonlinear Gravitodynamics: The Lense-Thirring Effect. Edited by
  RUFFINI REMO \& SIGISMONDI COSTANTINO. Published by World Scientific
  Publishing Co. Pte. Ltd}}}\ (\bibinfo {year} {2003})\ pp.\ \bibinfo {pages}
  {414--426}\BibitemShut {NoStop}%
\bibitem [{\citenamefont {Ciufolini}(2002)}]{ciufolini}%
  \BibitemOpen
  \bibfield  {author} {\bibinfo {author} {\bibfnamefont {I.}~\bibnamefont
  {Ciufolini}},\ }\href@noop {} {\bibfield  {journal} {\bibinfo  {journal}
  {eConf}\ }\textbf {\bibinfo {volume} {C020620}},\ \bibinfo {pages} {FRAT03}
  (\bibinfo {year} {2002})},\ \Eprint {http://arxiv.org/abs/gr-qc/0209109}
  {arXiv:gr-qc/0209109} \BibitemShut {NoStop}%
\bibitem [{\citenamefont {{Iorio}}(2011)}]{2011EL.....9630001I}%
  \BibitemOpen
  \bibfield  {author} {\bibinfo {author} {\bibfnamefont {L.}~\bibnamefont
  {{Iorio}}},\ }\href {\doibase 10.1209/0295-5075/96/30001} {\bibfield
  {journal} {\bibinfo  {journal} {EPL (Europhysics Letters)}\ }\textbf
  {\bibinfo {volume} {96}},\ \bibinfo {pages} {30001} (\bibinfo {year}
  {2011})},\ \Eprint {http://arxiv.org/abs/1105.4145} {arXiv:1105.4145 [gr-qc]}
  \BibitemShut {NoStop}%
\bibitem [{\citenamefont {Ries}\ \emph {et~al.}(2011)\citenamefont {Ries},
  \citenamefont {Ciufolini}, \citenamefont {Pavlis}, \citenamefont {Paolozzi},
  \citenamefont {Koenig}, \citenamefont {Matzner}, \citenamefont {Sindoni},\
  and\ \citenamefont {Neumayer}}]{Ries:2011qho}%
  \BibitemOpen
  \bibfield  {author} {\bibinfo {author} {\bibfnamefont {J.~C.}\ \bibnamefont
  {Ries}}, \bibinfo {author} {\bibfnamefont {I.}~\bibnamefont {Ciufolini}},
  \bibinfo {author} {\bibfnamefont {E.~C.}\ \bibnamefont {Pavlis}}, \bibinfo
  {author} {\bibfnamefont {A.}~\bibnamefont {Paolozzi}}, \bibinfo {author}
  {\bibfnamefont {R.}~\bibnamefont {Koenig}}, \bibinfo {author} {\bibfnamefont
  {R.~A.}\ \bibnamefont {Matzner}}, \bibinfo {author} {\bibfnamefont
  {G.}~\bibnamefont {Sindoni}}, \ and\ \bibinfo {author} {\bibfnamefont
  {H.}~\bibnamefont {Neumayer}},\ }\href {\doibase 10.1209/0295-5075/96/30002}
  {\bibfield  {journal} {\bibinfo  {journal} {Eur. Phys. Lett.}\ }\textbf
  {\bibinfo {volume} {96}},\ \bibinfo {pages} {1} (\bibinfo {year} {2011})},\
  \Eprint {http://arxiv.org/abs/1305.2908} {arXiv:1305.2908 [gr-qc]}
  \BibitemShut {NoStop}%
\bibitem [{\citenamefont {Iorio}\ \emph {et~al.}(2013)\citenamefont {Iorio},
  \citenamefont {Ruggiero},\ and\ \citenamefont {Corda}}]{Iorio:2013zha}%
  \BibitemOpen
  \bibfield  {author} {\bibinfo {author} {\bibfnamefont {L.}~\bibnamefont
  {Iorio}}, \bibinfo {author} {\bibfnamefont {M.~L.}\ \bibnamefont {Ruggiero}},
  \ and\ \bibinfo {author} {\bibfnamefont {C.}~\bibnamefont {Corda}},\ }\href
  {\doibase 10.1016/j.actaastro.2013.06.002} {\bibfield  {journal} {\bibinfo
  {journal} {Acta Astronaut.}\ }\textbf {\bibinfo {volume} {91}},\ \bibinfo
  {pages} {141} (\bibinfo {year} {2013})},\ \Eprint
  {http://arxiv.org/abs/1307.0753} {arXiv:1307.0753 [gr-qc]} \BibitemShut
  {NoStop}%
\bibitem [{\citenamefont {Iorio}(2016)}]{Iorio:2015wza}%
  \BibitemOpen
  \bibfield  {author} {\bibinfo {author} {\bibfnamefont {L.}~\bibnamefont
  {Iorio}},\ }\href {\doibase 10.1016/j.asr.2015.10.014} {\bibfield  {journal}
  {\bibinfo  {journal} {Adv. Space Res.}\ }\textbf {\bibinfo {volume} {57}},\
  \bibinfo {pages} {493} (\bibinfo {year} {2016})},\ \Eprint
  {http://arxiv.org/abs/1510.08585} {arXiv:1510.08585 [gr-qc]} \BibitemShut
  {NoStop}%
\bibitem [{\citenamefont {Iorio}(2018)}]{Iorio:2018ntb}%
  \BibitemOpen
  \bibfield  {author} {\bibinfo {author} {\bibfnamefont {L.}~\bibnamefont
  {Iorio}},\ }\href {\doibase 10.3390/universe4110113} {\bibfield  {journal}
  {\bibinfo  {journal} {Universe}\ }\textbf {\bibinfo {volume} {4}},\ \bibinfo
  {pages} {113} (\bibinfo {year} {2018})},\ \Eprint
  {http://arxiv.org/abs/1809.07620} {arXiv:1809.07620 [gr-qc]} \BibitemShut
  {NoStop}%
\bibitem [{\citenamefont {{Ciufolini}}\ \emph {et~al.}(2018)\citenamefont
  {{Ciufolini}}, \citenamefont {{Pavlis}}, \citenamefont {{Ries}},
  \citenamefont {{Matzner}}, \citenamefont {{Koenig}}, \citenamefont
  {{Paolozzi}}, \citenamefont {{Sindoni}}, \citenamefont {{Gurzadyan}},
  \citenamefont {{Penrose}},\ and\ \citenamefont {{Paris}}}]{ref2.2}%
  \BibitemOpen
  \bibfield  {author} {\bibinfo {author} {\bibfnamefont {I.}~\bibnamefont
  {{Ciufolini}}}, \bibinfo {author} {\bibfnamefont {E.~C.}\ \bibnamefont
  {{Pavlis}}}, \bibinfo {author} {\bibfnamefont {J.}~\bibnamefont {{Ries}}},
  \bibinfo {author} {\bibfnamefont {R.}~\bibnamefont {{Matzner}}}, \bibinfo
  {author} {\bibfnamefont {R.}~\bibnamefont {{Koenig}}}, \bibinfo {author}
  {\bibfnamefont {A.}~\bibnamefont {{Paolozzi}}}, \bibinfo {author}
  {\bibfnamefont {G.}~\bibnamefont {{Sindoni}}}, \bibinfo {author}
  {\bibfnamefont {V.}~\bibnamefont {{Gurzadyan}}}, \bibinfo {author}
  {\bibfnamefont {R.}~\bibnamefont {{Penrose}}}, \ and\ \bibinfo {author}
  {\bibfnamefont {C.}~\bibnamefont {{Paris}}},\ }\href {\doibase
  10.1140/epjc/s10052-018-6303-1} {\bibfield  {journal} {\bibinfo  {journal}
  {European Physical Journal C}\ }\textbf {\bibinfo {volume} {78}},\ \bibinfo
  {eid} {880} (\bibinfo {year} {2018})},\ \Eprint
  {http://arxiv.org/abs/1910.10224} {arXiv:1910.10224 [gr-qc]} \BibitemShut
  {NoStop}%
\bibitem [{\citenamefont {{Ciufolini}}\ \emph {et~al.}(2023)\citenamefont
  {{Ciufolini}}, \citenamefont {{Paolozzi}}, \citenamefont {{Pavlis}},
  \citenamefont {{Ries}}, \citenamefont {{Matzner}}, \citenamefont {{Paris}},
  \citenamefont {{Ortore}}, \citenamefont {{Gurzadyan}},\ and\ \citenamefont
  {{Penrose}}}]{ref2.3}%
  \BibitemOpen
  \bibfield  {author} {\bibinfo {author} {\bibfnamefont {I.}~\bibnamefont
  {{Ciufolini}}}, \bibinfo {author} {\bibfnamefont {A.}~\bibnamefont
  {{Paolozzi}}}, \bibinfo {author} {\bibfnamefont {E.~C.}\ \bibnamefont
  {{Pavlis}}}, \bibinfo {author} {\bibfnamefont {J.~C.}\ \bibnamefont
  {{Ries}}}, \bibinfo {author} {\bibfnamefont {R.}~\bibnamefont {{Matzner}}},
  \bibinfo {author} {\bibfnamefont {C.}~\bibnamefont {{Paris}}}, \bibinfo
  {author} {\bibfnamefont {E.}~\bibnamefont {{Ortore}}}, \bibinfo {author}
  {\bibfnamefont {V.}~\bibnamefont {{Gurzadyan}}}, \ and\ \bibinfo {author}
  {\bibfnamefont {R.}~\bibnamefont {{Penrose}}},\ }\href {\doibase
  10.1140/epjc/s10052-023-11230-6} {\bibfield  {journal} {\bibinfo  {journal}
  {European Physical Journal C}\ }\textbf {\bibinfo {volume} {83}},\ \bibinfo
  {eid} {87} (\bibinfo {year} {2023})}\BibitemShut {NoStop}%
\bibitem [{\citenamefont {{Williams}}\ \emph {et~al.}(2004)\citenamefont
  {{Williams}}, \citenamefont {{Turyshev}},\ and\ \citenamefont
  {{Boggs}}}]{LLR}%
  \BibitemOpen
  \bibfield  {author} {\bibinfo {author} {\bibfnamefont {J.~G.}\ \bibnamefont
  {{Williams}}}, \bibinfo {author} {\bibfnamefont {S.~G.}\ \bibnamefont
  {{Turyshev}}}, \ and\ \bibinfo {author} {\bibfnamefont {D.~H.}\ \bibnamefont
  {{Boggs}}},\ }\href {\doibase 10.1103/PhysRevLett.93.261101} {\bibfield
  {journal} {\bibinfo  {journal} {\prl}\ }\textbf {\bibinfo {volume} {93}},\
  \bibinfo {eid} {261101} (\bibinfo {year} {2004})},\ \Eprint
  {http://arxiv.org/abs/gr-qc/0411113} {arXiv:gr-qc/0411113 [gr-qc]}
  \BibitemShut {NoStop}%
\bibitem [{\citenamefont {Murphy}\ \emph
  {et~al.}(2007{\natexlab{a}})\citenamefont {Murphy}, \citenamefont
  {Nordtvedt},\ and\ \citenamefont {Turyshev}}]{Murphy:2007nt}%
  \BibitemOpen
  \bibfield  {author} {\bibinfo {author} {\bibfnamefont {T.~W.}\ \bibnamefont
  {Murphy}, \bibfnamefont {Jr.}}, \bibinfo {author} {\bibfnamefont
  {K.}~\bibnamefont {Nordtvedt}}, \ and\ \bibinfo {author} {\bibfnamefont
  {S.~G.}\ \bibnamefont {Turyshev}},\ }\href {\doibase
  10.1103/PhysRevLett.98.071102} {\bibfield  {journal} {\bibinfo  {journal}
  {Phys. Rev. Lett.}\ }\textbf {\bibinfo {volume} {98}},\ \bibinfo {pages}
  {071102} (\bibinfo {year} {2007}{\natexlab{a}})},\ \Eprint
  {http://arxiv.org/abs/gr-qc/0702028} {arXiv:gr-qc/0702028} \BibitemShut
  {NoStop}%
\bibitem [{\citenamefont {Ciufolini}(2008)}]{Ciufolini:2008ma}%
  \BibitemOpen
  \bibfield  {author} {\bibinfo {author} {\bibfnamefont {I.}~\bibnamefont
  {Ciufolini}},\ }\href@noop {} {\  (\bibinfo {year} {2008})},\ \Eprint
  {http://arxiv.org/abs/0809.3219} {arXiv:0809.3219 [gr-qc]} \BibitemShut
  {NoStop}%
\bibitem [{\citenamefont {Costa}\ \emph
  {et~al.}(2021{\natexlab{b}})\citenamefont {Costa}, \citenamefont {Wylleman},\
  and\ \citenamefont {Nat\'ario}}]{Costa:2016iwu}%
  \BibitemOpen
  \bibfield  {author} {\bibinfo {author} {\bibfnamefont {L.~F.~O.}\
  \bibnamefont {Costa}}, \bibinfo {author} {\bibfnamefont {L.}~\bibnamefont
  {Wylleman}}, \ and\ \bibinfo {author} {\bibfnamefont {J.}~\bibnamefont
  {Nat\'ario}},\ }\href {\doibase 10.1103/PhysRevD.104.084081} {\bibfield
  {journal} {\bibinfo  {journal} {Phys. Rev. D}\ }\textbf {\bibinfo {volume}
  {104}},\ \bibinfo {pages} {084081} (\bibinfo {year} {2021}{\natexlab{b}})},\
  \Eprint {http://arxiv.org/abs/1603.03143} {arXiv:1603.03143 [gr-qc]}
  \BibitemShut {NoStop}%
\bibitem [{\citenamefont {Kopeikin}(2007)}]{Kopeikin:2007sq}%
  \BibitemOpen
  \bibfield  {author} {\bibinfo {author} {\bibfnamefont {S.~M.}\ \bibnamefont
  {Kopeikin}},\ }\href {\doibase 10.1103/PhysRevLett.98.229001} {\bibfield
  {journal} {\bibinfo  {journal} {Phys. Rev. Lett.}\ }\textbf {\bibinfo
  {volume} {98}},\ \bibinfo {pages} {229001} (\bibinfo {year} {2007})},\
  \Eprint {http://arxiv.org/abs/gr-qc/0702120} {arXiv:gr-qc/0702120}
  \BibitemShut {NoStop}%
\bibitem [{\citenamefont {Murphy}\ \emph
  {et~al.}(2007{\natexlab{b}})\citenamefont {Murphy}, \citenamefont
  {Nordtvedt},\ and\ \citenamefont {Turyshev}}]{Murphy:2007qu}%
  \BibitemOpen
  \bibfield  {author} {\bibinfo {author} {\bibfnamefont {T.~W.}\ \bibnamefont
  {Murphy}, \bibfnamefont {Jr.}}, \bibinfo {author} {\bibfnamefont
  {K.}~\bibnamefont {Nordtvedt}}, \ and\ \bibinfo {author} {\bibfnamefont
  {S.~G.}\ \bibnamefont {Turyshev}},\ }\href {\doibase
  10.1103/PhysRevLett.98.229002} {\bibfield  {journal} {\bibinfo  {journal}
  {Phys. Rev. Lett.}\ }\textbf {\bibinfo {volume} {98}},\ \bibinfo {pages}
  {229002} (\bibinfo {year} {2007}{\natexlab{b}})},\ \Eprint
  {http://arxiv.org/abs/0705.0513} {arXiv:0705.0513 [gr-qc]} \BibitemShut
  {NoStop}%
\bibitem [{\citenamefont {Kopeikin}(2008)}]{Kopeikin:2008xa}%
  \BibitemOpen
  \bibfield  {author} {\bibinfo {author} {\bibfnamefont {S.}~\bibnamefont
  {Kopeikin}},\ }\href@noop {} {\  (\bibinfo {year} {2008})},\ \Eprint
  {http://arxiv.org/abs/0809.3392} {arXiv:0809.3392 [gr-qc]} \BibitemShut
  {NoStop}%
\bibitem [{\citenamefont {Xie}\ and\ \citenamefont
  {Kopeikin}(2009)}]{Xie:2009pv}%
  \BibitemOpen
  \bibfield  {author} {\bibinfo {author} {\bibfnamefont {Y.}~\bibnamefont
  {Xie}}\ and\ \bibinfo {author} {\bibfnamefont {S.}~\bibnamefont {Kopeikin}},\
  }in\ \href@noop {} {\emph {\bibinfo {booktitle} {{IAU Symposium 261}:
  {Relativity in Fundamental Astronomy: Dynamics, Reference Frames, and Data
  Analysis}}}}\ (\bibinfo {year} {2009})\ \Eprint
  {http://arxiv.org/abs/0905.2424} {arXiv:0905.2424 [gr-qc]} \BibitemShut
  {NoStop}%
\bibitem [{\citenamefont {Iorio}(2009)}]{Iorio:2008gr}%
  \BibitemOpen
  \bibfield  {author} {\bibinfo {author} {\bibfnamefont {L.}~\bibnamefont
  {Iorio}},\ }\href {\doibase 10.1142/S0218271809015114} {\bibfield  {journal}
  {\bibinfo  {journal} {Int. J. Mod. Phys. D}\ }\textbf {\bibinfo {volume}
  {18}},\ \bibinfo {pages} {1319} (\bibinfo {year} {2009})},\ \Eprint
  {http://arxiv.org/abs/0809.4014} {arXiv:0809.4014 [gr-qc]} \BibitemShut
  {NoStop}%
\bibitem [{\citenamefont {{Iorio}}\ and\ \citenamefont
  {{Ruggiero}}(2009)}]{2009JCAP...03..024I}%
  \BibitemOpen
  \bibfield  {author} {\bibinfo {author} {\bibfnamefont {L.}~\bibnamefont
  {{Iorio}}}\ and\ \bibinfo {author} {\bibfnamefont {M.~L.}\ \bibnamefont
  {{Ruggiero}}},\ }\href {\doibase 10.1088/1475-7516/2009/03/024} {\bibfield
  {journal} {\bibinfo  {journal} {\jcap}\ }\textbf {\bibinfo {volume} {2009}},\
  \bibinfo {eid} {024} (\bibinfo {year} {2009})},\ \Eprint
  {http://arxiv.org/abs/0810.0199} {arXiv:0810.0199 [gr-qc]} \BibitemShut
  {NoStop}%
\bibitem [{\citenamefont {Ruggiero}\ and\ \citenamefont
  {Iorio}(2010)}]{Ruggiero:2010yln}%
  \BibitemOpen
  \bibfield  {author} {\bibinfo {author} {\bibfnamefont {M.~L.}\ \bibnamefont
  {Ruggiero}}\ and\ \bibinfo {author} {\bibfnamefont {L.}~\bibnamefont
  {Iorio}},\ }\href {\doibase 10.1007/s10714-010-0987-3} {\bibfield  {journal}
  {\bibinfo  {journal} {Gen. Rel. Grav.}\ }\textbf {\bibinfo {volume} {42}},\
  \bibinfo {pages} {2393} (\bibinfo {year} {2010})},\ \Eprint
  {http://arxiv.org/abs/0906.1281} {arXiv:0906.1281 [gr-qc]} \BibitemShut
  {NoStop}%
\bibitem [{\citenamefont {Bini}\ \emph {et~al.}(2015)\citenamefont {Bini},
  \citenamefont {Iorio},\ and\ \citenamefont {Giordano}}]{Bini:2015pta}%
  \BibitemOpen
  \bibfield  {author} {\bibinfo {author} {\bibfnamefont {D.}~\bibnamefont
  {Bini}}, \bibinfo {author} {\bibfnamefont {L.}~\bibnamefont {Iorio}}, \ and\
  \bibinfo {author} {\bibfnamefont {D.}~\bibnamefont {Giordano}},\ }\href
  {\doibase 10.1007/s10714-015-1977-2} {\bibfield  {journal} {\bibinfo
  {journal} {Gen. Rel. Grav.}\ }\textbf {\bibinfo {volume} {47}},\ \bibinfo
  {pages} {130} (\bibinfo {year} {2015})},\ \Eprint
  {http://arxiv.org/abs/1510.02945} {arXiv:1510.02945 [gr-qc]} \BibitemShut
  {NoStop}%
\bibitem [{\citenamefont {Iorio}(2006)}]{Iorio:2006wr}%
  \BibitemOpen
  \bibfield  {author} {\bibinfo {author} {\bibfnamefont {L.}~\bibnamefont
  {Iorio}},\ }\href {\doibase 10.1088/0264-9381/23/17/N01} {\bibfield
  {journal} {\bibinfo  {journal} {Class. Quant. Grav.}\ }\textbf {\bibinfo
  {volume} {23}},\ \bibinfo {pages} {5451} (\bibinfo {year} {2006})},\ \Eprint
  {http://arxiv.org/abs/gr-qc/0606092} {arXiv:gr-qc/0606092} \BibitemShut
  {NoStop}%
\bibitem [{\citenamefont {{Krogh}}(2007)}]{ref2.4}%
  \BibitemOpen
  \bibfield  {author} {\bibinfo {author} {\bibfnamefont {K.}~\bibnamefont
  {{Krogh}}},\ }\href {\doibase 10.1088/0264-9381/24/22/N01} {\bibfield
  {journal} {\bibinfo  {journal} {Classical and Quantum Gravity}\ }\textbf
  {\bibinfo {volume} {24}},\ \bibinfo {pages} {5709} (\bibinfo {year}
  {2007})}\BibitemShut {NoStop}%
\bibitem [{\citenamefont {Iorio}(2019{\natexlab{a}})}]{Iorio:2018wwe}%
  \BibitemOpen
  \bibfield  {author} {\bibinfo {author} {\bibfnamefont {L.}~\bibnamefont
  {Iorio}},\ }\href {\doibase 10.3390/universe5040087} {\bibfield  {journal}
  {\bibinfo  {journal} {Universe}\ }\textbf {\bibinfo {volume} {5}},\ \bibinfo
  {pages} {87} (\bibinfo {year} {2019}{\natexlab{a}})},\ \Eprint
  {http://arxiv.org/abs/1810.00397} {arXiv:1810.00397 [gr-qc]} \BibitemShut
  {NoStop}%
\bibitem [{\citenamefont {Iorio}(2019{\natexlab{b}})}]{Iorio:2018jys}%
  \BibitemOpen
  \bibfield  {author} {\bibinfo {author} {\bibfnamefont {L.}~\bibnamefont
  {Iorio}},\ }\href {\doibase 10.1093/mnras/stz304} {\bibfield  {journal}
  {\bibinfo  {journal} {Mon. Not. Roy. Astron. Soc.}\ }\textbf {\bibinfo
  {volume} {484}},\ \bibinfo {pages} {4811} (\bibinfo {year}
  {2019}{\natexlab{b}})},\ \Eprint {http://arxiv.org/abs/1810.09288}
  {arXiv:1810.09288 [gr-qc]} \BibitemShut {NoStop}%
\bibitem [{\citenamefont {Capozziello}\ \emph {et~al.}(2009)\citenamefont
  {Capozziello}, \citenamefont {De~Laurentis}, \citenamefont {Garufi},\ and\
  \citenamefont {Milano}}]{Capozziello:2008dv}%
  \BibitemOpen
  \bibfield  {author} {\bibinfo {author} {\bibfnamefont {S.}~\bibnamefont
  {Capozziello}}, \bibinfo {author} {\bibfnamefont {M.}~\bibnamefont
  {De~Laurentis}}, \bibinfo {author} {\bibfnamefont {F.}~\bibnamefont
  {Garufi}}, \ and\ \bibinfo {author} {\bibfnamefont {L.}~\bibnamefont
  {Milano}},\ }\href {\doibase 10.1088/0031-8949/79/02/025901} {\bibfield
  {journal} {\bibinfo  {journal} {Phys. Scripta}\ }\textbf {\bibinfo {volume}
  {79}},\ \bibinfo {pages} {025901} (\bibinfo {year} {2009})},\ \Eprint
  {http://arxiv.org/abs/0812.4063} {arXiv:0812.4063 [gr-qc]} \BibitemShut
  {NoStop}%
\bibitem [{\citenamefont {Ruggiero}\ and\ \citenamefont
  {Tartaglia}(2019)}]{Ruggiero:2018jab}%
  \BibitemOpen
  \bibfield  {author} {\bibinfo {author} {\bibfnamefont {M.~L.}\ \bibnamefont
  {Ruggiero}}\ and\ \bibinfo {author} {\bibfnamefont {A.}~\bibnamefont
  {Tartaglia}},\ }\href {\doibase 10.1140/epjp/i2019-12602-6} {\bibfield
  {journal} {\bibinfo  {journal} {Eur. Phys. J. Plus}\ }\textbf {\bibinfo
  {volume} {134}},\ \bibinfo {pages} {205} (\bibinfo {year} {2019})},\ \Eprint
  {http://arxiv.org/abs/1810.11785} {arXiv:1810.11785 [gr-qc]} \BibitemShut
  {NoStop}%
\bibitem [{\citenamefont {Mirza}(2019)}]{Mirza:2019swf}%
  \BibitemOpen
  \bibfield  {author} {\bibinfo {author} {\bibfnamefont {B.~M.}\ \bibnamefont
  {Mirza}},\ }\href {\doibase 10.1093/mnras/stz2353} {\bibfield  {journal}
  {\bibinfo  {journal} {Mon. Not. Roy. Astron. Soc.}\ }\textbf {\bibinfo
  {volume} {489}},\ \bibinfo {pages} {3232} (\bibinfo {year} {2019})},\ \Eprint
  {http://arxiv.org/abs/1909.08083} {arXiv:1909.08083 [astro-ph.HE]}
  \BibitemShut {NoStop}%
\bibitem [{\citenamefont {Tartaglia}\ \emph {et~al.}(2021)\citenamefont
  {Tartaglia}, \citenamefont {Bassan}, \citenamefont {Pucacco}, \citenamefont
  {Ferroni},\ and\ \citenamefont {Vetrugno}}]{Tartaglia:2021idn}%
  \BibitemOpen
  \bibfield  {author} {\bibinfo {author} {\bibfnamefont {A.}~\bibnamefont
  {Tartaglia}}, \bibinfo {author} {\bibfnamefont {M.}~\bibnamefont {Bassan}},
  \bibinfo {author} {\bibfnamefont {G.}~\bibnamefont {Pucacco}}, \bibinfo
  {author} {\bibfnamefont {V.}~\bibnamefont {Ferroni}}, \ and\ \bibinfo
  {author} {\bibfnamefont {D.}~\bibnamefont {Vetrugno}},\ }in\ \href@noop {}
  {\emph {\bibinfo {booktitle} {{16th Marcel Grossmann Meeting on~Recent
  Developments in Theoretical and Experimental General Relativity, Astrophysics
  and Relativistic Field Theories}}}}\ (\bibinfo {year} {2021})\ \Eprint
  {http://arxiv.org/abs/2110.05135} {arXiv:2110.05135 [gr-qc]} \BibitemShut
  {NoStop}%
\bibitem [{\citenamefont {Sch\"arer}\ \emph {et~al.}(2017)\citenamefont
  {Sch\"arer}, \citenamefont {Bondarescu}, \citenamefont {Saha}, \citenamefont
  {Ang\'elil}, \citenamefont {Helled},\ and\ \citenamefont
  {Jetzer}}]{Scharer:2017zol}%
  \BibitemOpen
  \bibfield  {author} {\bibinfo {author} {\bibfnamefont {A.}~\bibnamefont
  {Sch\"arer}}, \bibinfo {author} {\bibfnamefont {R.}~\bibnamefont
  {Bondarescu}}, \bibinfo {author} {\bibfnamefont {P.}~\bibnamefont {Saha}},
  \bibinfo {author} {\bibfnamefont {R.}~\bibnamefont {Ang\'elil}}, \bibinfo
  {author} {\bibfnamefont {R.}~\bibnamefont {Helled}}, \ and\ \bibinfo {author}
  {\bibfnamefont {P.}~\bibnamefont {Jetzer}},\ }\href@noop {} {\  (\bibinfo
  {year} {2017})},\ \Eprint {http://arxiv.org/abs/1707.00319} {arXiv:1707.00319
  [gr-qc]} \BibitemShut {NoStop}%
\bibitem [{\citenamefont {Battista}\ \emph {et~al.}(2017)\citenamefont
  {Battista}, \citenamefont {Tartaglia}, \citenamefont {Esposito},
  \citenamefont {Lucchesi}, \citenamefont {Ruggiero}, \citenamefont {Valko},
  \citenamefont {Dell'Agnello}, \citenamefont {Di~Fiore}, \citenamefont
  {Simo},\ and\ \citenamefont {Grado}}]{esposito}%
  \BibitemOpen
  \bibfield  {author} {\bibinfo {author} {\bibfnamefont {E.}~\bibnamefont
  {Battista}}, \bibinfo {author} {\bibfnamefont {A.}~\bibnamefont {Tartaglia}},
  \bibinfo {author} {\bibfnamefont {G.}~\bibnamefont {Esposito}}, \bibinfo
  {author} {\bibfnamefont {D.}~\bibnamefont {Lucchesi}}, \bibinfo {author}
  {\bibfnamefont {M.~L.}\ \bibnamefont {Ruggiero}}, \bibinfo {author}
  {\bibfnamefont {P.}~\bibnamefont {Valko}}, \bibinfo {author} {\bibfnamefont
  {S.}~\bibnamefont {Dell'Agnello}}, \bibinfo {author} {\bibfnamefont
  {L.}~\bibnamefont {Di~Fiore}}, \bibinfo {author} {\bibfnamefont
  {J.}~\bibnamefont {Simo}}, \ and\ \bibinfo {author} {\bibfnamefont
  {A.}~\bibnamefont {Grado}},\ }\href {\doibase 10.1088/1361-6382/aa7f11}
  {\bibfield  {journal} {\bibinfo  {journal} {Class. Quant. Grav.}\ }\textbf
  {\bibinfo {volume} {34}},\ \bibinfo {pages} {165008} (\bibinfo {year}
  {2017})},\ \Eprint {http://arxiv.org/abs/1703.08095} {arXiv:1703.08095
  [gr-qc]} \BibitemShut {NoStop}%
\bibitem [{\citenamefont {Braginsky}\ \emph {et~al.}(1977)\citenamefont
  {Braginsky}, \citenamefont {Caves},\ and\ \citenamefont
  {Thorne}}]{PhysRevD.15.2047}%
  \BibitemOpen
  \bibfield  {author} {\bibinfo {author} {\bibfnamefont {V.~B.}\ \bibnamefont
  {Braginsky}}, \bibinfo {author} {\bibfnamefont {C.~M.}\ \bibnamefont
  {Caves}}, \ and\ \bibinfo {author} {\bibfnamefont {K.~S.}\ \bibnamefont
  {Thorne}},\ }\href {\doibase 10.1103/PhysRevD.15.2047} {\bibfield  {journal}
  {\bibinfo  {journal} {Phys. Rev. D}\ }\textbf {\bibinfo {volume} {15}},\
  \bibinfo {pages} {2047} (\bibinfo {year} {1977})}\BibitemShut {NoStop}%
\bibitem [{\citenamefont {Pascual-S\'anchez}(2003)}]{Pascual-Sanchez:2002rro}%
  \BibitemOpen
  \bibfield  {author} {\bibinfo {author} {\bibfnamefont {J.~F.}\ \bibnamefont
  {Pascual-S\'anchez}},\ }\href {\doibase 10.1007/3-540-36973-2_20} {\bibfield
  {journal} {\bibinfo  {journal} {Lect. Notes Phys.}\ }\textbf {\bibinfo
  {volume} {617}},\ \bibinfo {pages} {330} (\bibinfo {year} {2003})},\ \Eprint
  {http://arxiv.org/abs/gr-qc/0207122} {arXiv:gr-qc/0207122} \BibitemShut
  {NoStop}%
\bibitem [{\citenamefont {Braginsky}\ \emph {et~al.}(1984)\citenamefont
  {Braginsky}, \citenamefont {Polnarev},\ and\ \citenamefont
  {Thorne}}]{PhysRevLett.53.863}%
  \BibitemOpen
  \bibfield  {author} {\bibinfo {author} {\bibfnamefont {V.~B.}\ \bibnamefont
  {Braginsky}}, \bibinfo {author} {\bibfnamefont {A.~G.}\ \bibnamefont
  {Polnarev}}, \ and\ \bibinfo {author} {\bibfnamefont {K.~S.}\ \bibnamefont
  {Thorne}},\ }\href {\doibase 10.1103/PhysRevLett.53.863} {\bibfield
  {journal} {\bibinfo  {journal} {Phys. Rev. Lett.}\ }\textbf {\bibinfo
  {volume} {53}},\ \bibinfo {pages} {863} (\bibinfo {year} {1984})}\BibitemShut
  {NoStop}%
\bibitem [{\citenamefont {Modanese}(1996{\natexlab{a}})}]{Modanese:1995tx}%
  \BibitemOpen
  \bibfield  {author} {\bibinfo {author} {\bibfnamefont {G.}~\bibnamefont
  {Modanese}},\ }\href {\doibase 10.1209/epl/i1996-00129-8} {\bibfield
  {journal} {\bibinfo  {journal} {EPL}\ }\textbf {\bibinfo {volume} {35}},\
  \bibinfo {pages} {413} (\bibinfo {year} {1996}{\natexlab{a}})},\ \Eprint
  {http://arxiv.org/abs/hep-th/9505094} {arXiv:hep-th/9505094} \BibitemShut
  {NoStop}%
\bibitem [{\citenamefont {Modanese}(1996{\natexlab{b}})}]{Modanese:1996zm}%
  \BibitemOpen
  \bibfield  {author} {\bibinfo {author} {\bibfnamefont {G.}~\bibnamefont
  {Modanese}},\ }\href {\doibase 10.1103/PhysRevD.54.5002} {\bibfield
  {journal} {\bibinfo  {journal} {Phys. Rev. D}\ }\textbf {\bibinfo {volume}
  {54}},\ \bibinfo {pages} {5002} (\bibinfo {year} {1996}{\natexlab{b}})},\
  \Eprint {http://arxiv.org/abs/hep-th/9601160} {arXiv:hep-th/9601160}
  \BibitemShut {NoStop}%
\bibitem [{\citenamefont {Colella}\ \emph {et~al.}(1975)\citenamefont
  {Colella}, \citenamefont {Overhauser},\ and\ \citenamefont
  {Werner}}]{PhysRevLett.34.1472}%
  \BibitemOpen
  \bibfield  {author} {\bibinfo {author} {\bibfnamefont {R.}~\bibnamefont
  {Colella}}, \bibinfo {author} {\bibfnamefont {A.~W.}\ \bibnamefont
  {Overhauser}}, \ and\ \bibinfo {author} {\bibfnamefont {S.~A.}\ \bibnamefont
  {Werner}},\ }\href {\doibase 10.1103/PhysRevLett.34.1472} {\bibfield
  {journal} {\bibinfo  {journal} {Phys. Rev. Lett.}\ }\textbf {\bibinfo
  {volume} {34}},\ \bibinfo {pages} {1472} (\bibinfo {year}
  {1975})}\BibitemShut {NoStop}%
\bibitem [{\citenamefont {Gallerati}\ \emph {et~al.}(2022)\citenamefont
  {Gallerati}, \citenamefont {Modanese},\ and\ \citenamefont
  {Ummarino}}]{Gallerati:2022nwm}%
  \BibitemOpen
  \bibfield  {author} {\bibinfo {author} {\bibfnamefont {A.}~\bibnamefont
  {Gallerati}}, \bibinfo {author} {\bibfnamefont {G.}~\bibnamefont {Modanese}},
  \ and\ \bibinfo {author} {\bibfnamefont {G.}~\bibnamefont {Ummarino}},\
  }\href {\doibase 10.3389/fphy.2022.941858} {\bibfield  {journal} {\bibinfo
  {journal} {Front. in Phys.}\ }\textbf {\bibinfo {volume} {10}},\ \bibinfo
  {pages} {941858} (\bibinfo {year} {2022})},\ \Eprint
  {http://arxiv.org/abs/2206.07574} {arXiv:2206.07574 [gr-qc]} \BibitemShut
  {NoStop}%
\bibitem [{\citenamefont {Gallerati}\ and\ \citenamefont
  {Ummarino}(2022)}]{Gallerati:2022pgh}%
  \BibitemOpen
  \bibfield  {author} {\bibinfo {author} {\bibfnamefont {A.}~\bibnamefont
  {Gallerati}}\ and\ \bibinfo {author} {\bibfnamefont {G.~A.}\ \bibnamefont
  {Ummarino}},\ }\href {\doibase 10.3390/sym14030554} {\bibfield  {journal}
  {\bibinfo  {journal} {Symmetry}\ }\textbf {\bibinfo {volume} {14}},\ \bibinfo
  {pages} {554} (\bibinfo {year} {2022})},\ \Eprint
  {http://arxiv.org/abs/2203.09417} {arXiv:2203.09417 [gr-qc]} \BibitemShut
  {NoStop}%
\bibitem [{\citenamefont {Tate}\ \emph {et~al.}(1989)\citenamefont {Tate},
  \citenamefont {Cabrera}, \citenamefont {Felch},\ and\ \citenamefont
  {Anderson}}]{PhysRevLett.62.845}%
  \BibitemOpen
  \bibfield  {author} {\bibinfo {author} {\bibfnamefont {J.}~\bibnamefont
  {Tate}}, \bibinfo {author} {\bibfnamefont {B.}~\bibnamefont {Cabrera}},
  \bibinfo {author} {\bibfnamefont {S.~B.}\ \bibnamefont {Felch}}, \ and\
  \bibinfo {author} {\bibfnamefont {J.~T.}\ \bibnamefont {Anderson}},\ }\href
  {\doibase 10.1103/PhysRevLett.62.845} {\bibfield  {journal} {\bibinfo
  {journal} {Phys. Rev. Lett.}\ }\textbf {\bibinfo {volume} {62}},\ \bibinfo
  {pages} {845} (\bibinfo {year} {1989})}\BibitemShut {NoStop}%
\bibitem [{\citenamefont {Podkletnov}\ and\ \citenamefont
  {Nieminen}(1992)}]{pod}%
  \BibitemOpen
  \bibfield  {author} {\bibinfo {author} {\bibfnamefont {E.}~\bibnamefont
  {Podkletnov}}\ and\ \bibinfo {author} {\bibfnamefont {R.}~\bibnamefont
  {Nieminen}},\ }\href {\doibase https://doi.org/10.1016/0921-4534(92)90055-H}
  {\bibfield  {journal} {\bibinfo  {journal} {Physica C: Superconductivity}\
  }\textbf {\bibinfo {volume} {203}},\ \bibinfo {pages} {441} (\bibinfo {year}
  {1992})}\BibitemShut {NoStop}%
\bibitem [{\citenamefont {Tajmar}\ \emph {et~al.}(2007)\citenamefont {Tajmar},
  \citenamefont {Plesescu}, \citenamefont {Seifert},\ and\ \citenamefont
  {Marhold}}]{Tajmar:2006gh}%
  \BibitemOpen
  \bibfield  {author} {\bibinfo {author} {\bibfnamefont {M.}~\bibnamefont
  {Tajmar}}, \bibinfo {author} {\bibfnamefont {F.}~\bibnamefont {Plesescu}},
  \bibinfo {author} {\bibfnamefont {B.}~\bibnamefont {Seifert}}, \ and\
  \bibinfo {author} {\bibfnamefont {K.}~\bibnamefont {Marhold}},\ }\href
  {\doibase 10.1063/1.2437552} {\bibfield  {journal} {\bibinfo  {journal} {AIP
  Conf. Proc.}\ }\textbf {\bibinfo {volume} {880}},\ \bibinfo {pages} {1071}
  (\bibinfo {year} {2007})},\ \Eprint {http://arxiv.org/abs/gr-qc/0610015}
  {arXiv:gr-qc/0610015} \BibitemShut {NoStop}%
\bibitem [{\citenamefont {de~Matos}(2008)}]{deMatos:2006us}%
  \BibitemOpen
  \bibfield  {author} {\bibinfo {author} {\bibfnamefont {C.~J.}\ \bibnamefont
  {de~Matos}},\ }\href {\doibase 10.1142/S0218271807011292} {\bibfield
  {journal} {\bibinfo  {journal} {Int. J. Mod. Phys. D}\ }\textbf {\bibinfo
  {volume} {16}},\ \bibinfo {pages} {2599} (\bibinfo {year} {2008})},\ \Eprint
  {http://arxiv.org/abs/gr-qc/0607004} {arXiv:gr-qc/0607004} \BibitemShut
  {NoStop}%
\bibitem [{\citenamefont {de~Matos}\ and\ \citenamefont
  {Tajmar}(2005)}]{deMatos:2005tbu}%
  \BibitemOpen
  \bibfield  {author} {\bibinfo {author} {\bibfnamefont {C.~J.}\ \bibnamefont
  {de~Matos}}\ and\ \bibinfo {author} {\bibfnamefont {M.}~\bibnamefont
  {Tajmar}},\ }\href {\doibase 10.1016/j.physc.2005.08.004} {\bibfield
  {journal} {\bibinfo  {journal} {Physica C}\ }\textbf {\bibinfo {volume}
  {432}},\ \bibinfo {pages} {167} (\bibinfo {year} {2005})},\ \Eprint
  {http://arxiv.org/abs/cond-mat/0602591} {arXiv:cond-mat/0602591} \BibitemShut
  {NoStop}%
\bibitem [{\citenamefont {Bambi}(2008)}]{Bambi:2007vq}%
  \BibitemOpen
  \bibfield  {author} {\bibinfo {author} {\bibfnamefont {C.}~\bibnamefont
  {Bambi}},\ }\href {\doibase 10.1142/S0218271808012036} {\bibfield  {journal}
  {\bibinfo  {journal} {Int. J. Mod. Phys. D}\ }\textbf {\bibinfo {volume}
  {17}},\ \bibinfo {pages} {327} (\bibinfo {year} {2008})},\ \Eprint
  {http://arxiv.org/abs/0710.2042} {arXiv:0710.2042 [gr-qc]} \BibitemShut
  {NoStop}%
\bibitem [{\citenamefont {{Tajmar}}\ \emph {et~al.}(2022)\citenamefont
  {{Tajmar}}, \citenamefont {{Neunzig}},\ and\ \citenamefont
  {{K{\"o}{\ss}ling}}}]{ref3.1}%
  \BibitemOpen
  \bibfield  {author} {\bibinfo {author} {\bibfnamefont {M.}~\bibnamefont
  {{Tajmar}}}, \bibinfo {author} {\bibfnamefont {O.}~\bibnamefont {{Neunzig}}},
  \ and\ \bibinfo {author} {\bibfnamefont {M.}~\bibnamefont
  {{K{\"o}{\ss}ling}}},\ }\href {\doibase 10.3389/fphy.2022.892215} {\bibfield
  {journal} {\bibinfo  {journal} {Frontiers in Physics}\ }\textbf {\bibinfo
  {volume} {10}},\ \bibinfo {eid} {892215} (\bibinfo {year}
  {2022})}\BibitemShut {NoStop}%
\bibitem [{\citenamefont {Ummarino}\ and\ \citenamefont
  {Gallerati}(2017)}]{Ummarino:2017bvz}%
  \BibitemOpen
  \bibfield  {author} {\bibinfo {author} {\bibfnamefont {G.~A.}\ \bibnamefont
  {Ummarino}}\ and\ \bibinfo {author} {\bibfnamefont {A.}~\bibnamefont
  {Gallerati}},\ }\href {\doibase 10.1140/epjc/s10052-017-5116-y} {\bibfield
  {journal} {\bibinfo  {journal} {Eur. Phys. J. C}\ }\textbf {\bibinfo {volume}
  {77}},\ \bibinfo {pages} {549} (\bibinfo {year} {2017})},\ \Eprint
  {http://arxiv.org/abs/1710.01267} {arXiv:1710.01267 [gr-qc]} \BibitemShut
  {NoStop}%
\bibitem [{\citenamefont {Ummarino}\ and\ \citenamefont
  {Gallerati}(2020)}]{Ummarino:2020loo}%
  \BibitemOpen
  \bibfield  {author} {\bibinfo {author} {\bibfnamefont {G.~A.}\ \bibnamefont
  {Ummarino}}\ and\ \bibinfo {author} {\bibfnamefont {A.}~\bibnamefont
  {Gallerati}},\ }\href {\doibase 10.1088/1361-6382/abb57b} {\bibfield
  {journal} {\bibinfo  {journal} {Class. Quant. Grav.}\ }\textbf {\bibinfo
  {volume} {37}},\ \bibinfo {pages} {217001} (\bibinfo {year} {2020})},\
  \Eprint {http://arxiv.org/abs/2009.04967} {arXiv:2009.04967 [gr-qc]}
  \BibitemShut {NoStop}%
\bibitem [{\citenamefont {Ummarino}\ and\ \citenamefont
  {Gallerati}(2021)}]{Ummarino:2021tpz}%
  \BibitemOpen
  \bibfield  {author} {\bibinfo {author} {\bibfnamefont {G.~A.}\ \bibnamefont
  {Ummarino}}\ and\ \bibinfo {author} {\bibfnamefont {A.}~\bibnamefont
  {Gallerati}},\ }\href {\doibase 10.1016/j.rinp.2021.104838} {\bibfield
  {journal} {\bibinfo  {journal} {Results Phys.}\ }\textbf {\bibinfo {volume}
  {30}},\ \bibinfo {pages} {104838} (\bibinfo {year} {2021})},\ \Eprint
  {http://arxiv.org/abs/2110.07335} {arXiv:2110.07335 [gr-qc]} \BibitemShut
  {NoStop}%
\bibitem [{\citenamefont {{Nomura}}\ \emph {et~al.}(2012)\citenamefont
  {{Nomura}}, \citenamefont {{Ryu}}, \citenamefont {{Furusaki}},\ and\
  \citenamefont {{Nagaosa}}}]{2012PhRvL.108b6802N}%
  \BibitemOpen
  \bibfield  {author} {\bibinfo {author} {\bibfnamefont {K.}~\bibnamefont
  {{Nomura}}}, \bibinfo {author} {\bibfnamefont {S.}~\bibnamefont {{Ryu}}},
  \bibinfo {author} {\bibfnamefont {A.}~\bibnamefont {{Furusaki}}}, \ and\
  \bibinfo {author} {\bibfnamefont {N.}~\bibnamefont {{Nagaosa}}},\ }\href
  {\doibase 10.1103/PhysRevLett.108.026802} {\bibfield  {journal} {\bibinfo
  {journal} {\prl}\ }\textbf {\bibinfo {volume} {108}},\ \bibinfo {eid}
  {026802} (\bibinfo {year} {2012})},\ \Eprint {http://arxiv.org/abs/1108.5054}
  {arXiv:1108.5054 [cond-mat.supr-con]} \BibitemShut {NoStop}%
\bibitem [{\citenamefont {Sekine}(2016)}]{Sekine:2015bqa}%
  \BibitemOpen
  \bibfield  {author} {\bibinfo {author} {\bibfnamefont {A.}~\bibnamefont
  {Sekine}},\ }\href {\doibase 10.1103/PhysRevB.93.094510} {\bibfield
  {journal} {\bibinfo  {journal} {Phys. Rev. B}\ }\textbf {\bibinfo {volume}
  {93}},\ \bibinfo {pages} {094510} (\bibinfo {year} {2016})},\ \Eprint
  {http://arxiv.org/abs/1510.05903} {arXiv:1510.05903 [cond-mat.mes-hall]}
  \BibitemShut {NoStop}%
\bibitem [{\citenamefont {Camacho}\ and\ \citenamefont
  {Castellanos}(2012)}]{Camacho:2012ce}%
  \BibitemOpen
  \bibfield  {author} {\bibinfo {author} {\bibfnamefont {A.}~\bibnamefont
  {Camacho}}\ and\ \bibinfo {author} {\bibfnamefont {E.}~\bibnamefont
  {Castellanos}},\ }\href {\doibase 10.1142/S0217732312501982} {\bibfield
  {journal} {\bibinfo  {journal} {Mod. Phys. Lett. A}\ }\textbf {\bibinfo
  {volume} {27}},\ \bibinfo {pages} {1250198} (\bibinfo {year} {2012})},\
  \Eprint {http://arxiv.org/abs/1201.6672} {arXiv:1201.6672 [gr-qc]}
  \BibitemShut {NoStop}%
\bibitem [{\citenamefont {Sarkar}\ \emph {et~al.}(2018)\citenamefont {Sarkar},
  \citenamefont {Vaz},\ and\ \citenamefont {Wijewardhana}}]{Sarkar:2017aje}%
  \BibitemOpen
  \bibfield  {author} {\bibinfo {author} {\bibfnamefont {S.}~\bibnamefont
  {Sarkar}}, \bibinfo {author} {\bibfnamefont {C.}~\bibnamefont {Vaz}}, \ and\
  \bibinfo {author} {\bibfnamefont {L.~C.~R.}\ \bibnamefont {Wijewardhana}},\
  }\href {\doibase 10.1103/PhysRevD.97.103022} {\bibfield  {journal} {\bibinfo
  {journal} {Phys. Rev. D}\ }\textbf {\bibinfo {volume} {97}},\ \bibinfo
  {pages} {103022} (\bibinfo {year} {2018})},\ \Eprint
  {http://arxiv.org/abs/1711.01219} {arXiv:1711.01219 [astro-ph.GA]}
  \BibitemShut {NoStop}%
\bibitem [{\citenamefont {Dimopoulos}\ \emph {et~al.}(2007)\citenamefont
  {Dimopoulos}, \citenamefont {Graham}, \citenamefont {Hogan},\ and\
  \citenamefont {Kasevich}}]{Dimopoulos:2006nk}%
  \BibitemOpen
  \bibfield  {author} {\bibinfo {author} {\bibfnamefont {S.}~\bibnamefont
  {Dimopoulos}}, \bibinfo {author} {\bibfnamefont {P.~W.}\ \bibnamefont
  {Graham}}, \bibinfo {author} {\bibfnamefont {J.~M.}\ \bibnamefont {Hogan}}, \
  and\ \bibinfo {author} {\bibfnamefont {M.~A.}\ \bibnamefont {Kasevich}},\
  }\href {\doibase 10.1103/PhysRevLett.98.111102} {\bibfield  {journal}
  {\bibinfo  {journal} {Phys. Rev. Lett.}\ }\textbf {\bibinfo {volume} {98}},\
  \bibinfo {pages} {111102} (\bibinfo {year} {2007})},\ \Eprint
  {http://arxiv.org/abs/gr-qc/0610047} {arXiv:gr-qc/0610047} \BibitemShut
  {NoStop}%
\bibitem [{\citenamefont {Dimopoulos}\ \emph {et~al.}(2008)\citenamefont
  {Dimopoulos}, \citenamefont {Graham}, \citenamefont {Hogan},\ and\
  \citenamefont {Kasevich}}]{Dimopoulos:2008hx}%
  \BibitemOpen
  \bibfield  {author} {\bibinfo {author} {\bibfnamefont {S.}~\bibnamefont
  {Dimopoulos}}, \bibinfo {author} {\bibfnamefont {P.~W.}\ \bibnamefont
  {Graham}}, \bibinfo {author} {\bibfnamefont {J.~M.}\ \bibnamefont {Hogan}}, \
  and\ \bibinfo {author} {\bibfnamefont {M.~A.}\ \bibnamefont {Kasevich}},\
  }\href {\doibase 10.1103/PhysRevD.78.042003} {\bibfield  {journal} {\bibinfo
  {journal} {Phys. Rev. D}\ }\textbf {\bibinfo {volume} {78}},\ \bibinfo
  {pages} {042003} (\bibinfo {year} {2008})},\ \Eprint
  {http://arxiv.org/abs/0802.4098} {arXiv:0802.4098 [hep-ph]} \BibitemShut
  {NoStop}%
\bibitem [{\citenamefont {Angonin-Willaime}\ \emph {et~al.}(2004)\citenamefont
  {Angonin-Willaime}, \citenamefont {Ovido},\ and\ \citenamefont
  {Tourrenc}}]{Angonin-Willaime:2003rbe}%
  \BibitemOpen
  \bibfield  {author} {\bibinfo {author} {\bibfnamefont {M.~C.}\ \bibnamefont
  {Angonin-Willaime}}, \bibinfo {author} {\bibfnamefont {X.}~\bibnamefont
  {Ovido}}, \ and\ \bibinfo {author} {\bibfnamefont {P.}~\bibnamefont
  {Tourrenc}},\ }\href {\doibase 10.1023/B:GERG.0000010485.62147.0e} {\bibfield
   {journal} {\bibinfo  {journal} {Gen. Rel. Grav.}\ }\textbf {\bibinfo
  {volume} {36}},\ \bibinfo {pages} {411} (\bibinfo {year} {2004})},\ \Eprint
  {http://arxiv.org/abs/gr-qc/0310021} {arXiv:gr-qc/0310021} \BibitemShut
  {NoStop}%
\bibitem [{\citenamefont {Bosi}\ \emph {et~al.}(2011)\citenamefont {Bosi},
  \citenamefont {Cella}, \citenamefont {Di~Virgilio}, \citenamefont {Ortolan},
  \citenamefont {Porzio}, \citenamefont {Solimeno}, \citenamefont {Cerdonio},
  \citenamefont {Zendri}, \citenamefont {Allegrini}, \citenamefont {Belfi},
  \citenamefont {Beverini}, \citenamefont {Bouhadef}, \citenamefont {Carelli},
  \citenamefont {Ferrante}, \citenamefont {Maccioni}, \citenamefont
  {Passaquieti}, \citenamefont {Stefani}, \citenamefont {Ruggiero},
  \citenamefont {Tartaglia}, \citenamefont {Schreiber}, \citenamefont
  {Gebauer},\ and\ \citenamefont {Wells}}]{PhysRevD.84.122002}%
  \BibitemOpen
  \bibfield  {author} {\bibinfo {author} {\bibfnamefont {F.}~\bibnamefont
  {Bosi}}, \bibinfo {author} {\bibfnamefont {G.}~\bibnamefont {Cella}},
  \bibinfo {author} {\bibfnamefont {A.}~\bibnamefont {Di~Virgilio}}, \bibinfo
  {author} {\bibfnamefont {A.}~\bibnamefont {Ortolan}}, \bibinfo {author}
  {\bibfnamefont {A.}~\bibnamefont {Porzio}}, \bibinfo {author} {\bibfnamefont
  {S.}~\bibnamefont {Solimeno}}, \bibinfo {author} {\bibfnamefont
  {M.}~\bibnamefont {Cerdonio}}, \bibinfo {author} {\bibfnamefont {J.~P.}\
  \bibnamefont {Zendri}}, \bibinfo {author} {\bibfnamefont {M.}~\bibnamefont
  {Allegrini}}, \bibinfo {author} {\bibfnamefont {J.}~\bibnamefont {Belfi}},
  \bibinfo {author} {\bibfnamefont {N.}~\bibnamefont {Beverini}}, \bibinfo
  {author} {\bibfnamefont {B.}~\bibnamefont {Bouhadef}}, \bibinfo {author}
  {\bibfnamefont {G.}~\bibnamefont {Carelli}}, \bibinfo {author} {\bibfnamefont
  {I.}~\bibnamefont {Ferrante}}, \bibinfo {author} {\bibfnamefont
  {E.}~\bibnamefont {Maccioni}}, \bibinfo {author} {\bibfnamefont
  {R.}~\bibnamefont {Passaquieti}}, \bibinfo {author} {\bibfnamefont
  {F.}~\bibnamefont {Stefani}}, \bibinfo {author} {\bibfnamefont {M.~L.}\
  \bibnamefont {Ruggiero}}, \bibinfo {author} {\bibfnamefont {A.}~\bibnamefont
  {Tartaglia}}, \bibinfo {author} {\bibfnamefont {K.~U.}\ \bibnamefont
  {Schreiber}}, \bibinfo {author} {\bibfnamefont {A.}~\bibnamefont {Gebauer}},
  \ and\ \bibinfo {author} {\bibfnamefont {J.-P.~R.}\ \bibnamefont {Wells}},\
  }\href {\doibase 10.1103/PhysRevD.84.122002} {\bibfield  {journal} {\bibinfo
  {journal} {Phys. Rev. D}\ }\textbf {\bibinfo {volume} {84}},\ \bibinfo
  {pages} {122002} (\bibinfo {year} {2011})}\BibitemShut {NoStop}%
\bibitem [{\citenamefont {Tartaglia}\ \emph {et~al.}(2017)\citenamefont
  {Tartaglia}, \citenamefont {Di~Virgilio}, \citenamefont {Belfi},
  \citenamefont {Beverini},\ and\ \citenamefont
  {Ruggiero}}]{Tartaglia:2016jfo}%
  \BibitemOpen
  \bibfield  {author} {\bibinfo {author} {\bibfnamefont {A.}~\bibnamefont
  {Tartaglia}}, \bibinfo {author} {\bibfnamefont {A.}~\bibnamefont
  {Di~Virgilio}}, \bibinfo {author} {\bibfnamefont {J.}~\bibnamefont {Belfi}},
  \bibinfo {author} {\bibfnamefont {N.}~\bibnamefont {Beverini}}, \ and\
  \bibinfo {author} {\bibfnamefont {M.~L.}\ \bibnamefont {Ruggiero}},\ }\href
  {\doibase 10.1140/epjp/i2017-11372-5} {\bibfield  {journal} {\bibinfo
  {journal} {Eur. Phys. J. Plus}\ }\textbf {\bibinfo {volume} {132}},\ \bibinfo
  {pages} {73} (\bibinfo {year} {2017})},\ \Eprint
  {http://arxiv.org/abs/1612.09099} {arXiv:1612.09099 [gr-qc]} \BibitemShut
  {NoStop}%
\bibitem [{\citenamefont {Di~Virgilio}\ \emph {et~al.}(2020)\citenamefont
  {Di~Virgilio} \emph {et~al.}}]{DiVirgilio:2020ior}%
  \BibitemOpen
  \bibfield  {author} {\bibinfo {author} {\bibfnamefont {A.~D.~V.}\
  \bibnamefont {Di~Virgilio}} \emph {et~al.},\ }\href {\doibase
  10.1103/PhysRevResearch.2.032069} {\bibfield  {journal} {\bibinfo  {journal}
  {Phys. Rev. Res.}\ }\textbf {\bibinfo {volume} {2}},\ \bibinfo {pages}
  {032069} (\bibinfo {year} {2020})},\ \Eprint
  {http://arxiv.org/abs/2003.11819} {arXiv:2003.11819 [astro-ph.IM]}
  \BibitemShut {NoStop}%
\bibitem [{\citenamefont {Altucci}\ \emph {et~al.}(2022)\citenamefont {Altucci}
  \emph {et~al.}}]{Altucci:2022rxr}%
  \BibitemOpen
  \bibfield  {author} {\bibinfo {author} {\bibfnamefont {C.}~\bibnamefont
  {Altucci}} \emph {et~al.},\ }\href@noop {} {\  (\bibinfo {year} {2022})},\
  \Eprint {http://arxiv.org/abs/2209.09328} {arXiv:2209.09328 [gr-qc]}
  \BibitemShut {NoStop}%
\bibitem [{\citenamefont {Di~Virgilio}\ \emph {et~al.}(2023)\citenamefont
  {Di~Virgilio} \emph {et~al.}}]{DiVirgilio:2023nrc}%
  \BibitemOpen
  \bibfield  {author} {\bibinfo {author} {\bibfnamefont {A.~D.~V.}\
  \bibnamefont {Di~Virgilio}} \emph {et~al.},\ }\href@noop {} {\  (\bibinfo
  {year} {2023})},\ \Eprint {http://arxiv.org/abs/2301.01386} {arXiv:2301.01386
  [quant-ph]} \BibitemShut {NoStop}%
\bibitem [{\citenamefont {Kannan}\ and\ \citenamefont
  {Saha}(2009)}]{Kannan:2008nm}%
  \BibitemOpen
  \bibfield  {author} {\bibinfo {author} {\bibfnamefont {R.}~\bibnamefont
  {Kannan}}\ and\ \bibinfo {author} {\bibfnamefont {P.}~\bibnamefont {Saha}},\
  }\href {\doibase 10.1088/0004-637X/690/2/1553} {\bibfield  {journal}
  {\bibinfo  {journal} {Astrophys. J.}\ }\textbf {\bibinfo {volume} {690}},\
  \bibinfo {pages} {1553} (\bibinfo {year} {2009})},\ \Eprint
  {http://arxiv.org/abs/0809.1866} {arXiv:0809.1866 [astro-ph]} \BibitemShut
  {NoStop}%
\bibitem [{\citenamefont {Iorio}(2011)}]{Iorio:2010hw}%
  \BibitemOpen
  \bibfield  {author} {\bibinfo {author} {\bibfnamefont {L.}~\bibnamefont
  {Iorio}},\ }\href {\doibase 10.1111/j.1365-2966.2010.17701.x} {\bibfield
  {journal} {\bibinfo  {journal} {Mon. Not. Roy. Astron. Soc.}\ }\textbf
  {\bibinfo {volume} {411}},\ \bibinfo {pages} {453} (\bibinfo {year}
  {2011})},\ \Eprint {http://arxiv.org/abs/1008.1720} {arXiv:1008.1720 [gr-qc]}
  \BibitemShut {NoStop}%
\bibitem [{\citenamefont {Poirier}\ and\ \citenamefont
  {Mathews}(2015)}]{Poirier:2015hga}%
  \BibitemOpen
  \bibfield  {author} {\bibinfo {author} {\bibfnamefont {J.}~\bibnamefont
  {Poirier}}\ and\ \bibinfo {author} {\bibfnamefont {G.~J.}\ \bibnamefont
  {Mathews}},\ }\href@noop {} {\  (\bibinfo {year} {2015})},\ \Eprint
  {http://arxiv.org/abs/1504.02499} {arXiv:1504.02499 [gr-qc]} \BibitemShut
  {NoStop}%
\bibitem [{\citenamefont {Rueda}\ \emph {et~al.}(2022)\citenamefont {Rueda},
  \citenamefont {Ruffini},\ and\ \citenamefont {Kerr}}]{Rueda:2022fgz}%
  \BibitemOpen
  \bibfield  {author} {\bibinfo {author} {\bibfnamefont {J.~A.}\ \bibnamefont
  {Rueda}}, \bibinfo {author} {\bibfnamefont {R.}~\bibnamefont {Ruffini}}, \
  and\ \bibinfo {author} {\bibfnamefont {R.~P.}\ \bibnamefont {Kerr}},\ }\href
  {\doibase 10.3847/1538-4357/ac5b6e} {\bibfield  {journal} {\bibinfo
  {journal} {Astrophys. J.}\ }\textbf {\bibinfo {volume} {929}},\ \bibinfo
  {pages} {56} (\bibinfo {year} {2022})},\ \Eprint
  {http://arxiv.org/abs/2203.03471} {arXiv:2203.03471 [astro-ph.HE]}
  \BibitemShut {NoStop}%
\bibitem [{\citenamefont {Ricarte}\ \emph {et~al.}(2022)\citenamefont
  {Ricarte}, \citenamefont {Palumbo}, \citenamefont {Narayan}, \citenamefont
  {Roelofs},\ and\ \citenamefont {Emami}}]{Ricarte:2022wpd}%
  \BibitemOpen
  \bibfield  {author} {\bibinfo {author} {\bibfnamefont {A.}~\bibnamefont
  {Ricarte}}, \bibinfo {author} {\bibfnamefont {D.~C.~M.}\ \bibnamefont
  {Palumbo}}, \bibinfo {author} {\bibfnamefont {R.}~\bibnamefont {Narayan}},
  \bibinfo {author} {\bibfnamefont {F.}~\bibnamefont {Roelofs}}, \ and\
  \bibinfo {author} {\bibfnamefont {R.}~\bibnamefont {Emami}},\ }\href
  {\doibase 10.3847/2041-8213/aca087} {\bibfield  {journal} {\bibinfo
  {journal} {Astrophys. J. Lett.}\ }\textbf {\bibinfo {volume} {941}},\
  \bibinfo {pages} {L12} (\bibinfo {year} {2022})},\ \Eprint
  {http://arxiv.org/abs/2211.01810} {arXiv:2211.01810 [gr-qc]} \BibitemShut
  {NoStop}%
\bibitem [{\citenamefont {Herdeiro}\ \emph {et~al.}(2009)\citenamefont
  {Herdeiro}, \citenamefont {Rebelo},\ and\ \citenamefont
  {Warnick}}]{Herdeiro:2009qy}%
  \BibitemOpen
  \bibfield  {author} {\bibinfo {author} {\bibfnamefont {C.~A.~R.}\
  \bibnamefont {Herdeiro}}, \bibinfo {author} {\bibfnamefont {C.}~\bibnamefont
  {Rebelo}}, \ and\ \bibinfo {author} {\bibfnamefont {C.~M.}\ \bibnamefont
  {Warnick}},\ }\href {\doibase 10.1103/PhysRevD.80.084037} {\bibfield
  {journal} {\bibinfo  {journal} {Phys. Rev. D}\ }\textbf {\bibinfo {volume}
  {80}},\ \bibinfo {pages} {084037} (\bibinfo {year} {2009})},\ \Eprint
  {http://arxiv.org/abs/0907.5104} {arXiv:0907.5104 [gr-qc]} \BibitemShut
  {NoStop}%
\bibitem [{\citenamefont {Ruggiero}(2015)}]{Ruggiero:2015pva}%
  \BibitemOpen
  \bibfield  {author} {\bibinfo {author} {\bibfnamefont {M.~L.}\ \bibnamefont
  {Ruggiero}},\ }\href {\doibase 10.1142/S0218271815500601} {\bibfield
  {journal} {\bibinfo  {journal} {Int. J. Mod. Phys. D}\ }\textbf {\bibinfo
  {volume} {24}},\ \bibinfo {pages} {1550060} (\bibinfo {year} {2015})},\
  \Eprint {http://arxiv.org/abs/1502.01473} {arXiv:1502.01473 [gr-qc]}
  \BibitemShut {NoStop}%
\bibitem [{\citenamefont {Ruggiero}(2016)}]{Ruggiero:2015ima}%
  \BibitemOpen
  \bibfield  {author} {\bibinfo {author} {\bibfnamefont {M.~L.}\ \bibnamefont
  {Ruggiero}},\ }\href {\doibase 10.1007/s10509-016-2723-2} {\bibfield
  {journal} {\bibinfo  {journal} {Astrophys. Space Sci.}\ }\textbf {\bibinfo
  {volume} {361}},\ \bibinfo {pages} {140} (\bibinfo {year} {2016})},\ \Eprint
  {http://arxiv.org/abs/1507.06797} {arXiv:1507.06797 [gr-qc]} \BibitemShut
  {NoStop}%
\bibitem [{\citenamefont {Ruggiero}\ and\ \citenamefont
  {Tartaglia}(2005)}]{Ruggiero:2005dg}%
  \BibitemOpen
  \bibfield  {author} {\bibinfo {author} {\bibfnamefont {M.~L.}\ \bibnamefont
  {Ruggiero}}\ and\ \bibinfo {author} {\bibfnamefont {A.}~\bibnamefont
  {Tartaglia}},\ }\href {\doibase 10.1103/PhysRevD.72.084030} {\bibfield
  {journal} {\bibinfo  {journal} {Phys. Rev. D}\ }\textbf {\bibinfo {volume}
  {72}},\ \bibinfo {pages} {084030} (\bibinfo {year} {2005})},\ \Eprint
  {http://arxiv.org/abs/gr-qc/0509098} {arXiv:gr-qc/0509098} \BibitemShut
  {NoStop}%
\bibitem [{\citenamefont {Kocherlakota}\ \emph {et~al.}(2019)\citenamefont
  {Kocherlakota}, \citenamefont {Biswas}, \citenamefont {Joshi}, \citenamefont
  {Bhattacharyya}, \citenamefont {Chakraborty},\ and\ \citenamefont
  {Ray}}]{Kocherlakota:2017hkn}%
  \BibitemOpen
  \bibfield  {author} {\bibinfo {author} {\bibfnamefont {P.}~\bibnamefont
  {Kocherlakota}}, \bibinfo {author} {\bibfnamefont {S.}~\bibnamefont
  {Biswas}}, \bibinfo {author} {\bibfnamefont {P.~S.}\ \bibnamefont {Joshi}},
  \bibinfo {author} {\bibfnamefont {S.}~\bibnamefont {Bhattacharyya}}, \bibinfo
  {author} {\bibfnamefont {C.}~\bibnamefont {Chakraborty}}, \ and\ \bibinfo
  {author} {\bibfnamefont {A.}~\bibnamefont {Ray}},\ }\href {\doibase
  10.1093/mnras/stz2538} {\bibfield  {journal} {\bibinfo  {journal} {Mon. Not.
  Roy. Astron. Soc.}\ }\textbf {\bibinfo {volume} {490}},\ \bibinfo {pages}
  {3262} (\bibinfo {year} {2019})},\ \Eprint {http://arxiv.org/abs/1711.04053}
  {arXiv:1711.04053 [astro-ph.HE]} \BibitemShut {NoStop}%
\bibitem [{\citenamefont {Venkatraman~Krishnan}\ \emph
  {et~al.}(2020)\citenamefont {Venkatraman~Krishnan} \emph
  {et~al.}}]{VenkatramanKrishnan:2020pbi}%
  \BibitemOpen
  \bibfield  {author} {\bibinfo {author} {\bibfnamefont {V.}~\bibnamefont
  {Venkatraman~Krishnan}} \emph {et~al.},\ }\href {\doibase
  10.1126/science.aax7007} {\bibfield  {journal} {\bibinfo  {journal}
  {Science}\ }\textbf {\bibinfo {volume} {367}},\ \bibinfo {pages} {577}
  (\bibinfo {year} {2020})},\ \Eprint {http://arxiv.org/abs/2001.11405}
  {arXiv:2001.11405 [astro-ph.HE]} \BibitemShut {NoStop}%
\bibitem [{\citenamefont {Iorio}(2020)}]{Iorio:2020xos}%
  \BibitemOpen
  \bibfield  {author} {\bibinfo {author} {\bibfnamefont {L.}~\bibnamefont
  {Iorio}},\ }\href {\doibase 10.1093/mnras/staa1322} {\bibfield  {journal}
  {\bibinfo  {journal} {Mon. Not. Roy. Astron. Soc.}\ }\textbf {\bibinfo
  {volume} {495}},\ \bibinfo {pages} {2777} (\bibinfo {year} {2020})},\ \Eprint
  {http://arxiv.org/abs/2003.08244} {arXiv:2003.08244 [gr-qc]} \BibitemShut
  {NoStop}%
\bibitem [{\citenamefont {{Flanagan}}\ and\ \citenamefont
  {{Racine}}(2007)}]{flanagan}%
  \BibitemOpen
  \bibfield  {author} {\bibinfo {author} {\bibfnamefont {{\'E}.~{\'E}.}\
  \bibnamefont {{Flanagan}}}\ and\ \bibinfo {author} {\bibfnamefont
  {{\'E}.}~\bibnamefont {{Racine}}},\ }\href {\doibase
  10.1103/PhysRevD.75.044001} {\bibfield  {journal} {\bibinfo  {journal}
  {\prd}\ }\textbf {\bibinfo {volume} {75}},\ \bibinfo {eid} {044001} (\bibinfo
  {year} {2007})},\ \Eprint {http://arxiv.org/abs/gr-qc/0601029}
  {arXiv:gr-qc/0601029 [gr-qc]} \BibitemShut {NoStop}%
\bibitem [{\citenamefont {Favata}(2006)}]{Favata:2005da}%
  \BibitemOpen
  \bibfield  {author} {\bibinfo {author} {\bibfnamefont {M.}~\bibnamefont
  {Favata}},\ }\href {\doibase 10.1103/PhysRevD.73.104005} {\bibfield
  {journal} {\bibinfo  {journal} {Phys. Rev. D}\ }\textbf {\bibinfo {volume}
  {73}},\ \bibinfo {pages} {104005} (\bibinfo {year} {2006})},\ \Eprint
  {http://arxiv.org/abs/astro-ph/0510668} {arXiv:astro-ph/0510668} \BibitemShut
  {NoStop}%
\bibitem [{\citenamefont {Poisson}(2020)}]{Poisson:2020eki}%
  \BibitemOpen
  \bibfield  {author} {\bibinfo {author} {\bibfnamefont {E.}~\bibnamefont
  {Poisson}},\ }\href {\doibase 10.1103/PhysRevD.101.104028} {\bibfield
  {journal} {\bibinfo  {journal} {Phys. Rev. D}\ }\textbf {\bibinfo {volume}
  {101}},\ \bibinfo {pages} {104028} (\bibinfo {year} {2020})},\ \Eprint
  {http://arxiv.org/abs/2003.10427} {arXiv:2003.10427 [gr-qc]} \BibitemShut
  {NoStop}%
\bibitem [{\citenamefont {Iorio}(2022)}]{Iorio:2022raj}%
  \BibitemOpen
  \bibfield  {author} {\bibinfo {author} {\bibfnamefont {L.}~\bibnamefont
  {Iorio}},\ }\href {\doibase 10.3390/universe8100546} {\bibfield  {journal}
  {\bibinfo  {journal} {Universe}\ }\textbf {\bibinfo {volume} {8}},\ \bibinfo
  {pages} {546} (\bibinfo {year} {2022})},\ \Eprint
  {http://arxiv.org/abs/2210.09154} {arXiv:2210.09154 [gr-qc]} \BibitemShut
  {NoStop}%
\bibitem [{\citenamefont {Kopeikin}\ and\ \citenamefont
  {Mashhoon}(2002)}]{Kopeikin:2001dz}%
  \BibitemOpen
  \bibfield  {author} {\bibinfo {author} {\bibfnamefont {S.}~\bibnamefont
  {Kopeikin}}\ and\ \bibinfo {author} {\bibfnamefont {B.}~\bibnamefont
  {Mashhoon}},\ }\href {\doibase 10.1103/PhysRevD.65.064025} {\bibfield
  {journal} {\bibinfo  {journal} {Phys. Rev. D}\ }\textbf {\bibinfo {volume}
  {65}},\ \bibinfo {pages} {064025} (\bibinfo {year} {2002})},\ \Eprint
  {http://arxiv.org/abs/gr-qc/0110101} {arXiv:gr-qc/0110101} \BibitemShut
  {NoStop}%
\bibitem [{\citenamefont {Ciufolini}\ \emph {et~al.}(2003)\citenamefont
  {Ciufolini}, \citenamefont {Kopeikin}, \citenamefont {Mashhoon},\ and\
  \citenamefont {Ricci}}]{Ciufolini:2002iq}%
  \BibitemOpen
  \bibfield  {author} {\bibinfo {author} {\bibfnamefont {I.}~\bibnamefont
  {Ciufolini}}, \bibinfo {author} {\bibfnamefont {S.}~\bibnamefont {Kopeikin}},
  \bibinfo {author} {\bibfnamefont {B.}~\bibnamefont {Mashhoon}}, \ and\
  \bibinfo {author} {\bibfnamefont {F.}~\bibnamefont {Ricci}},\ }\href
  {\doibase 10.1016/S0375-9601(02)01804-2} {\bibfield  {journal} {\bibinfo
  {journal} {Phys. Lett. A}\ }\textbf {\bibinfo {volume} {308}},\ \bibinfo
  {pages} {101} (\bibinfo {year} {2003})},\ \Eprint
  {http://arxiv.org/abs/gr-qc/0210015} {arXiv:gr-qc/0210015} \BibitemShut
  {NoStop}%
\bibitem [{\citenamefont {Ciufolini}\ and\ \citenamefont
  {Ricci}(2003)}]{Ciufolini:2003yy}%
  \BibitemOpen
  \bibfield  {author} {\bibinfo {author} {\bibfnamefont {I.}~\bibnamefont
  {Ciufolini}}\ and\ \bibinfo {author} {\bibfnamefont {F.}~\bibnamefont
  {Ricci}},\ }\href@noop {} {\  (\bibinfo {year} {2003})},\ \Eprint
  {http://arxiv.org/abs/gr-qc/0301030} {arXiv:gr-qc/0301030} \BibitemShut
  {NoStop}%
\bibitem [{\citenamefont {Sereno}(2003{\natexlab{a}})}]{Sereno:2003tk}%
  \BibitemOpen
  \bibfield  {author} {\bibinfo {author} {\bibfnamefont {M.}~\bibnamefont
  {Sereno}},\ }\href {\doibase 10.1103/PhysRevD.67.064007} {\bibfield
  {journal} {\bibinfo  {journal} {Phys. Rev. D}\ }\textbf {\bibinfo {volume}
  {67}},\ \bibinfo {pages} {064007} (\bibinfo {year} {2003}{\natexlab{a}})},\
  \Eprint {http://arxiv.org/abs/astro-ph/0301290} {arXiv:astro-ph/0301290}
  \BibitemShut {NoStop}%
\bibitem [{\citenamefont {Sereno}(2003{\natexlab{b}})}]{Sereno:2003kx}%
  \BibitemOpen
  \bibfield  {author} {\bibinfo {author} {\bibfnamefont {M.}~\bibnamefont
  {Sereno}},\ }\href {\doibase 10.1046/j.1365-8711.2003.06881.x} {\bibfield
  {journal} {\bibinfo  {journal} {Mon. Not. Roy. Astron. Soc.}\ }\textbf
  {\bibinfo {volume} {344}},\ \bibinfo {pages} {942} (\bibinfo {year}
  {2003}{\natexlab{b}})},\ \Eprint {http://arxiv.org/abs/astro-ph/0307243}
  {arXiv:astro-ph/0307243} \BibitemShut {NoStop}%
\bibitem [{\citenamefont {Capozziello}\ \emph {et~al.}(2003)\citenamefont
  {Capozziello}, \citenamefont {Cardone}, \citenamefont {Re},\ and\
  \citenamefont {Sereno}}]{Capozziello:2003hh}%
  \BibitemOpen
  \bibfield  {author} {\bibinfo {author} {\bibfnamefont {S.}~\bibnamefont
  {Capozziello}}, \bibinfo {author} {\bibfnamefont {V.~F.}\ \bibnamefont
  {Cardone}}, \bibinfo {author} {\bibfnamefont {V.}~\bibnamefont {Re}}, \ and\
  \bibinfo {author} {\bibfnamefont {M.}~\bibnamefont {Sereno}},\ }\href
  {\doibase 10.1046/j.1365-8711.2003.06671.x} {\bibfield  {journal} {\bibinfo
  {journal} {Mon. Not. Roy. Astron. Soc.}\ }\textbf {\bibinfo {volume} {343}},\
  \bibinfo {pages} {360} (\bibinfo {year} {2003})},\ \Eprint
  {http://arxiv.org/abs/astro-ph/0304272} {arXiv:astro-ph/0304272} \BibitemShut
  {NoStop}%
\bibitem [{\citenamefont {Sereno}(2005{\natexlab{a}})}]{Sereno:2004nc}%
  \BibitemOpen
  \bibfield  {author} {\bibinfo {author} {\bibfnamefont {M.}~\bibnamefont
  {Sereno}},\ }\href {\doibase 10.1111/j.1365-2966.2005.08709.x} {\bibfield
  {journal} {\bibinfo  {journal} {Mon. Not. Roy. Astron. Soc.}\ }\textbf
  {\bibinfo {volume} {357}},\ \bibinfo {pages} {1205} (\bibinfo {year}
  {2005}{\natexlab{a}})},\ \Eprint {http://arxiv.org/abs/astro-ph/0412108}
  {arXiv:astro-ph/0412108} \BibitemShut {NoStop}%
\bibitem [{\citenamefont {Sereno}(2005{\natexlab{b}})}]{Sereno:2005wr}%
  \BibitemOpen
  \bibfield  {author} {\bibinfo {author} {\bibfnamefont {M.}~\bibnamefont
  {Sereno}},\ }\href {\doibase 10.1111/j.1745-3933.2005.00026.x} {\bibfield
  {journal} {\bibinfo  {journal} {Mon. Not. Roy. Astron. Soc.}\ }\textbf
  {\bibinfo {volume} {359}},\ \bibinfo {pages} {19} (\bibinfo {year}
  {2005}{\natexlab{b}})},\ \Eprint {http://arxiv.org/abs/astro-ph/0501605}
  {arXiv:astro-ph/0501605} \BibitemShut {NoStop}%
\bibitem [{\citenamefont {Sereno}(2004)}]{Sereno:2004jx}%
  \BibitemOpen
  \bibfield  {author} {\bibinfo {author} {\bibfnamefont {M.}~\bibnamefont
  {Sereno}},\ }\href {\doibase 10.1103/PhysRevD.69.087501} {\bibfield
  {journal} {\bibinfo  {journal} {Phys. Rev. D}\ }\textbf {\bibinfo {volume}
  {69}},\ \bibinfo {pages} {087501} (\bibinfo {year} {2004})},\ \Eprint
  {http://arxiv.org/abs/astro-ph/0401295} {arXiv:astro-ph/0401295} \BibitemShut
  {NoStop}%
\bibitem [{\citenamefont {Ruggiero}\ and\ \citenamefont
  {Tartaglia}(2007)}]{Ruggiero:2006rh}%
  \BibitemOpen
  \bibfield  {author} {\bibinfo {author} {\bibfnamefont {M.~L.}\ \bibnamefont
  {Ruggiero}}\ and\ \bibinfo {author} {\bibfnamefont {A.}~\bibnamefont
  {Tartaglia}},\ }\href {\doibase 10.1111/j.1365-2966.2006.11187.x} {\bibfield
  {journal} {\bibinfo  {journal} {Mon. Not. Roy. Astron. Soc.}\ }\textbf
  {\bibinfo {volume} {374}},\ \bibinfo {pages} {847} (\bibinfo {year}
  {2007})},\ \Eprint {http://arxiv.org/abs/astro-ph/0609712}
  {arXiv:astro-ph/0609712} \BibitemShut {NoStop}%
\bibitem [{\citenamefont {Kraniotis}(2005)}]{Kraniotis:2005zm}%
  \BibitemOpen
  \bibfield  {author} {\bibinfo {author} {\bibfnamefont {G.~V.}\ \bibnamefont
  {Kraniotis}},\ }\href {\doibase 10.1088/0264-9381/22/21/001} {\bibfield
  {journal} {\bibinfo  {journal} {Class. Quant. Grav.}\ }\textbf {\bibinfo
  {volume} {22}},\ \bibinfo {pages} {4391} (\bibinfo {year} {2005})},\ \Eprint
  {http://arxiv.org/abs/gr-qc/0507056} {arXiv:gr-qc/0507056} \BibitemShut
  {NoStop}%
\bibitem [{\citenamefont {Kraniotis}(2014)}]{Kraniotis:2014paa}%
  \BibitemOpen
  \bibfield  {author} {\bibinfo {author} {\bibfnamefont {G.~V.}\ \bibnamefont
  {Kraniotis}},\ }\href {\doibase 10.1007/s10714-014-1818-8} {\bibfield
  {journal} {\bibinfo  {journal} {Gen. Rel. Grav.}\ }\textbf {\bibinfo {volume}
  {46}},\ \bibinfo {pages} {1818} (\bibinfo {year} {2014})},\ \Eprint
  {http://arxiv.org/abs/1401.7118} {arXiv:1401.7118 [gr-qc]} \BibitemShut
  {NoStop}%
\bibitem [{\citenamefont {Kraniotis}(2021)}]{Kraniotis:2019ked}%
  \BibitemOpen
  \bibfield  {author} {\bibinfo {author} {\bibfnamefont {G.~V.}\ \bibnamefont
  {Kraniotis}},\ }\href {\doibase 10.1140/epjc/s10052-021-08911-5} {\bibfield
  {journal} {\bibinfo  {journal} {Eur. Phys. J. C}\ }\textbf {\bibinfo {volume}
  {81}},\ \bibinfo {pages} {147} (\bibinfo {year} {2021})},\ \Eprint
  {http://arxiv.org/abs/1912.10320} {arXiv:1912.10320 [gr-qc]} \BibitemShut
  {NoStop}%
\bibitem [{\citenamefont {Kraniotis}(2007)}]{Kraniotis:2007zz}%
  \BibitemOpen
  \bibfield  {author} {\bibinfo {author} {\bibfnamefont {G.~V.}\ \bibnamefont
  {Kraniotis}},\ }\href {\doibase 10.1088/0264-9381/24/7/007} {\bibfield
  {journal} {\bibinfo  {journal} {Class. Quant. Grav.}\ }\textbf {\bibinfo
  {volume} {24}},\ \bibinfo {pages} {1775} (\bibinfo {year} {2007})},\ \Eprint
  {http://arxiv.org/abs/gr-qc/0602056} {arXiv:gr-qc/0602056} \BibitemShut
  {NoStop}%
\bibitem [{\citenamefont {Arakida}(2021)}]{Arakida:2018szn}%
  \BibitemOpen
  \bibfield  {author} {\bibinfo {author} {\bibfnamefont {H.}~\bibnamefont
  {Arakida}},\ }\href {\doibase 10.1142/S0218271821500450} {\bibfield
  {journal} {\bibinfo  {journal} {Int. J. Mod. Phys. D}\ }\textbf {\bibinfo
  {volume} {30}},\ \bibinfo {pages} {2150045} (\bibinfo {year} {2021})},\
  \Eprint {http://arxiv.org/abs/1808.03418} {arXiv:1808.03418 [gr-qc]}
  \BibitemShut {NoStop}%
\bibitem [{\citenamefont {Iyer}(2018)}]{Iyer:2018omj}%
  \BibitemOpen
  \bibfield  {author} {\bibinfo {author} {\bibfnamefont {S.~V.}\ \bibnamefont
  {Iyer}},\ }\href@noop {} {\  (\bibinfo {year} {2018})},\ \Eprint
  {http://arxiv.org/abs/1808.06630} {arXiv:1808.06630 [gr-qc]} \BibitemShut
  {NoStop}%
\bibitem [{\citenamefont {Iyer}\ and\ \citenamefont
  {Hansen}(2009)}]{PhysRevD.80.124023}%
  \BibitemOpen
  \bibfield  {author} {\bibinfo {author} {\bibfnamefont {S.~V.}\ \bibnamefont
  {Iyer}}\ and\ \bibinfo {author} {\bibfnamefont {E.~C.}\ \bibnamefont
  {Hansen}},\ }\href {\doibase 10.1103/PhysRevD.80.124023} {\bibfield
  {journal} {\bibinfo  {journal} {Phys. Rev. D}\ }\textbf {\bibinfo {volume}
  {80}},\ \bibinfo {pages} {124023} (\bibinfo {year} {2009})}\BibitemShut
  {NoStop}%
\bibitem [{\citenamefont {Aghanim}\ \emph {et~al.}(2020)\citenamefont {Aghanim}
  \emph {et~al.}}]{planck2018b}%
  \BibitemOpen
  \bibfield  {author} {\bibinfo {author} {\bibfnamefont {N.}~\bibnamefont
  {Aghanim}} \emph {et~al.} (\bibinfo {collaboration} {Planck}),\ }\href
  {\doibase 10.1051/0004-6361/201833910} {\bibfield  {journal} {\bibinfo
  {journal} {Astron. Astrophys.}\ }\textbf {\bibinfo {volume} {641}},\ \bibinfo
  {pages} {A6} (\bibinfo {year} {2020})},\ \bibinfo {note} {[Erratum:
  Astron.Astrophys. 652, C4 (2021)]},\ \Eprint
  {http://arxiv.org/abs/1807.06209} {arXiv:1807.06209 [astro-ph.CO]}
  \BibitemShut {NoStop}%
\bibitem [{\citenamefont {{Zwicky}}(1933)}]{zwicky}%
  \BibitemOpen
  \bibfield  {author} {\bibinfo {author} {\bibfnamefont {F.}~\bibnamefont
  {{Zwicky}}},\ }\href@noop {} {\bibfield  {journal} {\bibinfo  {journal}
  {Helvetica Physica Acta}\ }\textbf {\bibinfo {volume} {6}},\ \bibinfo {pages}
  {110} (\bibinfo {year} {1933})}\BibitemShut {NoStop}%
\bibitem [{\citenamefont {{Corbelli}}\ and\ \citenamefont
  {{Salucci}}(2000)}]{rotation_curves}%
  \BibitemOpen
  \bibfield  {author} {\bibinfo {author} {\bibfnamefont {E.}~\bibnamefont
  {{Corbelli}}}\ and\ \bibinfo {author} {\bibfnamefont {P.}~\bibnamefont
  {{Salucci}}},\ }\href {\doibase 10.1046/j.1365-8711.2000.03075.x} {\bibfield
  {journal} {\bibinfo  {journal} {\mnras}\ }\textbf {\bibinfo {volume} {311}},\
  \bibinfo {pages} {441} (\bibinfo {year} {2000})},\ \Eprint
  {http://arxiv.org/abs/astro-ph/9909252} {arXiv:astro-ph/9909252 [astro-ph]}
  \BibitemShut {NoStop}%
\bibitem [{\citenamefont {{Allen}}\ \emph {et~al.}(2011)\citenamefont
  {{Allen}}, \citenamefont {{Evrard}},\ and\ \citenamefont {{Mantz}}}]{gas}%
  \BibitemOpen
  \bibfield  {author} {\bibinfo {author} {\bibfnamefont {S.~W.}\ \bibnamefont
  {{Allen}}}, \bibinfo {author} {\bibfnamefont {A.~E.}\ \bibnamefont
  {{Evrard}}}, \ and\ \bibinfo {author} {\bibfnamefont {A.~B.}\ \bibnamefont
  {{Mantz}}},\ }\href {\doibase 10.1146/annurev-astro-081710-102514} {\bibfield
   {journal} {\bibinfo  {journal} {\araa}\ }\textbf {\bibinfo {volume} {49}},\
  \bibinfo {pages} {409} (\bibinfo {year} {2011})},\ \Eprint
  {http://arxiv.org/abs/1103.4829} {arXiv:1103.4829 [astro-ph.CO]} \BibitemShut
  {NoStop}%
\bibitem [{\citenamefont {{Taylor}}(1999)}]{lensing}%
  \BibitemOpen
  \bibfield  {author} {\bibinfo {author} {\bibfnamefont {A.}~\bibnamefont
  {{Taylor}}},\ }in\ \href@noop {} {\emph {\bibinfo {booktitle} {Evolution of
  Large Scale Structure : From Recombination to Garching}}},\ \bibinfo {editor}
  {edited by\ \bibinfo {editor} {\bibfnamefont {A.~J.}\ \bibnamefont
  {{Banday}}}, \bibinfo {editor} {\bibfnamefont {R.~K.}\ \bibnamefont
  {{Sheth}}}, \ and\ \bibinfo {editor} {\bibfnamefont {L.~N.}\ \bibnamefont
  {{da Costa}}}}\ (\bibinfo {year} {1999})\ p.\ \bibinfo {pages}
  {226}\BibitemShut {NoStop}%
\bibitem [{\citenamefont {Clowe}\ \emph {et~al.}(2006)\citenamefont {Clowe},
  \citenamefont {Bradac}, \citenamefont {Gonzalez}, \citenamefont {Markevitch},
  \citenamefont {Randall}, \citenamefont {Jones},\ and\ \citenamefont
  {Zaritsky}}]{bullet1}%
  \BibitemOpen
  \bibfield  {author} {\bibinfo {author} {\bibfnamefont {D.}~\bibnamefont
  {Clowe}}, \bibinfo {author} {\bibfnamefont {M.}~\bibnamefont {Bradac}},
  \bibinfo {author} {\bibfnamefont {A.~H.}\ \bibnamefont {Gonzalez}}, \bibinfo
  {author} {\bibfnamefont {M.}~\bibnamefont {Markevitch}}, \bibinfo {author}
  {\bibfnamefont {S.~W.}\ \bibnamefont {Randall}}, \bibinfo {author}
  {\bibfnamefont {C.}~\bibnamefont {Jones}}, \ and\ \bibinfo {author}
  {\bibfnamefont {D.}~\bibnamefont {Zaritsky}},\ }\href {\doibase
  10.1086/508162} {\bibfield  {journal} {\bibinfo  {journal} {Astrophys. J.
  Lett.}\ }\textbf {\bibinfo {volume} {648}},\ \bibinfo {pages} {L109}
  (\bibinfo {year} {2006})},\ \Eprint {http://arxiv.org/abs/astro-ph/0608407}
  {arXiv:astro-ph/0608407} \BibitemShut {NoStop}%
\bibitem [{\citenamefont {Brada{\v c}}\ \emph {et~al.}(2008)\citenamefont
  {Brada{\v c}}, \citenamefont {Allen}, \citenamefont {Treu}, \citenamefont
  {Ebeling}, \citenamefont {Massey}, \citenamefont {Morris}, \citenamefont
  {von~der Linden},\ and\ \citenamefont {Applegate}}]{bullet2}%
  \BibitemOpen
  \bibfield  {author} {\bibinfo {author} {\bibfnamefont {M.}~\bibnamefont
  {Brada{\v c}}}, \bibinfo {author} {\bibfnamefont {S.~W.}\ \bibnamefont
  {Allen}}, \bibinfo {author} {\bibfnamefont {T.}~\bibnamefont {Treu}},
  \bibinfo {author} {\bibfnamefont {H.}~\bibnamefont {Ebeling}}, \bibinfo
  {author} {\bibfnamefont {R.}~\bibnamefont {Massey}}, \bibinfo {author}
  {\bibfnamefont {R.~G.}\ \bibnamefont {Morris}}, \bibinfo {author}
  {\bibfnamefont {A.}~\bibnamefont {von~der Linden}}, \ and\ \bibinfo {author}
  {\bibfnamefont {D.}~\bibnamefont {Applegate}},\ }\href {\doibase
  10.1086/591246} {\bibfield  {journal} {\bibinfo  {journal} {The Astrophysical
  Journal}\ }\textbf {\bibinfo {volume} {687}},\ \bibinfo {pages} {959}
  (\bibinfo {year} {2008})}\BibitemShut {NoStop}%
\bibitem [{\citenamefont {Hinshaw}\ \emph {et~al.}(2009)\citenamefont {Hinshaw}
  \emph {et~al.}}]{CMB1}%
  \BibitemOpen
  \bibfield  {author} {\bibinfo {author} {\bibfnamefont {G.}~\bibnamefont
  {Hinshaw}} \emph {et~al.} (\bibinfo {collaboration} {WMAP}),\ }\href
  {\doibase 10.1088/0067-0049/180/2/225} {\bibfield  {journal} {\bibinfo
  {journal} {Astrophys. J. Suppl.}\ }\textbf {\bibinfo {volume} {180}},\
  \bibinfo {pages} {225} (\bibinfo {year} {2009})},\ \Eprint
  {http://arxiv.org/abs/0803.0732} {arXiv:0803.0732 [astro-ph]} \BibitemShut
  {NoStop}%
\bibitem [{\citenamefont {Ade}\ \emph {et~al.}(2016)\citenamefont {Ade} \emph
  {et~al.}}]{CMB2}%
  \BibitemOpen
  \bibfield  {author} {\bibinfo {author} {\bibfnamefont {P.~A.~R.}\
  \bibnamefont {Ade}} \emph {et~al.} (\bibinfo {collaboration} {Planck}),\
  }\href {\doibase 10.1051/0004-6361/201525830} {\bibfield  {journal} {\bibinfo
   {journal} {Astron. Astrophys.}\ }\textbf {\bibinfo {volume} {594}},\
  \bibinfo {pages} {A13} (\bibinfo {year} {2016})},\ \Eprint
  {http://arxiv.org/abs/1502.01589} {arXiv:1502.01589 [astro-ph.CO]}
  \BibitemShut {NoStop}%
\bibitem [{\citenamefont {{Conroy}}\ \emph {et~al.}(2006)\citenamefont
  {{Conroy}}, \citenamefont {{Wechsler}},\ and\ \citenamefont
  {{Kravtsov}}}]{perturb1}%
  \BibitemOpen
  \bibfield  {author} {\bibinfo {author} {\bibfnamefont {C.}~\bibnamefont
  {{Conroy}}}, \bibinfo {author} {\bibfnamefont {R.~H.}\ \bibnamefont
  {{Wechsler}}}, \ and\ \bibinfo {author} {\bibfnamefont {A.~V.}\ \bibnamefont
  {{Kravtsov}}},\ }\href {\doibase 10.1086/503602} {\bibfield  {journal}
  {\bibinfo  {journal} {\apj}\ }\textbf {\bibinfo {volume} {647}},\ \bibinfo
  {pages} {201} (\bibinfo {year} {2006})},\ \Eprint
  {http://arxiv.org/abs/astro-ph/0512234} {arXiv:astro-ph/0512234 [astro-ph]}
  \BibitemShut {NoStop}%
\bibitem [{\citenamefont {{Piattella}}(2018)}]{perturb2}%
  \BibitemOpen
  \bibfield  {author} {\bibinfo {author} {\bibfnamefont {O.}~\bibnamefont
  {{Piattella}}},\ }\href {\doibase 10.1007/978-3-319-95570-4} {\emph {\bibinfo
  {title} {{Lecture Notes in Cosmology}}}}\ (\bibinfo {year}
  {2018})\BibitemShut {NoStop}%
\bibitem [{\citenamefont {Strigari}(2013)}]{strigari2013galactic}%
  \BibitemOpen
  \bibfield  {author} {\bibinfo {author} {\bibfnamefont {L.~E.}\ \bibnamefont
  {Strigari}},\ }\href@noop {} {\bibfield  {journal} {\bibinfo  {journal}
  {Physics Reports}\ }\textbf {\bibinfo {volume} {531}},\ \bibinfo {pages} {1}
  (\bibinfo {year} {2013})}\BibitemShut {NoStop}%
\bibitem [{\citenamefont {{Cooperstock}}\ and\ \citenamefont
  {{Tieu}}(2005)}]{Cooperstock:2005qw}%
  \BibitemOpen
  \bibfield  {author} {\bibinfo {author} {\bibfnamefont {F.~I.}\ \bibnamefont
  {{Cooperstock}}}\ and\ \bibinfo {author} {\bibfnamefont {S.}~\bibnamefont
  {{Tieu}}},\ }\href@noop {} {\bibfield  {journal} {\bibinfo  {journal} {arXiv
  e-prints}\ ,\ \bibinfo {eid} {astro-ph/0507619}} (\bibinfo {year} {2005})},\
  \Eprint {http://arxiv.org/abs/astro-ph/0507619} {arXiv:astro-ph/0507619
  [astro-ph]} \BibitemShut {NoStop}%
\bibitem [{\citenamefont {Cooperstock}\ and\ \citenamefont
  {Tieu}(2007)}]{Cooperstock:2006dt}%
  \BibitemOpen
  \bibfield  {author} {\bibinfo {author} {\bibfnamefont {F.~I.}\ \bibnamefont
  {Cooperstock}}\ and\ \bibinfo {author} {\bibfnamefont {S.}~\bibnamefont
  {Tieu}},\ }\href {\doibase 10.1142/S0217751X0703666X} {\bibfield  {journal}
  {\bibinfo  {journal} {Int. J. Mod. Phys. A}\ }\textbf {\bibinfo {volume}
  {22}},\ \bibinfo {pages} {2293} (\bibinfo {year} {2007})},\ \Eprint
  {http://arxiv.org/abs/astro-ph/0610370} {arXiv:astro-ph/0610370} \BibitemShut
  {NoStop}%
\bibitem [{\citenamefont {{Carrick}}\ and\ \citenamefont
  {{Cooperstock}}(2012)}]{Carrick:2011ac}%
  \BibitemOpen
  \bibfield  {author} {\bibinfo {author} {\bibfnamefont {J.~D.}\ \bibnamefont
  {{Carrick}}}\ and\ \bibinfo {author} {\bibfnamefont {F.~I.}\ \bibnamefont
  {{Cooperstock}}},\ }\href {\doibase 10.1007/s10509-011-0854-z} {\bibfield
  {journal} {\bibinfo  {journal} {\apss}\ }\textbf {\bibinfo {volume} {337}},\
  \bibinfo {pages} {321} (\bibinfo {year} {2012})},\ \Eprint
  {http://arxiv.org/abs/1101.3224} {arXiv:1101.3224 [astro-ph.GA]} \BibitemShut
  {NoStop}%
\bibitem [{\citenamefont {Cross}(2006)}]{Cross:2006rx}%
  \BibitemOpen
  \bibfield  {author} {\bibinfo {author} {\bibfnamefont {D.~J.}\ \bibnamefont
  {Cross}},\ }\href@noop {} {\  (\bibinfo {year} {2006})},\ \Eprint
  {http://arxiv.org/abs/astro-ph/0601191} {arXiv:astro-ph/0601191} \BibitemShut
  {NoStop}%
\bibitem [{\citenamefont {Menzies}\ and\ \citenamefont
  {Mathews}(2007)}]{Menzies:2007dm}%
  \BibitemOpen
  \bibfield  {author} {\bibinfo {author} {\bibfnamefont {D.}~\bibnamefont
  {Menzies}}\ and\ \bibinfo {author} {\bibfnamefont {G.~J.}\ \bibnamefont
  {Mathews}},\ }\href@noop {} {\  (\bibinfo {year} {2007})},\ \Eprint
  {http://arxiv.org/abs/astro-ph/0701019} {arXiv:astro-ph/0701019} \BibitemShut
  {NoStop}%
\bibitem [{\citenamefont {Balasin}\ and\ \citenamefont
  {Grumiller}(2008)}]{Balasin:2006cg}%
  \BibitemOpen
  \bibfield  {author} {\bibinfo {author} {\bibfnamefont {H.}~\bibnamefont
  {Balasin}}\ and\ \bibinfo {author} {\bibfnamefont {D.}~\bibnamefont
  {Grumiller}},\ }\href {\doibase 10.1142/S0218271808012140} {\bibfield
  {journal} {\bibinfo  {journal} {Int. J. Mod. Phys. D}\ }\textbf {\bibinfo
  {volume} {17}},\ \bibinfo {pages} {475} (\bibinfo {year} {2008})},\ \Eprint
  {http://arxiv.org/abs/astro-ph/0602519} {arXiv:astro-ph/0602519} \BibitemShut
  {NoStop}%
\bibitem [{\citenamefont {Crosta}\ \emph {et~al.}(2020)\citenamefont {Crosta},
  \citenamefont {Giammaria}, \citenamefont {Lattanzi},\ and\ \citenamefont
  {Poggio}}]{crosta2020testing}%
  \BibitemOpen
  \bibfield  {author} {\bibinfo {author} {\bibfnamefont {M.}~\bibnamefont
  {Crosta}}, \bibinfo {author} {\bibfnamefont {M.}~\bibnamefont {Giammaria}},
  \bibinfo {author} {\bibfnamefont {M.~G.}\ \bibnamefont {Lattanzi}}, \ and\
  \bibinfo {author} {\bibfnamefont {E.}~\bibnamefont {Poggio}},\ }\href@noop {}
  {\bibfield  {journal} {\bibinfo  {journal} {Monthly Notices of the Royal
  Astronomical Society}\ }\textbf {\bibinfo {volume} {496}},\ \bibinfo {pages}
  {2107} (\bibinfo {year} {2020})}\BibitemShut {NoStop}%
\bibitem [{\citenamefont {{Gaia Collaboration}}(2016)}]{GAIA1}%
  \BibitemOpen
  \bibfield  {author} {\bibinfo {author} {\bibnamefont {{Gaia
  Collaboration}}},\ }\href@noop {} {\bibfield  {journal} {\bibinfo  {journal}
  {Astronomy \& Astrophysics}\ }\textbf {\bibinfo {volume} {595}},\ \bibinfo
  {pages} {A1} (\bibinfo {year} {2016})}\BibitemShut {NoStop}%
\bibitem [{\citenamefont {{Gaia Collaboration}}(2018)}]{GAIA2}%
  \BibitemOpen
  \bibfield  {author} {\bibinfo {author} {\bibnamefont {{Gaia
  Collaboration}}},\ }\href@noop {} {\bibfield  {journal} {\bibinfo  {journal}
  {Astronomy \& Astrophysics}\ }\textbf {\bibinfo {volume} {616}},\ \bibinfo
  {pages} {A1} (\bibinfo {year} {2018})}\BibitemShut {NoStop}%
\bibitem [{\citenamefont {{Crosta}}(2019)}]{crosta_astrometry}%
  \BibitemOpen
  \bibfield  {author} {\bibinfo {author} {\bibfnamefont {M.}~\bibnamefont
  {{Crosta}}},\ }\href {\doibase 10.1393/ncr/i2019-10164-2} {\bibfield
  {journal} {\bibinfo  {journal} {Nuovo Cimento Rivista Serie}\ }\textbf
  {\bibinfo {volume} {42}},\ \bibinfo {pages} {443} (\bibinfo {year}
  {2019})}\BibitemShut {NoStop}%
\bibitem [{\citenamefont {Costa}\ \emph {et~al.}(2023)\citenamefont {Costa},
  \citenamefont {Nat\'ario}, \citenamefont {Frutos-Alfaro},\ and\ \citenamefont
  {Soffel}}]{Costa:2023lrn}%
  \BibitemOpen
  \bibfield  {author} {\bibinfo {author} {\bibfnamefont {L.~F.~O.}\
  \bibnamefont {Costa}}, \bibinfo {author} {\bibfnamefont {J.}~\bibnamefont
  {Nat\'ario}}, \bibinfo {author} {\bibfnamefont {F.}~\bibnamefont
  {Frutos-Alfaro}}, \ and\ \bibinfo {author} {\bibfnamefont {M.}~\bibnamefont
  {Soffel}},\ }\href@noop {} {\  (\bibinfo {year} {2023})},\ \Eprint
  {http://arxiv.org/abs/2303.17516} {arXiv:2303.17516 [gr-qc]} \BibitemShut
  {NoStop}%
\bibitem [{\citenamefont {Astesiano}\ \emph {et~al.}(2022)\citenamefont
  {Astesiano}, \citenamefont {Cacciatori}, \citenamefont {Gorini},\ and\
  \citenamefont {Re}}]{Astesiano:2021ren}%
  \BibitemOpen
  \bibfield  {author} {\bibinfo {author} {\bibfnamefont {D.}~\bibnamefont
  {Astesiano}}, \bibinfo {author} {\bibfnamefont {S.~L.}\ \bibnamefont
  {Cacciatori}}, \bibinfo {author} {\bibfnamefont {V.}~\bibnamefont {Gorini}},
  \ and\ \bibinfo {author} {\bibfnamefont {F.}~\bibnamefont {Re}},\ }\href
  {\doibase 10.1140/epjc/s10052-022-10506-7} {\bibfield  {journal} {\bibinfo
  {journal} {Eur. Phys. J. C}\ }\textbf {\bibinfo {volume} {82}},\ \bibinfo
  {pages} {554} (\bibinfo {year} {2022})},\ \Eprint
  {http://arxiv.org/abs/2106.12818} {arXiv:2106.12818 [gr-qc]} \BibitemShut
  {NoStop}%
\bibitem [{\citenamefont {Astesiano}\ and\ \citenamefont
  {Ruggiero}(2022{\natexlab{a}})}]{astesiano}%
  \BibitemOpen
  \bibfield  {author} {\bibinfo {author} {\bibfnamefont {D.}~\bibnamefont
  {Astesiano}}\ and\ \bibinfo {author} {\bibfnamefont {M.~L.}\ \bibnamefont
  {Ruggiero}},\ }\href {\doibase 10.1103/PhysRevD.106.044061} {\bibfield
  {journal} {\bibinfo  {journal} {Phys. Rev. D}\ }\textbf {\bibinfo {volume}
  {106}},\ \bibinfo {pages} {044061} (\bibinfo {year}
  {2022}{\natexlab{a}})}\BibitemShut {NoStop}%
\bibitem [{\citenamefont {{Gupta}}\ and\ \citenamefont
  {{Lohiya}}(2020)}]{2020arXiv200914553G}%
  \BibitemOpen
  \bibfield  {author} {\bibinfo {author} {\bibfnamefont {A.}~\bibnamefont
  {{Gupta}}}\ and\ \bibinfo {author} {\bibfnamefont {D.}~\bibnamefont
  {{Lohiya}}},\ }\href@noop {} {\bibfield  {journal} {\bibinfo  {journal}
  {arXiv e-prints}\ ,\ \bibinfo {eid} {arXiv:2009.14553}} (\bibinfo {year}
  {2020})},\ \Eprint {http://arxiv.org/abs/2009.14553} {arXiv:2009.14553
  [astro-ph.GA]} \BibitemShut {NoStop}%
\bibitem [{\citenamefont {{Ludwig}}(2021)}]{Ludwig}%
  \BibitemOpen
  \bibfield  {author} {\bibinfo {author} {\bibfnamefont {G.~O.}\ \bibnamefont
  {{Ludwig}}},\ }\href {\doibase 10.1140/epjc/s10052-021-08967-3} {\bibfield
  {journal} {\bibinfo  {journal} {European Physical Journal C}\ }\textbf
  {\bibinfo {volume} {81}},\ \bibinfo {eid} {186} (\bibinfo {year}
  {2021})}\BibitemShut {NoStop}%
\bibitem [{\citenamefont {Ruggiero}\ \emph {et~al.}(2022)\citenamefont
  {Ruggiero}, \citenamefont {Ortolan},\ and\ \citenamefont
  {Speake}}]{Ruggiero:2021lpf}%
  \BibitemOpen
  \bibfield  {author} {\bibinfo {author} {\bibfnamefont {M.~L.}\ \bibnamefont
  {Ruggiero}}, \bibinfo {author} {\bibfnamefont {A.}~\bibnamefont {Ortolan}}, \
  and\ \bibinfo {author} {\bibfnamefont {C.~C.}\ \bibnamefont {Speake}},\
  }\href {\doibase 10.1088/1361-6382/ac9949} {\bibfield  {journal} {\bibinfo
  {journal} {Class. Quant. Grav.}\ }\textbf {\bibinfo {volume} {39}},\ \bibinfo
  {pages} {225015} (\bibinfo {year} {2022})},\ \Eprint
  {http://arxiv.org/abs/2112.08290} {arXiv:2112.08290 [gr-qc]} \BibitemShut
  {NoStop}%
\bibitem [{\citenamefont {Srivastava}\ \emph {et~al.}(2023)\citenamefont
  {Srivastava}, \citenamefont {Immirzi}, \citenamefont {Swain}, \citenamefont
  {Panella},\ and\ \citenamefont {Pacetti}}]{Srivastava:2022eza}%
  \BibitemOpen
  \bibfield  {author} {\bibinfo {author} {\bibfnamefont {Y.}~\bibnamefont
  {Srivastava}}, \bibinfo {author} {\bibfnamefont {G.}~\bibnamefont {Immirzi}},
  \bibinfo {author} {\bibfnamefont {J.}~\bibnamefont {Swain}}, \bibinfo
  {author} {\bibfnamefont {O.}~\bibnamefont {Panella}}, \ and\ \bibinfo
  {author} {\bibfnamefont {S.}~\bibnamefont {Pacetti}},\ }\href {\doibase
  10.1140/epjc/s10052-022-11031-3} {\bibfield  {journal} {\bibinfo  {journal}
  {Eur. Phys. J. C}\ }\textbf {\bibinfo {volume} {83}},\ \bibinfo {pages} {100}
  (\bibinfo {year} {2023})},\ \Eprint {http://arxiv.org/abs/2207.04279}
  {arXiv:2207.04279 [gr-qc]} \BibitemShut {NoStop}%
\bibitem [{\citenamefont {Ciotti}(2022)}]{Ciotti_2022}%
  \BibitemOpen
  \bibfield  {author} {\bibinfo {author} {\bibfnamefont {L.}~\bibnamefont
  {Ciotti}},\ }\href {\doibase 10.3847/1538-4357/ac82b3} {\bibfield  {journal}
  {\bibinfo  {journal} {The Astrophysical Journal}\ }\textbf {\bibinfo {volume}
  {936}},\ \bibinfo {pages} {180} (\bibinfo {year} {2022})}\BibitemShut
  {NoStop}%
\bibitem [{\citenamefont {Astesiano}\ and\ \citenamefont
  {Ruggiero}(2022{\natexlab{b}})}]{Astesiano:2022ghr}%
  \BibitemOpen
  \bibfield  {author} {\bibinfo {author} {\bibfnamefont {D.}~\bibnamefont
  {Astesiano}}\ and\ \bibinfo {author} {\bibfnamefont {M.~L.}\ \bibnamefont
  {Ruggiero}},\ }\href {\doibase 10.1103/PhysRevD.106.L121501} {\bibfield
  {journal} {\bibinfo  {journal} {Phys. Rev. D}\ }\textbf {\bibinfo {volume}
  {106}},\ \bibinfo {pages} {L121501} (\bibinfo {year} {2022}{\natexlab{b}})},\
  \Eprint {http://arxiv.org/abs/2211.11815} {arXiv:2211.11815 [gr-qc]}
  \BibitemShut {NoStop}%
\bibitem [{\citenamefont {Lasenby}\ \emph {et~al.}(2023)\citenamefont
  {Lasenby}, \citenamefont {Hobson},\ and\ \citenamefont
  {Barker}}]{Lasenby:2023zdo}%
  \BibitemOpen
  \bibfield  {author} {\bibinfo {author} {\bibfnamefont {A.~N.}\ \bibnamefont
  {Lasenby}}, \bibinfo {author} {\bibfnamefont {M.~P.}\ \bibnamefont {Hobson}},
  \ and\ \bibinfo {author} {\bibfnamefont {W.~E.~V.}\ \bibnamefont {Barker}},\
  }\href@noop {} {\  (\bibinfo {year} {2023})},\ \Eprint
  {http://arxiv.org/abs/2303.06115} {arXiv:2303.06115 [gr-qc]} \BibitemShut
  {NoStop}%
\bibitem [{\citenamefont {Ramos-Caro}\ \emph {et~al.}(2012)\citenamefont
  {Ramos-Caro}, \citenamefont {Agon},\ and\ \citenamefont
  {Pedraza}}]{Ramos-Caro:2012ren}%
  \BibitemOpen
  \bibfield  {author} {\bibinfo {author} {\bibfnamefont {J.}~\bibnamefont
  {Ramos-Caro}}, \bibinfo {author} {\bibfnamefont {C.~A.}\ \bibnamefont
  {Agon}}, \ and\ \bibinfo {author} {\bibfnamefont {J.~F.}\ \bibnamefont
  {Pedraza}},\ }\href {\doibase 10.1103/PhysRevD.86.043008} {\bibfield
  {journal} {\bibinfo  {journal} {Phys. Rev. D}\ }\textbf {\bibinfo {volume}
  {86}},\ \bibinfo {pages} {043008} (\bibinfo {year} {2012})},\ \Eprint
  {http://arxiv.org/abs/1206.5804} {arXiv:1206.5804 [gr-qc]} \BibitemShut
  {NoStop}%
\bibitem [{\citenamefont {{Lobodzinski}}(2014)}]{2014arXiv1406.6082L}%
  \BibitemOpen
  \bibfield  {author} {\bibinfo {author} {\bibfnamefont {B.}~\bibnamefont
  {{Lobodzinski}}},\ }\href@noop {} {\bibfield  {journal} {\bibinfo  {journal}
  {arXiv e-prints}\ ,\ \bibinfo {eid} {arXiv:1406.6082}} (\bibinfo {year}
  {2014})},\ \Eprint {http://arxiv.org/abs/1406.6082} {arXiv:1406.6082
  [astro-ph.GA]} \BibitemShut {NoStop}%
\bibitem [{\citenamefont {Astesiano}(2022)}]{Astesiano:2022gph}%
  \BibitemOpen
  \bibfield  {author} {\bibinfo {author} {\bibfnamefont {D.}~\bibnamefont
  {Astesiano}},\ }\href {\doibase 10.1007/s10714-022-02947-y} {\bibfield
  {journal} {\bibinfo  {journal} {Gen. Rel. Grav.}\ }\textbf {\bibinfo {volume}
  {54}},\ \bibinfo {pages} {63} (\bibinfo {year} {2022})},\ \Eprint
  {http://arxiv.org/abs/2201.03959} {arXiv:2201.03959 [gr-qc]} \BibitemShut
  {NoStop}%
\bibitem [{\citenamefont {{Sebens}}(2018)}]{2018arXiv181110602S}%
  \BibitemOpen
  \bibfield  {author} {\bibinfo {author} {\bibfnamefont {C.~T.}\ \bibnamefont
  {{Sebens}}},\ }\href {\doibase 10.48550/arXiv.1811.10602} {\bibfield
  {journal} {\bibinfo  {journal} {arXiv e-prints}\ ,\ \bibinfo {eid}
  {arXiv:1811.10602}} (\bibinfo {year} {2018})},\ \Eprint
  {http://arxiv.org/abs/1811.10602} {arXiv:1811.10602 [physics.class-ph]}
  \BibitemShut {NoStop}%
\bibitem [{\citenamefont {Abbott}\ \emph {et~al.}(2016)\citenamefont {Abbott},
  \citenamefont {Abbott}, \citenamefont {Abbott}, \citenamefont {Abernathy},
  \citenamefont {Acernese}, \citenamefont {Ackley}, \citenamefont {Adams},
  \citenamefont {Adams}, \citenamefont {Addesso}, \citenamefont {Adhikari}
  \emph {et~al.}}]{abbott2016observation}%
  \BibitemOpen
  \bibfield  {author} {\bibinfo {author} {\bibfnamefont {B.~P.}\ \bibnamefont
  {Abbott}}, \bibinfo {author} {\bibfnamefont {R.}~\bibnamefont {Abbott}},
  \bibinfo {author} {\bibfnamefont {T.}~\bibnamefont {Abbott}}, \bibinfo
  {author} {\bibfnamefont {M.}~\bibnamefont {Abernathy}}, \bibinfo {author}
  {\bibfnamefont {F.}~\bibnamefont {Acernese}}, \bibinfo {author}
  {\bibfnamefont {K.}~\bibnamefont {Ackley}}, \bibinfo {author} {\bibfnamefont
  {C.}~\bibnamefont {Adams}}, \bibinfo {author} {\bibfnamefont
  {T.}~\bibnamefont {Adams}}, \bibinfo {author} {\bibfnamefont
  {P.}~\bibnamefont {Addesso}}, \bibinfo {author} {\bibfnamefont
  {R.}~\bibnamefont {Adhikari}},  \emph {et~al.},\ }\href@noop {} {\bibfield
  {journal} {\bibinfo  {journal} {Physical review letters}\ }\textbf {\bibinfo
  {volume} {116}},\ \bibinfo {pages} {061102} (\bibinfo {year}
  {2016})}\BibitemShut {NoStop}%
\bibitem [{\citenamefont {Ruggiero}(2021{\natexlab{b}})}]{Ruggiero:2021qnu}%
  \BibitemOpen
  \bibfield  {author} {\bibinfo {author} {\bibfnamefont {M.~L.}\ \bibnamefont
  {Ruggiero}},\ }\href {\doibase 10.1119/10.0003513} {\bibfield  {journal}
  {\bibinfo  {journal} {Am. J. Phys.}\ }\textbf {\bibinfo {volume} {89}},\
  \bibinfo {pages} {639} (\bibinfo {year} {2021}{\natexlab{b}})},\ \Eprint
  {http://arxiv.org/abs/2101.06746} {arXiv:2101.06746 [gr-qc]} \BibitemShut
  {NoStop}%
\bibitem [{\citenamefont {Ruggiero}\ and\ \citenamefont
  {Ortolan}(2020{\natexlab{b}})}]{Ruggiero_2020b}%
  \BibitemOpen
  \bibfield  {author} {\bibinfo {author} {\bibfnamefont {M.~L.}\ \bibnamefont
  {Ruggiero}}\ and\ \bibinfo {author} {\bibfnamefont {A.}~\bibnamefont
  {Ortolan}},\ }\href {\doibase 10.1103/PhysRevD.102.101501} {\bibfield
  {journal} {\bibinfo  {journal} {Phys. Rev. D}\ }\textbf {\bibinfo {volume}
  {102}},\ \bibinfo {pages} {101501} (\bibinfo {year}
  {2020}{\natexlab{b}})}\BibitemShut {NoStop}%
\bibitem [{\citenamefont {Ruggiero}\ and\ \citenamefont
  {Ortolan}(2021)}]{Ruggiero:2021dri}%
  \BibitemOpen
  \bibfield  {author} {\bibinfo {author} {\bibfnamefont {M.~L.}\ \bibnamefont
  {Ruggiero}}\ and\ \bibinfo {author} {\bibfnamefont {A.}~\bibnamefont
  {Ortolan}},\ }in\ \href {\doibase 10.1142/9789811269776_0335} {\emph
  {\bibinfo {booktitle} {{16th Marcel Grossmann Meeting on~Recent Developments
  in Theoretical and Experimental General Relativity, Astrophysics and
  Relativistic Field Theories}}}}\ (\bibinfo {year} {2021})\ \Eprint
  {http://arxiv.org/abs/2111.00502} {arXiv:2111.00502 [gr-qc]} \BibitemShut
  {NoStop}%
\bibitem [{\citenamefont {Ramos}\ and\ \citenamefont
  {Mashhoon}(2006)}]{PhysRevD.73.084003}%
  \BibitemOpen
  \bibfield  {author} {\bibinfo {author} {\bibfnamefont {J.}~\bibnamefont
  {Ramos}}\ and\ \bibinfo {author} {\bibfnamefont {B.}~\bibnamefont
  {Mashhoon}},\ }\href {\doibase 10.1103/PhysRevD.73.084003} {\bibfield
  {journal} {\bibinfo  {journal} {Phys. Rev. D}\ }\textbf {\bibinfo {volume}
  {73}},\ \bibinfo {pages} {084003} (\bibinfo {year} {2006})}\BibitemShut
  {NoStop}%
\bibitem [{\citenamefont {{Baskaran}}\ and\ \citenamefont
  {{Grishchuk}}(2004)}]{baskaran}%
  \BibitemOpen
  \bibfield  {author} {\bibinfo {author} {\bibfnamefont {D.}~\bibnamefont
  {{Baskaran}}}\ and\ \bibinfo {author} {\bibfnamefont {L.~P.}\ \bibnamefont
  {{Grishchuk}}},\ }\href {\doibase 10.1088/0264-9381/21/17/003} {\bibfield
  {journal} {\bibinfo  {journal} {Classical and Quantum Gravity}\ }\textbf
  {\bibinfo {volume} {21}},\ \bibinfo {pages} {4041} (\bibinfo {year}
  {2004})},\ \Eprint {http://arxiv.org/abs/gr-qc/0309058} {arXiv:gr-qc/0309058
  [gr-qc]} \BibitemShut {NoStop}%
\bibitem [{\citenamefont {Iorio}\ and\ \citenamefont
  {Corda}(2009)}]{Iorio:2009gn}%
  \BibitemOpen
  \bibfield  {author} {\bibinfo {author} {\bibfnamefont {L.}~\bibnamefont
  {Iorio}}\ and\ \bibinfo {author} {\bibfnamefont {C.}~\bibnamefont {Corda}},\
  }\href {\doibase 10.1063/1.3241239} {\bibfield  {journal} {\bibinfo
  {journal} {AIP Conf. Proc.}\ }\textbf {\bibinfo {volume} {1168}},\ \bibinfo
  {pages} {1072} (\bibinfo {year} {2009})},\ \Eprint
  {http://arxiv.org/abs/0907.2154} {arXiv:0907.2154 [gr-qc]} \BibitemShut
  {NoStop}%
\bibitem [{\citenamefont {{Capozziello}}\ \emph {et~al.}(2008)\citenamefont
  {{Capozziello}}, \citenamefont {{Corda}},\ and\ \citenamefont {{de
  Laurentis}}}]{capozziello1}%
  \BibitemOpen
  \bibfield  {author} {\bibinfo {author} {\bibfnamefont {S.}~\bibnamefont
  {{Capozziello}}}, \bibinfo {author} {\bibfnamefont {C.}~\bibnamefont
  {{Corda}}}, \ and\ \bibinfo {author} {\bibfnamefont {M.~F.}\ \bibnamefont
  {{de Laurentis}}},\ }\href {\doibase 10.1016/j.physletb.2008.10.001}
  {\bibfield  {journal} {\bibinfo  {journal} {Physics Letters B}\ }\textbf
  {\bibinfo {volume} {669}},\ \bibinfo {pages} {255} (\bibinfo {year}
  {2008})},\ \Eprint {http://arxiv.org/abs/0812.2272} {arXiv:0812.2272
  [astro-ph]} \BibitemShut {NoStop}%
\bibitem [{\citenamefont {{Capozziello}}\ and\ \citenamefont {{de
  Laurentis}}(2011)}]{capozziello2}%
  \BibitemOpen
  \bibfield  {author} {\bibinfo {author} {\bibfnamefont {S.}~\bibnamefont
  {{Capozziello}}}\ and\ \bibinfo {author} {\bibfnamefont {M.}~\bibnamefont
  {{de Laurentis}}},\ }\href {\doibase 10.1016/j.physrep.2011.09.003}
  {\bibfield  {journal} {\bibinfo  {journal} {Physics Reports}\ }\textbf
  {\bibinfo {volume} {509}},\ \bibinfo {pages} {167} (\bibinfo {year}
  {2011})},\ \Eprint {http://arxiv.org/abs/1108.6266} {arXiv:1108.6266 [gr-qc]}
  \BibitemShut {NoStop}%
\bibitem [{\citenamefont {{Corda}}(2011)}]{corda1}%
  \BibitemOpen
  \bibfield  {author} {\bibinfo {author} {\bibfnamefont {C.}~\bibnamefont
  {{Corda}}},\ }\href {\doibase 10.1103/PhysRevD.83.062002} {\bibfield
  {journal} {\bibinfo  {journal} {\prd}\ }\textbf {\bibinfo {volume} {83}},\
  \bibinfo {eid} {062002} (\bibinfo {year} {2011})},\ \Eprint
  {http://arxiv.org/abs/1102.0619} {arXiv:1102.0619 [gr-qc]} \BibitemShut
  {NoStop}%
\bibitem [{\citenamefont {{Corda}}\ \emph {et~al.}(2010)\citenamefont
  {{Corda}}, \citenamefont {{Ali}},\ and\ \citenamefont {{Cafaro}}}]{corda2}%
  \BibitemOpen
  \bibfield  {author} {\bibinfo {author} {\bibfnamefont {C.}~\bibnamefont
  {{Corda}}}, \bibinfo {author} {\bibfnamefont {S.~A.}\ \bibnamefont {{Ali}}},
  \ and\ \bibinfo {author} {\bibfnamefont {C.}~\bibnamefont {{Cafaro}}},\
  }\href {\doibase 10.1142/S0218271810018219} {\bibfield  {journal} {\bibinfo
  {journal} {International Journal of Modern Physics D}\ }\textbf {\bibinfo
  {volume} {19}},\ \bibinfo {pages} {2095} (\bibinfo {year} {2010})},\ \Eprint
  {http://arxiv.org/abs/0902.0093} {arXiv:0902.0093 [gr-qc]} \BibitemShut
  {NoStop}%
\bibitem [{\citenamefont {Barcelo}\ \emph {et~al.}(2005)\citenamefont
  {Barcelo}, \citenamefont {Liberati},\ and\ \citenamefont
  {Visser}}]{Barcelo:2005fc}%
  \BibitemOpen
  \bibfield  {author} {\bibinfo {author} {\bibfnamefont {C.}~\bibnamefont
  {Barcelo}}, \bibinfo {author} {\bibfnamefont {S.}~\bibnamefont {Liberati}}, \
  and\ \bibinfo {author} {\bibfnamefont {M.}~\bibnamefont {Visser}},\ }\href
  {\doibase 10.12942/lrr-2005-12} {\bibfield  {journal} {\bibinfo  {journal}
  {Living Rev. Rel.}\ }\textbf {\bibinfo {volume} {8}},\ \bibinfo {pages} {12}
  (\bibinfo {year} {2005})},\ \Eprint {http://arxiv.org/abs/gr-qc/0505065}
  {arXiv:gr-qc/0505065} \BibitemShut {NoStop}%
\bibitem [{\citenamefont {Unruh}(1981)}]{PhysRevLett.46.1351}%
  \BibitemOpen
  \bibfield  {author} {\bibinfo {author} {\bibfnamefont {W.~G.}\ \bibnamefont
  {Unruh}},\ }\href {\doibase 10.1103/PhysRevLett.46.1351} {\bibfield
  {journal} {\bibinfo  {journal} {Phys. Rev. Lett.}\ }\textbf {\bibinfo
  {volume} {46}},\ \bibinfo {pages} {1351} (\bibinfo {year}
  {1981})}\BibitemShut {NoStop}%
\bibitem [{\citenamefont {Visser}(1998)}]{Visser:1998qn}%
  \BibitemOpen
  \bibfield  {author} {\bibinfo {author} {\bibfnamefont {M.}~\bibnamefont
  {Visser}},\ }in\ \href@noop {} {\emph {\bibinfo {booktitle} {{Advanced School
  on Cosmology and Particle Physics (ASCPP 98)}}}}\ (\bibinfo {year} {1998})\
  \Eprint {http://arxiv.org/abs/gr-qc/9901047} {arXiv:gr-qc/9901047}
  \BibitemShut {NoStop}%
\bibitem [{\citenamefont {{Puthoff}}(2008)}]{2008arXiv0808.3404P}%
  \BibitemOpen
  \bibfield  {author} {\bibinfo {author} {\bibfnamefont {H.~E.}\ \bibnamefont
  {{Puthoff}}},\ }\href@noop {} {\bibfield  {journal} {\bibinfo  {journal}
  {arXiv e-prints}\ ,\ \bibinfo {eid} {arXiv:0808.3404}} (\bibinfo {year}
  {2008})},\ \Eprint {http://arxiv.org/abs/0808.3404} {arXiv:0808.3404
  [physics.gen-ph]} \BibitemShut {NoStop}%
\bibitem [{\citenamefont {Kivotides}(2021)}]{Kivotides:2020nue}%
  \BibitemOpen
  \bibfield  {author} {\bibinfo {author} {\bibfnamefont {D.}~\bibnamefont
  {Kivotides}},\ }\href {\doibase 10.1142/S0129183121501291} {\bibfield
  {journal} {\bibinfo  {journal} {Int. J. Mod. Phys. C}\ }\textbf {\bibinfo
  {volume} {32}},\ \bibinfo {pages} {2150129} (\bibinfo {year} {2021})},\
  \Eprint {http://arxiv.org/abs/2005.10871} {arXiv:2005.10871 [gr-qc]}
  \BibitemShut {NoStop}%
\bibitem [{\citenamefont {Chakraborty}\ \emph {et~al.}(2018)\citenamefont
  {Chakraborty}, \citenamefont {Ganguly},\ and\ \citenamefont
  {Majumdar}}]{Chakraborty:2015ioa}%
  \BibitemOpen
  \bibfield  {author} {\bibinfo {author} {\bibfnamefont {C.}~\bibnamefont
  {Chakraborty}}, \bibinfo {author} {\bibfnamefont {O.}~\bibnamefont
  {Ganguly}}, \ and\ \bibinfo {author} {\bibfnamefont {P.}~\bibnamefont
  {Majumdar}},\ }\href {\doibase 10.1002/andp.201700231} {\bibfield  {journal}
  {\bibinfo  {journal} {Annalen Phys.}\ }\textbf {\bibinfo {volume} {530}},\
  \bibinfo {pages} {1700231} (\bibinfo {year} {2018})},\ \Eprint
  {http://arxiv.org/abs/1510.01436} {arXiv:1510.01436 [gr-qc]} \BibitemShut
  {NoStop}%
\bibitem [{\citenamefont {Banerjee}\ \emph {et~al.}(2018)\citenamefont
  {Banerjee}, \citenamefont {Koley},\ and\ \citenamefont
  {Majumdar}}]{Banerjee:2018wev}%
  \BibitemOpen
  \bibfield  {author} {\bibinfo {author} {\bibfnamefont {A.}~\bibnamefont
  {Banerjee}}, \bibinfo {author} {\bibfnamefont {R.}~\bibnamefont {Koley}}, \
  and\ \bibinfo {author} {\bibfnamefont {P.}~\bibnamefont {Majumdar}},\
  }\href@noop {} {\  (\bibinfo {year} {2018})},\ \Eprint
  {http://arxiv.org/abs/1808.01828} {arXiv:1808.01828 [gr-qc]} \BibitemShut
  {NoStop}%
\bibitem [{\citenamefont {Ratzel}\ \emph {et~al.}(2016)\citenamefont {Ratzel},
  \citenamefont {Wilkens},\ and\ \citenamefont {Menzel}}]{Ratzel_2016}%
  \BibitemOpen
  \bibfield  {author} {\bibinfo {author} {\bibfnamefont {D.}~\bibnamefont
  {Ratzel}}, \bibinfo {author} {\bibfnamefont {M.}~\bibnamefont {Wilkens}}, \
  and\ \bibinfo {author} {\bibfnamefont {R.}~\bibnamefont {Menzel}},\ }\href
  {\doibase 10.1088/1367-2630/18/2/023009} {\bibfield  {journal} {\bibinfo
  {journal} {New Journal of Physics}\ }\textbf {\bibinfo {volume} {18}},\
  \bibinfo {pages} {023009} (\bibinfo {year} {2016})}\BibitemShut {NoStop}%
\bibitem [{\citenamefont {Spengler}\ \emph {et~al.}(2022)\citenamefont
  {Spengler}, \citenamefont {R‚{\`a}{\"o}{\S}tzel},\ and\ \citenamefont
  {Braun}}]{Spengler_2022}%
  \BibitemOpen
  \bibfield  {author} {\bibinfo {author} {\bibfnamefont {F.}~\bibnamefont
  {Spengler}}, \bibinfo {author} {\bibfnamefont {D.}~\bibnamefont
  {R‚{\`a}{\"o}{\S}tzel}}, \ and\ \bibinfo {author} {\bibfnamefont
  {D.}~\bibnamefont {Braun}},\ }\href {\doibase 10.1088/1367-2630/ac5372}
  {\bibfield  {journal} {\bibinfo  {journal} {New Journal of Physics}\ }\textbf
  {\bibinfo {volume} {24}},\ \bibinfo {pages} {053021} (\bibinfo {year}
  {2022})}\BibitemShut {NoStop}%
\bibitem [{\citenamefont {Padgett}(2017)}]{Padgett:17}%
  \BibitemOpen
  \bibfield  {author} {\bibinfo {author} {\bibfnamefont {M.~J.}\ \bibnamefont
  {Padgett}},\ }\href {\doibase 10.1364/OE.25.011265} {\bibfield  {journal}
  {\bibinfo  {journal} {Opt. Express}\ }\textbf {\bibinfo {volume} {25}},\
  \bibinfo {pages} {11265} (\bibinfo {year} {2017})}\BibitemShut {NoStop}%
\bibitem [{\citenamefont {Strohaber}(2013)}]{Strohaber:2011aa}%
  \BibitemOpen
  \bibfield  {author} {\bibinfo {author} {\bibfnamefont {J.}~\bibnamefont
  {Strohaber}},\ }\href {\doibase 10.1007/s10714-013-1596-8} {\bibfield
  {journal} {\bibinfo  {journal} {Gen. Rel. Grav.}\ }\textbf {\bibinfo {volume}
  {45}},\ \bibinfo {pages} {2457} (\bibinfo {year} {2013})},\ \Eprint
  {http://arxiv.org/abs/1112.3414} {arXiv:1112.3414 [gr-qc]} \BibitemShut
  {NoStop}%
\bibitem [{\citenamefont {{Schneiter}}\ \emph {et~al.}(2018)\citenamefont
  {{Schneiter}}, \citenamefont {{R{\"a}tzel}},\ and\ \citenamefont
  {{Braun}}}]{2018CQGra..35s5007S}%
  \BibitemOpen
  \bibfield  {author} {\bibinfo {author} {\bibfnamefont {F.}~\bibnamefont
  {{Schneiter}}}, \bibinfo {author} {\bibfnamefont {D.}~\bibnamefont
  {{R{\"a}tzel}}}, \ and\ \bibinfo {author} {\bibfnamefont {D.}~\bibnamefont
  {{Braun}}},\ }\href {\doibase 10.1088/1361-6382/aadc81} {\bibfield  {journal}
  {\bibinfo  {journal} {Classical and Quantum Gravity}\ }\textbf {\bibinfo
  {volume} {35}},\ \bibinfo {eid} {195007} (\bibinfo {year} {2018})},\ \Eprint
  {http://arxiv.org/abs/1804.08706} {arXiv:1804.08706 [gr-qc]} \BibitemShut
  {NoStop}%
\bibitem [{\citenamefont {{Schneiter}}\ \emph
  {et~al.}(2019{\natexlab{a}})\citenamefont {{Schneiter}}, \citenamefont
  {{R{\"a}tzel}},\ and\ \citenamefont {{Braun}}}]{2019CQGra..36t5007S}%
  \BibitemOpen
  \bibfield  {author} {\bibinfo {author} {\bibfnamefont {F.}~\bibnamefont
  {{Schneiter}}}, \bibinfo {author} {\bibfnamefont {D.}~\bibnamefont
  {{R{\"a}tzel}}}, \ and\ \bibinfo {author} {\bibfnamefont {D.}~\bibnamefont
  {{Braun}}},\ }\href {\doibase 10.1088/1361-6382/ab3523} {\bibfield  {journal}
  {\bibinfo  {journal} {Classical and Quantum Gravity}\ }\textbf {\bibinfo
  {volume} {36}},\ \bibinfo {eid} {205007} (\bibinfo {year}
  {2019}{\natexlab{a}})},\ \Eprint {http://arxiv.org/abs/1812.04505}
  {arXiv:1812.04505 [gr-qc]} \BibitemShut {NoStop}%
\bibitem [{\citenamefont {{Schneiter}}\ \emph
  {et~al.}(2019{\natexlab{b}})\citenamefont {{Schneiter}}, \citenamefont
  {{R{\"a}tzel}},\ and\ \citenamefont {{Braun}}}]{2019CQGra..36k9501S}%
  \BibitemOpen
  \bibfield  {author} {\bibinfo {author} {\bibfnamefont {F.}~\bibnamefont
  {{Schneiter}}}, \bibinfo {author} {\bibfnamefont {D.}~\bibnamefont
  {{R{\"a}tzel}}}, \ and\ \bibinfo {author} {\bibfnamefont {D.}~\bibnamefont
  {{Braun}}},\ }\href {\doibase 10.1088/1361-6382/ab1ae3} {\bibfield  {journal}
  {\bibinfo  {journal} {Classical and Quantum Gravity}\ }\textbf {\bibinfo
  {volume} {36}},\ \bibinfo {eid} {119501} (\bibinfo {year}
  {2019}{\natexlab{b}})}\BibitemShut {NoStop}%
\bibitem [{\citenamefont {Page}(2005)}]{Page:2004xp}%
  \BibitemOpen
  \bibfield  {author} {\bibinfo {author} {\bibfnamefont {D.~N.}\ \bibnamefont
  {Page}},\ }\href {\doibase 10.1088/1367-2630/7/1/203} {\bibfield  {journal}
  {\bibinfo  {journal} {New J. Phys.}\ }\textbf {\bibinfo {volume} {7}},\
  \bibinfo {pages} {203} (\bibinfo {year} {2005})},\ \Eprint
  {http://arxiv.org/abs/hep-th/0409024} {arXiv:hep-th/0409024} \BibitemShut
  {NoStop}%
\bibitem [{\citenamefont {Kiefer}\ and\ \citenamefont
  {Weber}(2005)}]{Kiefer:2004hv}%
  \BibitemOpen
  \bibfield  {author} {\bibinfo {author} {\bibfnamefont {C.}~\bibnamefont
  {Kiefer}}\ and\ \bibinfo {author} {\bibfnamefont {C.}~\bibnamefont {Weber}},\
  }\href {\doibase 10.1002/andp.200410119} {\bibfield  {journal} {\bibinfo
  {journal} {Annalen Phys.}\ }\textbf {\bibinfo {volume} {14}},\ \bibinfo
  {pages} {253} (\bibinfo {year} {2005})},\ \Eprint
  {http://arxiv.org/abs/gr-qc/0408010} {arXiv:gr-qc/0408010} \BibitemShut
  {NoStop}%
\bibitem [{\citenamefont {Hammad}\ \emph {et~al.}(2021)\citenamefont {Hammad},
  \citenamefont {Landry},\ and\ \citenamefont {Mathieu}}]{Hammad:2019udg}%
  \BibitemOpen
  \bibfield  {author} {\bibinfo {author} {\bibfnamefont {F.}~\bibnamefont
  {Hammad}}, \bibinfo {author} {\bibfnamefont {A.}~\bibnamefont {Landry}}, \
  and\ \bibinfo {author} {\bibfnamefont {K.}~\bibnamefont {Mathieu}},\ }\href
  {\doibase 10.1142/S0218271821500048} {\bibfield  {journal} {\bibinfo
  {journal} {Int. J. Mod. Phys. D}\ }\textbf {\bibinfo {volume} {30}},\
  \bibinfo {pages} {2150004} (\bibinfo {year} {2021})},\ \Eprint
  {http://arxiv.org/abs/1910.13814} {arXiv:1910.13814 [gr-qc]} \BibitemShut
  {NoStop}%
\bibitem [{\citenamefont {Cong}\ \emph {et~al.}(2021)\citenamefont {Cong},
  \citenamefont {Bicak}, \citenamefont {Kubiznak},\ and\ \citenamefont
  {Mann}}]{Cong:2020xlt}%
  \BibitemOpen
  \bibfield  {author} {\bibinfo {author} {\bibfnamefont {W.}~\bibnamefont
  {Cong}}, \bibinfo {author} {\bibfnamefont {J.}~\bibnamefont {Bicak}},
  \bibinfo {author} {\bibfnamefont {D.}~\bibnamefont {Kubiznak}}, \ and\
  \bibinfo {author} {\bibfnamefont {R.~B.}\ \bibnamefont {Mann}},\ }\href
  {\doibase 10.1103/PhysRevD.103.024027} {\bibfield  {journal} {\bibinfo
  {journal} {Phys. Rev. D}\ }\textbf {\bibinfo {volume} {103}},\ \bibinfo
  {pages} {024027} (\bibinfo {year} {2021})},\ \Eprint
  {http://arxiv.org/abs/2009.10584} {arXiv:2009.10584 [gr-qc]} \BibitemShut
  {NoStop}%
\bibitem [{\citenamefont {Adler}\ and\ \citenamefont
  {Chen}(2010)}]{Adler:2009du}%
  \BibitemOpen
  \bibfield  {author} {\bibinfo {author} {\bibfnamefont {R.~J.}\ \bibnamefont
  {Adler}}\ and\ \bibinfo {author} {\bibfnamefont {P.}~\bibnamefont {Chen}},\
  }\href {\doibase 10.1103/PhysRevD.82.025004} {\bibfield  {journal} {\bibinfo
  {journal} {Phys. Rev. D}\ }\textbf {\bibinfo {volume} {82}},\ \bibinfo
  {pages} {025004} (\bibinfo {year} {2010})},\ \Eprint
  {http://arxiv.org/abs/0912.2814} {arXiv:0912.2814 [gr-qc]} \BibitemShut
  {NoStop}%
\bibitem [{\citenamefont {Adler}\ \emph {et~al.}(2012)\citenamefont {Adler},
  \citenamefont {Chen},\ and\ \citenamefont {Varani}}]{Adler:2011bg}%
  \BibitemOpen
  \bibfield  {author} {\bibinfo {author} {\bibfnamefont {R.~J.}\ \bibnamefont
  {Adler}}, \bibinfo {author} {\bibfnamefont {P.}~\bibnamefont {Chen}}, \ and\
  \bibinfo {author} {\bibfnamefont {E.}~\bibnamefont {Varani}},\ }\href
  {\doibase 10.1103/PhysRevD.85.025016} {\bibfield  {journal} {\bibinfo
  {journal} {Phys. Rev. D}\ }\textbf {\bibinfo {volume} {85}},\ \bibinfo
  {pages} {025016} (\bibinfo {year} {2012})},\ \Eprint
  {http://arxiv.org/abs/1108.3859} {arXiv:1108.3859 [gr-qc]} \BibitemShut
  {NoStop}%
\bibitem [{\citenamefont {Bahrami}\ \emph {et~al.}(2014)\citenamefont
  {Bahrami}, \citenamefont {Gro\ss{}ardt}, \citenamefont {Donadi},\ and\
  \citenamefont {Bassi}}]{Bahrami:2014gwa}%
  \BibitemOpen
  \bibfield  {author} {\bibinfo {author} {\bibfnamefont {M.}~\bibnamefont
  {Bahrami}}, \bibinfo {author} {\bibfnamefont {A.}~\bibnamefont
  {Gro\ss{}ardt}}, \bibinfo {author} {\bibfnamefont {S.}~\bibnamefont
  {Donadi}}, \ and\ \bibinfo {author} {\bibfnamefont {A.}~\bibnamefont
  {Bassi}},\ }\href {\doibase 10.1088/1367-2630/16/11/115007} {\bibfield
  {journal} {\bibinfo  {journal} {New J. Phys.}\ }\textbf {\bibinfo {volume}
  {16}},\ \bibinfo {pages} {115007} (\bibinfo {year} {2014})},\ \Eprint
  {http://arxiv.org/abs/1407.4370} {arXiv:1407.4370 [quant-ph]} \BibitemShut
  {NoStop}%
\bibitem [{\citenamefont {Ruffini}\ and\ \citenamefont
  {Bonazzola}(1969)}]{PhysRev.187.1767}%
  \BibitemOpen
  \bibfield  {author} {\bibinfo {author} {\bibfnamefont {R.}~\bibnamefont
  {Ruffini}}\ and\ \bibinfo {author} {\bibfnamefont {S.}~\bibnamefont
  {Bonazzola}},\ }\href {\doibase 10.1103/PhysRev.187.1767} {\bibfield
  {journal} {\bibinfo  {journal} {Phys. Rev.}\ }\textbf {\bibinfo {volume}
  {187}},\ \bibinfo {pages} {1767} (\bibinfo {year} {1969})}\BibitemShut
  {NoStop}%
\bibitem [{\citenamefont {Brizuela}\ and\ \citenamefont
  {Duran-Cabac\'es}(2022)}]{Brizuela:2022fcb}%
  \BibitemOpen
  \bibfield  {author} {\bibinfo {author} {\bibfnamefont {D.}~\bibnamefont
  {Brizuela}}\ and\ \bibinfo {author} {\bibfnamefont {A.}~\bibnamefont
  {Duran-Cabac\'es}},\ }\href {\doibase 10.1103/PhysRevD.106.124038} {\bibfield
   {journal} {\bibinfo  {journal} {Phys. Rev. D}\ }\textbf {\bibinfo {volume}
  {106}},\ \bibinfo {pages} {124038} (\bibinfo {year} {2022})},\ \Eprint
  {http://arxiv.org/abs/2210.06195} {arXiv:2210.06195 [gr-qc]} \BibitemShut
  {NoStop}%
\bibitem [{\citenamefont {Manfredi}(2015)}]{Manfredi:2014hna}%
  \BibitemOpen
  \bibfield  {author} {\bibinfo {author} {\bibfnamefont {G.}~\bibnamefont
  {Manfredi}},\ }\href {\doibase 10.1007/s10714-014-1846-4} {\bibfield
  {journal} {\bibinfo  {journal} {Gen. Rel. Grav.}\ }\textbf {\bibinfo {volume}
  {47}},\ \bibinfo {pages} {1} (\bibinfo {year} {2015})},\ \Eprint
  {http://arxiv.org/abs/1412.1662} {arXiv:1412.1662 [gr-qc]} \BibitemShut
  {NoStop}%
\bibitem [{\citenamefont {Zhao}\ \emph {et~al.}(2017)\citenamefont {Zhao},
  \citenamefont {Faizal},\ and\ \citenamefont {Zaz}}]{Zhao:2017xjj}%
  \BibitemOpen
  \bibfield  {author} {\bibinfo {author} {\bibfnamefont {Q.}~\bibnamefont
  {Zhao}}, \bibinfo {author} {\bibfnamefont {M.}~\bibnamefont {Faizal}}, \ and\
  \bibinfo {author} {\bibfnamefont {Z.}~\bibnamefont {Zaz}},\ }\href {\doibase
  10.1016/j.physletb.2017.01.029} {\bibfield  {journal} {\bibinfo  {journal}
  {Phys. Lett. B}\ }\textbf {\bibinfo {volume} {770}},\ \bibinfo {pages} {564}
  (\bibinfo {year} {2017})},\ \Eprint {http://arxiv.org/abs/1707.00636}
  {arXiv:1707.00636 [hep-th]} \BibitemShut {NoStop}%
\bibitem [{\citenamefont {Lanzagorta}(2012)}]{Lanzagorta:2012tn}%
  \BibitemOpen
  \bibfield  {author} {\bibinfo {author} {\bibfnamefont {M.}~\bibnamefont
  {Lanzagorta}},\ }\href@noop {} {\  (\bibinfo {year} {2012})},\ \Eprint
  {http://arxiv.org/abs/1212.2200} {arXiv:1212.2200 [quant-ph]} \BibitemShut
  {NoStop}%
\bibitem [{\citenamefont {Lanzagorta}\ and\ \citenamefont
  {Salgado}(2016)}]{Lanzagorta:2016mhh}%
  \BibitemOpen
  \bibfield  {author} {\bibinfo {author} {\bibfnamefont {M.}~\bibnamefont
  {Lanzagorta}}\ and\ \bibinfo {author} {\bibfnamefont {M.}~\bibnamefont
  {Salgado}},\ }\href {\doibase 10.1088/0264-9381/33/10/105013} {\bibfield
  {journal} {\bibinfo  {journal} {Class. Quant. Grav.}\ }\textbf {\bibinfo
  {volume} {33}},\ \bibinfo {pages} {105013} (\bibinfo {year}
  {2016})}\BibitemShut {NoStop}%
\bibitem [{\citenamefont {Jesus}\ \emph {et~al.}(2022)\citenamefont {Jesus},
  \citenamefont {Souza}, \citenamefont {Santos},\ and\ \citenamefont
  {Khanna}}]{Jesus:2022ywg}%
  \BibitemOpen
  \bibfield  {author} {\bibinfo {author} {\bibfnamefont {W.~D.~R.}\
  \bibnamefont {Jesus}}, \bibinfo {author} {\bibfnamefont {P.~R.~A.}\
  \bibnamefont {Souza}}, \bibinfo {author} {\bibfnamefont {A.~F.}\ \bibnamefont
  {Santos}}, \ and\ \bibinfo {author} {\bibfnamefont {F.~C.}\ \bibnamefont
  {Khanna}},\ }\href {\doibase 10.1140/epjp/s13360-022-02479-z} {\bibfield
  {journal} {\bibinfo  {journal} {Eur. Phys. J. Plus}\ }\textbf {\bibinfo
  {volume} {137}},\ \bibinfo {pages} {260} (\bibinfo {year} {2022})},\ \Eprint
  {http://arxiv.org/abs/2202.06778} {arXiv:2202.06778 [hep-th]} \BibitemShut
  {NoStop}%
\bibitem [{\citenamefont {Kim}(2022)}]{Kim:2022iub}%
  \BibitemOpen
  \bibfield  {author} {\bibinfo {author} {\bibfnamefont {J.-W.}\ \bibnamefont
  {Kim}},\ }\href {\doibase 10.1103/PhysRevD.106.L081901} {\bibfield  {journal}
  {\bibinfo  {journal} {Phys. Rev. D}\ }\textbf {\bibinfo {volume} {106}},\
  \bibinfo {pages} {L081901} (\bibinfo {year} {2022})},\ \Eprint
  {http://arxiv.org/abs/2207.04970} {arXiv:2207.04970 [hep-th]} \BibitemShut
  {NoStop}%
\bibitem [{\citenamefont {{Bonnor}}(1964)}]{olbers}%
  \BibitemOpen
  \bibfield  {author} {\bibinfo {author} {\bibfnamefont {W.~B.}\ \bibnamefont
  {{Bonnor}}},\ }\href {\doibase 10.1093/mnras/128.1.33} {\bibfield  {journal}
  {\bibinfo  {journal} {\mnras}\ }\textbf {\bibinfo {volume} {128}},\ \bibinfo
  {pages} {33} (\bibinfo {year} {1964})}\BibitemShut {NoStop}%
\bibitem [{\citenamefont {Mach}(1960)}]{mach}%
  \BibitemOpen
  \bibfield  {author} {\bibinfo {author} {\bibfnamefont {E.}~\bibnamefont
  {Mach}},\ }\href@noop {} {\emph {\bibinfo {title} {The Science of Mechanics:
  A Critical and Historical Account of Its Development}}}\ (\bibinfo
  {publisher} {Open Court, La Salle},\ \bibinfo {year} {1960})\BibitemShut
  {NoStop}%
\bibitem [{\citenamefont {{Vassallo}}\ and\ \citenamefont
  {{Hoefer}}(2019)}]{2019arXiv190110766V}%
  \BibitemOpen
  \bibfield  {author} {\bibinfo {author} {\bibfnamefont {A.}~\bibnamefont
  {{Vassallo}}}\ and\ \bibinfo {author} {\bibfnamefont {C.}~\bibnamefont
  {{Hoefer}}},\ }\href {\doibase 10.48550/arXiv.1901.10766} {\bibfield
  {journal} {\bibinfo  {journal} {arXiv e-prints}\ ,\ \bibinfo {eid}
  {arXiv:1901.10766}} (\bibinfo {year} {2019})},\ \Eprint
  {http://arxiv.org/abs/1901.10766} {arXiv:1901.10766 [physics.hist-ph]}
  \BibitemShut {NoStop}%
\bibitem [{\citenamefont {{Christillin}}\ and\ \citenamefont
  {{Barattini}}(2012)}]{2012arXiv1206.4593C}%
  \BibitemOpen
  \bibfield  {author} {\bibinfo {author} {\bibfnamefont {P.}~\bibnamefont
  {{Christillin}}}\ and\ \bibinfo {author} {\bibfnamefont {L.}~\bibnamefont
  {{Barattini}}},\ }\href {\doibase 10.48550/arXiv.1206.4593} {\bibfield
  {journal} {\bibinfo  {journal} {arXiv e-prints}\ ,\ \bibinfo {eid}
  {arXiv:1206.4593}} (\bibinfo {year} {2012})},\ \Eprint
  {http://arxiv.org/abs/1206.4593} {arXiv:1206.4593 [physics.gen-ph]}
  \BibitemShut {NoStop}%
\bibitem [{\citenamefont {Schmid}(2009)}]{Schmid:2008np}%
  \BibitemOpen
  \bibfield  {author} {\bibinfo {author} {\bibfnamefont {C.}~\bibnamefont
  {Schmid}},\ }\href {\doibase 10.1103/PhysRevD.79.064007} {\bibfield
  {journal} {\bibinfo  {journal} {Phys. Rev. D}\ }\textbf {\bibinfo {volume}
  {79}},\ \bibinfo {pages} {064007} (\bibinfo {year} {2009})},\ \Eprint
  {http://arxiv.org/abs/0801.2907} {arXiv:0801.2907 [astro-ph]} \BibitemShut
  {NoStop}%
\end{thebibliography}

%

\end{document}